\documentclass[longauth,dvipsnames,]{aa}

\usepackage{xcolor}
\usepackage{url}
\usepackage{hyperref}

\usepackage{amsmath,amssymb}
\hypersetup{
    colorlinks=true,
    linkcolor=Blue,
    urlcolor=Magenta,
    citecolor=Blue
}
\usepackage{placeins}
\usepackage[utf8]{inputenc}
\usepackage{natbib}
\usepackage{graphicx}
\usepackage{color}
\usepackage{xcolor}
\usepackage{txfonts}
\usepackage{tabularx}
\usepackage{booktabs}
\usepackage{amsmath}
\usepackage{amssymb}
\usepackage{mathtools}
\usepackage[normalem]{ulem}
\usepackage{tikz}
\usepackage{longtable}
\usepackage{caption}
\usetikzlibrary{arrows,arrows.meta,shapes,decorations.pathmorphing,positioning,intersections,calc,backgrounds}
\tikzstyle{line} = [draw, -latex']
\tikzstyle{line2} = [draw, -|]
\tikzstyle{marginalized}=[black, {Circle[color=black,length=3pt]-latex[]}]
\usepackage{multirow}
\usepackage{bm}

\providecommand{\gmag}{\ensuremath{G}}

\newcommand\gaia{\textit{Gaia~}}

\newcommand\gdrthree{\gaia~DR3~}

\newcommand*{\myalign}[2]{\multicolumn{1}{#1}{#2}}

\providecommand{\kms}{\ensuremath{\rm \,km\,s^{-1}}\xspace}

\providecommand{\teff}{\ensuremath{T_{\mathrm{{eff}}}}\xspace}

\providecommand{\logg}{\ensuremath{\log\,g}\xspace}

\providecommand{\gmag}{\ensuremath{G}}

\providecommand{\nm}{\ensuremath{\,\mathrm{nm}}\xspace}

\providecommand{\kms}{\ensuremath{\textrm{km\,s}^{-1}}}

\providecommand{\modulename}[1]{#1\xspace}

\providecommand{\gspphot}{\modulename{GSP-Phot}}

\providecommand{\flame}{\modulename{FLAME}}

\providecommand{\espucd}{\modulename{ESP-UCD}}

\providecommand\gaia{\textit{Gaia}\xspace}

\providecommand{\linktodoc}{https://geapre.esac.esa.int/archive/documentation/GDR3}

\providecommand{\linksec}[2]{\href{\linktodoc/Data_analysis/chap_cu8par/#1}{#2\xspace}}

\providecommand{\linkfig}[1]{\href{\linktodoc/Data_analysis/chap_cu8par/#1.html}{see table\xspace}}

\newcommand{\orcit}[1]{\protect\href{https://orcid.org/#1}{\protect\includegraphics[width=8pt]{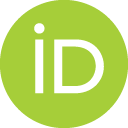}}}

\makeatletter
\DeclareRobustCommand*{\fieldName}[1]{%
  \begingroup\@fieldName\scantokens{\texttt{\small {#1}}\noexpand}\endgroup}
\begingroup\lccode`\~=`\_\relax
   \lowercase{\endgroup\def\@fieldName{\catcode`\_=\active \let~\_}}
\makeatother

\begin{document}

\title{Ultracool dwarfs in \gaia DR3}
\titlerunning{Ultra-cool dwarfs from \gaia DR3}

\author{ L.M.~ Sarro\orcit{0000-0002-5622-5191}\inst{\ref{inst:0017}}
  \and A.~Berihuete\orcit{0000-0002-8589-4423 }\inst{\ref{cadiz}} \and
  R.L.~ Smart\orcit{0000-0002-4424-4766}\inst{\ref{inst:0019}} \and C.
  ~Reyl\'{e} \inst{\ref{inst:0141}} \and
  D. Barrado\inst{\ref{inst:barrado}} \and M.~
  Garc\'{i}a-Torres\orcit{0000-0002-6867-7080}\inst{\ref{inst:0047}}
  \and W.J.~
  Cooper\orcit{0000-0003-3501-8967}\inst{\ref{inst:0038},\ref{inst:0019}}
  \and H.R.A.~Jones\orcit{0000-0003-0433-3665}\inst{\ref{inst:0038}}
  \and F.~ Marocco\orcit{0000-0001-7519-1700}\inst{\ref{inst:0067}}
  \and O.L.~ Creevey\orcit{0000-0003-1853-6631}\inst{\ref{inst:0001}}
  \and R.~ Sordo\orcit{0000-0003-4979-0659}\inst{\ref{inst:0002}} \and
  C.A.L.~ Bailer-Jones\inst{\ref{inst:0007}} \and P.~
  Montegriffo\orcit{0000-0001-5013-5948}\inst{\ref{inst:0026}} \and
  R.~ Carballo\orcit{0000-0001-7412-2498}\inst{\ref{inst:0032}} \and
  R.~ Andrae\orcit{0000-0001-8006-6365}\inst{\ref{inst:0007}} \and M.~
  Fouesneau\orcit{0000-0001-9256-5516}\inst{\ref{inst:0007}} \and
  A.C.~
  Lanzafame\orcit{0000-0002-2697-3607}\inst{\ref{inst:0013},\ref{inst:0057}}
  \and F.~ Pailler\orcit{0000-0002-4834-481X}\inst{\ref{inst:0003}}
  \and F.~ Th\'{e}venin\inst{\ref{inst:0001}} \and A.~
  Lobel\orcit{0000-0001-5030-019X}\inst{\ref{inst:0004}} \and L.~
  Delchambre\orcit{0000-0003-2559-408X}\inst{\ref{inst:0018}} \and
  A.J.~ Korn\orcit{0000-0002-3881-6756}\inst{\ref{inst:0005}} \and A.~
  Recio-Blanco\orcit{0000-0002-6550-7377}\inst{\ref{inst:0001}} \and
  M.S.~ Schultheis\orcit{0000-0002-6590-1657}\inst{\ref{inst:0001}}
  \and F.~ De Angeli\orcit{0000-0003-1879-0488}\inst{\ref{inst:0025}}
  \and N.~ Brouillet\orcit{0000-0002-3274-7024}\inst{\ref{inst:0030}}
  \and L.~
  Casamiquela\orcit{0000-0001-5238-8674}\inst{\ref{inst:0030},\ref{inst:0034}}
  \and G.~ Contursi\orcit{0000-0001-5370-1511}\inst{\ref{inst:0001}}
  \and P.~ de Laverny\orcit{0000-0002-2817-4104}\inst{\ref{inst:0001}}
  \and P.~
  Garc\'{i}a-Lario\orcit{0000-0003-4039-8212}\inst{\ref{inst:0046}}
  \and G.~ Kordopatis\orcit{0000-0002-9035-3920}\inst{\ref{inst:0001}}
  \and Y.~
  Lebreton\orcit{0000-0002-4834-2144}\inst{\ref{inst:0058},\ref{inst:0059}}
  \and E.~ Livanou\orcit{0000-0003-0628-2347}\inst{\ref{inst:0051}}
  \and A.~ Lorca\orcit{0000-0002-7985-250X}\inst{\ref{inst:0014}} \and
  P.A.~ Palicio\orcit{0000-0002-7432-8709}\inst{\ref{inst:0001}} \and
  I.~ Slezak-Oreshina\inst{\ref{inst:0001}} \and C.~
  Soubiran\orcit{0000-0003-3304-8134}\inst{\ref{inst:0030}} \and A.~
  Ulla\orcit{0000-0001-6424-5005}\inst{\ref{inst:0085}} \and H.~
  Zhao\orcit{0000-0003-2645-6869}\inst{\ref{inst:0001}} }

\institute{Dpto. de Inteligencia Artificial, UNED, c/ Juan del Rosal
  16, 28040 Madrid, Spain\relax \label{inst:0017}\vfill \and INAF -
  Osservatorio Astrofisico di Torino, via Osservatorio 20, 10025 Pino
  Torinese (TO), Italy\relax \label{inst:0019}\vfill \and
  Universit\'{e} C\^{o}te d'Azur, Observatoire de la C\^{o}te d'Azur,
  CNRS, Laboratoire Lagrange, Bd de l'Observatoire, CS 34229, 06304
  Nice Cedex 4, France\relax \label{inst:0001} \and INAF -
  Osservatorio astronomico di Padova, Vicolo Osservatorio 5, 35122
  Padova, Italy\relax \label{inst:0002}\vfill \and CNES Centre Spatial
  de Toulouse, 18 avenue Edouard Belin, 31401 Toulouse Cedex 9,
  France\relax \label{inst:0003}\vfill \and Royal Observatory of
  Belgium, Ringlaan 3, 1180 Brussels,
  Belgium\relax \label{inst:0004}\vfill \and Observational
  Astrophysics, Division of Astronomy and Space Physics, Department of
  Physics and Astronomy, Uppsala University, Box 516, 751 20 Uppsala,
  Sweden\relax \label{inst:0005}\vfill \and Max Planck Institute for
  Astronomy, K\"{ o}nigstuhl 17, 69117 Heidelberg,
  Germany\relax \label{inst:0007}\vfill \and INAF - Osservatorio
  Astrofisico di Catania, via S. Sofia 78, 95123 Catania,
  Italy\relax \label{inst:0013}\vfill \and Aurora Technology for
  European Space Agency (ESA), Camino bajo del Castillo, s/n,
  Urbanizacion Villafranca del Castillo, Villanueva de la Ca\~{n}ada,
  28692 Madrid, Spain\relax \label{inst:0014}\vfill \and Institut
  d'Astrophysique et de G\'{e}ophysique, Universit\'{e} de Li\`{e}ge,
  19c, All\'{e}e du 6 Ao\^{u}t, B-4000 Li\`{e}ge,
  Belgium\relax \label{inst:0018}\vfill \and Institute of Astronomy,
  University of Cambridge, Madingley Road, Cambridge CB3 0HA, United
  Kingdom\relax \label{inst:0025}\vfill \and INAF - Osservatorio di
  Astrofisica e Scienza dello Spazio di Bologna, via Piero Gobetti
  93/3, 40129 Bologna, Italy\relax \label{inst:0026}\vfill \and
  Laboratoire d'astrophysique de Bordeaux, Univ. Bordeaux, CNRS, B18N,
  all{\'e}e Geoffroy Saint-Hilaire, 33615 Pessac,
  France\relax \label{inst:0030}\vfill \and Dpto. de Matem\'{a}tica
  Aplicada y Ciencias de la Computaci\'{o}n, Univ. de Cantabria, ETS
  Ingenieros de Caminos, Canales y Puertos, Avda. de los Castros s/n,
  39005 Santander, Spain\relax \label{inst:0032}\vfill \and GEPI,
  Observatoire de Paris, Universit\'{e} PSL, CNRS, 5 Place Jules
  Janssen, 92190 Meudon, France\relax \label{inst:0034}\vfill \and
  Centre for Astrophysics Research, University of Hertfordshire,
  College Lane, AL10 9AB, Hatfield, United
  Kingdom\relax \label{inst:0038}\vfill \and European Space Agency
  (ESA), European Space Astronomy Centre (ESAC), Camino bajo del
  Castillo, s/n, Urbanizacion Villafranca del Castillo, Villanueva de
  la Ca\~{n}ada, 28692 Madrid, Spain\relax \label{inst:0046}\vfill
  \and Data Science and Big Data Lab, Pablo de Olavide University,
  41013, Seville, Spain\relax \label{inst:0047}\vfill \and Department
  of Astrophysics, Astronomy and Mechanics, National and Kapodistrian
  University of Athens, Panepistimiopolis, Zografos, 15783 Athens,
  Greece\relax \label{inst:0051}\vfill \and Dipartimento di Fisica e
  Astronomia ""Ettore Majorana"", Universit\`{a} di Catania, Via
  S. Sofia 64, 95123 Catania, Italy\relax \label{inst:0057}\vfill \and
  LESIA, Observatoire de Paris, Universit\'{e} PSL, CNRS, Sorbonne
  Universit\'{e}, Universit\'{e} de Paris, 5 Place Jules Janssen,
  92190 Meudon, France\relax \label{inst:0058}\vfill \and
  Universit\'{e} Rennes, CNRS, IPR (Institut de Physique de Rennes) -
  UMR 6251, 35000 Rennes, France\relax \label{inst:0059}\vfill \and
  IPAC, Mail Code 100-22, California Institute of Technology, 1200
  E. California Blvd., Pasadena, CA 91125,
  USA\relax \label{inst:0067}\vfill \and Applied Physics Department,
  Universidade de Vigo, 36310 Vigo,
  Spain\relax \label{inst:0085}\vfill \and Departamento de
  Astrof\'{i}sica, Centro de Astrobiolog\'{i}a (CSIC-INTA),
  ESA-ESAC. Camino Bajo del Castillo s/n. 28692 Villanueva de la
  Ca\~{n}ada, Madrid, Spain\relax \label{inst:barrado}\vfill \and
  Institut UTINAM CNRS UMR6213, Universit\'{e} Bourgogne
  Franche-Comt\'{e}, OSU THETA Franche-Comt\'{e} Bourgogne,
  Observatoire de Besan\c{c}on, BP1615, 25010 Besan\c{c}on Cedex,
  France\relax \label{inst:0141} \and School of Physics, Astronomy and
  Mathematics, University of Hertfordshire, College Lane, Hatfield
  AL10 9AB, UK\label{inst:jones} \and Depto. Estad\'{i}stica e
  Investigaci\'{o}n Operativa, Universidad de C\'{a}diz,
  Avda. Rep\'{u}blica Saharaui s/n, 11510 Puerto Real, C\'{a}diz,
  Spain \label{cadiz} \vfill}

\date{Received date /
Accepted date}

\abstract{
  Previous \gaia Data Releases offered the opportunity to uncover ultracool dwarfs (UCDs) through astrometric, rather than purely photometric selection. The most recent, third data release offers in addition the opportunity to use low-resolution spectra to refine and widen the selection.}
{
In this work we use the \gaia DR3 set of ultracool dwarf candidates and complement the \gaia spectrophotometry with additional photometry in order to characterise its global properties. This includes the inference of the distances, their locus in the \gaia colour-absolute magnitude diagram and the (biased through selection) luminosity function in the faint end of the Main Sequence. We study the overall changes in the \gaia RP spectra as a function of spectral type. We study the UCDs in binary systems, attempt to identify low-mass members of nearby young associations, star forming regions and clusters, and analyse their variability properties.
}{
We use a forward model and the Bayesian inference framework to produce posterior probabilities for the distribution parameters and a calibration of the colour index as a function of the absolute magnitude in the form of a Gaussian Process. Additionally we apply the HMAC unsupervised classification algorithm for the detection and characterisation of overdensities in the space of celestial coordinates, projected velocities and parallaxes. 
}{
We detect 57 young, kinematically homogeneous groups some of which are identified as well known star forming regions, associations and clusters of different ages. We find that the primary members of 880 binary systems with a UCD belong mainly to the thin and thick disk components of the Milky Way. We identify 1109 variable UCDs using the variability tables in the \gaia archive, 728 of which belong to the star forming regions defined by HMAC. We define two groups of variable UCDs with extreme bright or faint outliers. 
}{
The set of sources identified as UCDs in the \gaia archive contains a wealth of information that will require focused follow-up studies and observations. It will help to advance our understanding of the nature of the faint end of the Main Sequence and the stellar/substellar transition.  
}

\maketitle

\section{Introduction \label{sec:intro}}

Since its launch in December 2013, the European Space Agency
astrometric mission \gaia \citep{2016A&A...595A...1G} has already
produced three data releases
\citep{2016A&A...595A...2G,2018A&A...616A...1G,2021A&A...649A...1G},
and is already changing our perception of the Galaxy and the
properties of different type of astronomical objects, among them
substellar objects. Brown dwarfs (BDs) are characterised by masses
below $\sim$0.072 $M_\odot$ \citep[depending on the specific models
  and the metallicity; see, for instance,][]{2015A&A...577A..42B}, and
very low temperatures and luminosities, which depend strongly on the
age (being significantly brighter and warmer at young ages). Although
they are very numerous, these properties make them hard to find and
characterise. Within the \gaia framework, Ultra Cool Dwarfs (UCDs) are
defined as objects (both stellar and substellar) with spectral types
M7 or later, and they include the spectral types L, T and Y,
characterised by strong and wide potassium lines in the optical, and
water, methane, and ammonia absorption bands in the near-IR
\citep[see][amongst others]
      {1999ApJ...519..802K,1997A&A...327L..29M,2002ApJ...564..421B,2008MNRAS.391..320B}.

The stellar-substellar transition is still poorly understood and new
processes come into play at these low temperatures, such as the
condensation of solids inside clouds in the atmospheres. In general,
the BDs properties are bracketed between those of low-mass stars and
massive hot planets. However, their formation mechanism is still under
debate, since they can be formed in star-like scenarios (turbulent
fragmentation as in \citeauthor{2004ApJ...617..559P}
\citeyear{2004ApJ...617..559P} or \citeauthor{2008ApJ...684..395H}
\citeyear{2008ApJ...684..395H}; gravitational collapse and
fragmentation \citep{2018MNRAS.478.5460R}; ejection from multiple
protostellar systems as in \citeauthor{2012MNRAS.419.3115B}
\citeyear{2012MNRAS.419.3115B}); or in a planet-like scenario
\citep[such as disk fragmentation as in][]{2006A&A...458..817W},
followed by ejection by dynamical interactions
\citep{2012MNRAS.421L.117V}. Additional mechanisms are possible, such
as aborted stellar embryos \citep{2001AJ....122..432R},
photo-evaporation of cores near massive stars
\citep{2004A&A...427..299W}, gravitational fragmentation of dense
filaments formed in young stellar association
\citep{2008MNRAS.389.1556B} or compression by turbulent flows in
molecular clouds \citep{2019MNRAS.488.2644S}. An overview can be found
in \cite{2014prpl.conf..619C}.  In any case, the photometric and
spectroscopic properties of BDs fully bridge the gap between those of
stars and planets and, in fact, can be used to improve our models of
exoplanetary atmospheres (grain scattering and absorption,
pressure-temperature profiles, chemistry, molecular opacities,
condensation, cloud formation and rainout) since they are easier to
observe and generally lack the effects provoked by the irradiation
from a host star. Among the most interesting characteristics are the
formation of cloud decks (iron and silicates for L dwarfs; chlorides
and sulphides for T and early-type Y dwarfs; water clouds in the
coolest Y dwarfs) and non-equilibrium chemistry
\citep{2000ApJ...541..374S,2014ApJ...797...41Z}.

\cite{2013A&A...550A..44S} estimated the expected end-of-mission
number of ultra-cool dwarfs in the \gaia archive: 600 objects between
L0 and L5, 30 objects between L5 and T0, and 10 objects between T0 and
T8.  Later on, in \cite{2017MNRAS.469..401S}, we cross-matched the
known UCDs with the first Gaia Data Release, and identified 321 dwarfs
with spectral types between L and T (the Gaia Ultracool Dwarf Sample
or GUCDS), which were used as a starting point by the Data Processing
and Analysis Consortium pipeline for parameter estimation purposes
based on the \gaia RP spectra (resolution 50-30 between 630 and 1090
nm, see \citealt{2021A&A...652A..86C} and
\citealt{EDR3-DPACP-118}). \cite{2018A&A...616A..10G} identified 601
UCDs by comparing very restrictive subsets of the second \gaia Data
Release (DR2) with several all-sky catalogues and
\cite{2018A&A...619L...8R} extended this work and identified $\approx$
13000 sources in the \gaia DR2 with spectral types $\ge$M7 (631 with
spectral type L).

Here we discuss the UCD candidate content in the \gaia catalog. This
UCD content is defined and characterised using the RP spectra
\citep{EDR3-DPACP-118} in addition to all other \gaia measurements
used in previous data releases. We define this catalogue of UCDs as
composed of sources in the \gaia DR3 archive with estimates of the
effective temperature \teff. This implies that they were identified as
UCDs by the software module \espucd, which is in charge of selecting
and characterising UCD candidates in the context of the Gaia Data
Processing and Analysis Consortium (DPAC). In the following, we will
refer to the set of \gaia sources processed by the \espucd module and
therefore with \teff estimates from it as the \gaia UCD catalogue or
UCD catalogue for the sake of conciseness.

In Section \ref{sec:candidatesel} we characterise the distribution of
sources in several diagrams including external photometry and compare
it to previous compilations based on \gaia data. In Section
\ref{sec:rpspectra} we discuss the RP spectra of the UCD candidates
and compare them with ground-based high-resolution spectra of a few
examples. In Section \ref{sec:LF} we describe a probabilistic model to
infer distances and luminosities (absolute magnitudes) under very
simple prior assumptions. These results are affected by the selection
function and can only be taken as useful first approximations to the
unbiased distributions. The characterisation of the selection function
and the inference of debiased distributions will be the subject of a
subsequent article. In Section \ref{sec:binaries} we examine potential
wide binary pairs in our sample with UCD components. In Section
\ref{sec:young} we study overdensities of UCDs in the celestial sphere
and identify them with young stellar associations and clusters. This
allows us to study the changes in RP spectra as a function of age that
can potentially serve as a tool to identify youth indicators in future
Data Releases. In Section \ref{sec:vari} we check the variability
properties (produced as part of \gaia DR3) of the UCD
candidates. Finally, in Section \ref{sec:conclusions} we summarise the
results.

\section{The Gaia catalogue of UCD candidates}
\label{sec:candidatesel}

The \gaia DR3 archive contains entries for $\sim$ 1.8 billion sources
including astrophysical parameters produced by the Coordination Unit 8
(CU8). This is accomplished with Apsis, the CU8 software chain
composed of thirteen modules, which includes the Extended Stellar
Parametrizer for Ultra-Cool Dwarfs \espucd. The \espucd software module
produced \teff estimates for 94\,158 \gaia sources that constitute the
subject of analysis of this work. The data processing in CU8 and the
results included in \gaia DR3 are described in \linksec{}{the CU8
  Chapter of the official documentation}, and in \cite{DR3-DPACP-157},
\cite{DR3-DPACP-160}, and \cite{DR3-DPACP-158}.

\subsection{Characterisation of the \gaia UCD catalogue}
\label{sec:character}

\begin{figure}[ht]
	\centering
	\includegraphics[width=\linewidth]{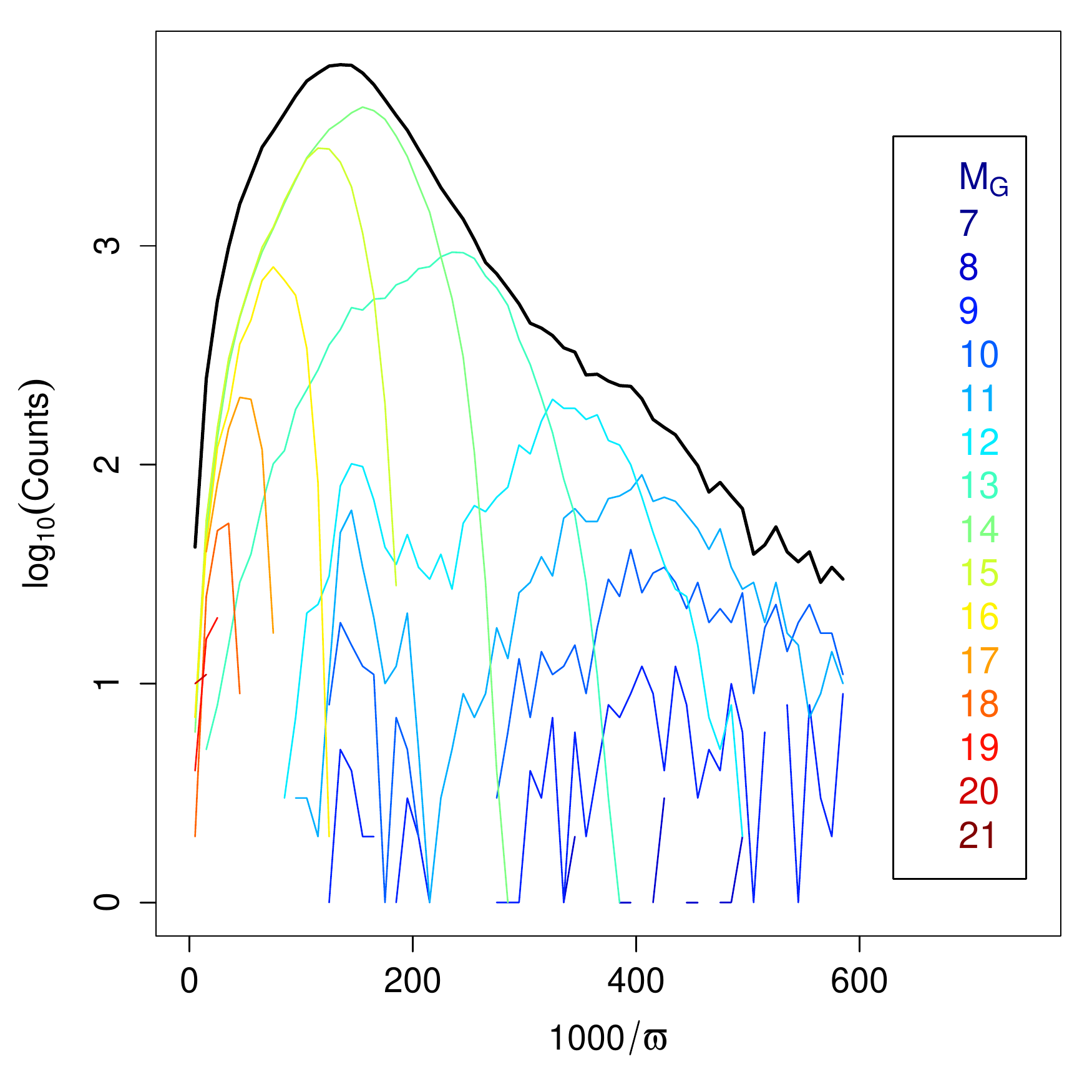}
	\caption{\label{fig:hist-varpi}Histogram of $1000/\varpi$
          (with the parallax expressed in milliarcseconds) in several
          bins of $G+5\cdot \log_{10}(\varpi/1000)+5$. The $y$ axis
          represents the decadic logarithm of the counts in each
          bin. The black line represents the histogram for all sources
          in the \gaia UCD catalogue regardless of their brightness.}
	
\end{figure}

Figure \ref{fig:hist-varpi} shows the histogram (the decadic logarithm
of the counts in each bin) of $1000/\varpi$ (with $\varpi$ in units of
milliarcseconds) in different absolute magnitude bins for all sources
in the \gaia UCD catalogue. In the following we will use the term
absolute G magnitude ($M_G$) to refer to $G +
5\cdot\log(\varpi/1000)+5$ and likewise for other photometric
bands. We are aware that this represents an oversimplification that
neglects the potential effects of extinction/reddening and the naive
derivation of distances (in parsecs) as the reciprocal of the parallax
\citep[again, assumed in units of milliarcseconds; see][for a
  discussion of the proper inference of distances from measured
  parallaxes]{2018A&A...616A...9L,2015PASP..127..994B}. A proper
treatment of the inference of distances from observed parallaxes is
included in Sect. \ref{sec:LF}. For absolute magnitudes fainter than
13 mag the histogram shows a coherent picture of decreasing typical
distances for fainter sources. For sources brighter than $M_G=13$ mag
the counts are dominated by a few stellar associations and star
forming regions discussed in Section \ref{sec:young}.

In the following we will discuss the distribution of the \gaia UCD
catalogue in the space of multi-band photometry built by adding 2MASS
\citep{2006AJ....131.1163S} and AllWISE
\citep{WISE,2011ApJ...731...53M} measurements. For this purpose we
used the pre-computed cross-matches allwise\_best\_neighbour and
tmass\_psc\_xsc\_best\_neighbour in Gaia DR3.

\begin{figure*}[ht]
	\centering
	\includegraphics[width=\textwidth]{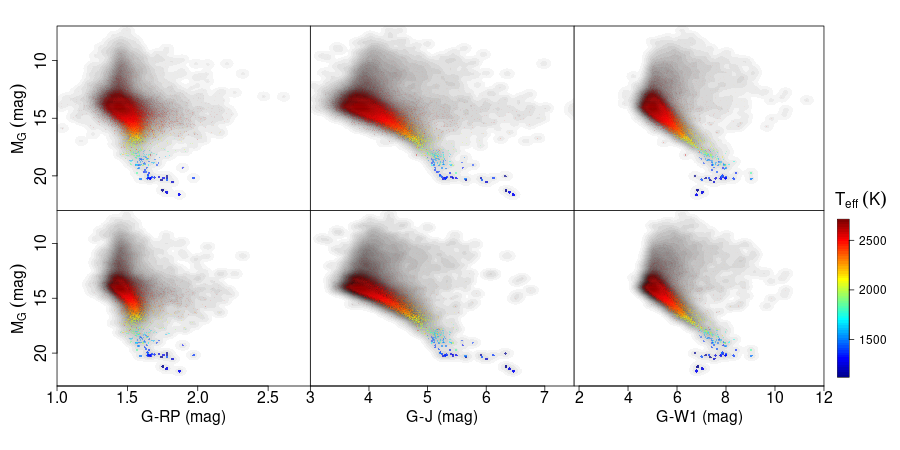}
	\caption{Colour-absolute magnitude diagrams (CAMDs) combining
          several \gaia, 2MASS and WISE magnitudes. The top row
          represents all sources with available measurements and the
          bottom row shows the subset with quality flags A or B in the
          corresponding 2MASS and AllWISE catalogues. A kernel density
          estimate is shown using a grey scale. The transparency and
          the symbol sizes were chosen to enhance visibility of the
          main densities. The colour code reflects the estimated
          \teff\ as indicated by the colour bar at the lower right
          edge of the Figure.}
	\label{fig:camds-dense}
\end{figure*}

Figure \ref{fig:camds-dense} shows the distribution in three
colour-absolute magnitude diagrams (hereafter CAMD) using the
\textit{Gaia}, 2MASS $J$ and AllWISE $W1$ passbands.  The symbol sizes
and transparencies have been chosen to enhance visibility. We plot the
cube root of a kernel density estimate of the distribution of sources
using a gray scale. The top row represents all sources with available
measurements and the bottom row shows the subset with quality flags A
or B in the corresponding 2MASS and AllWISE catalogues. The axes span
the range of values of the colours and absolute magnitude. Appendix
\ref{appendix:altCAMDs} includes other 2D projections of the
multi-band space.

The various CAMDs show two parallel sequences at the hotter end: the
Main Sequence and the equal mass binaries sequence (EMBS), 0.75 mag
brighter. However, the equal mass binary sequence seems to vanish for
temperatures around 2000 K.  We interpret it as the result of the drop
in the number of sources fainter than $M_G\approx 13$ mag detected by
\gaia, due to the intrinsic faintness and the shift of the peak
emission towards longer wavelengths. It affects both the much denser
Main Sequence and EMBS but given that the latter is only a fraction of
the former, the drop in density results in the vanishing of the EMBS
as previously observed in Gaia data \citep{2018A&A...616A..10G}. The
overdensity above the locus of these parallel sequences is mainly due
to the stellar associations and star forming regions discussed in
Section \ref{sec:young}.

The left most panels of Fig. \ref{fig:camds-dense} show a significant
scatter to the right of the Main Sequence. This is a well known
problem with the \gaia photometry of some sources discussed in
\cite{2021A&A...649A...6G} and \cite{EDR3-DPACP-117}. It is explained
by the presence of more than one source in the BP and RP windows which
adds contaminating flux from the secondary source(s). This results in
more flux in BP or RP, a decrease in the $G_{BP}$ and $G_{RP}$
magnitudes and the subsequent increase in colour indices that include
the (unaffected) $G$ magnitude.

\subsection{Comparison with other catalogues}
\label{sec:comparisonCats}

As mentioned above, the \espucd module produced 94158 UCD candidates
with \teff estimates below 2700 K. Given the luminosity function
\citep[see for example][and references therein, for a recent
  estimate]{2019ApJ...883..205B}, the actual number of UCD candidates
in the faint end represents only a small fraction of these 94158
sources which are dominated by the brighter regime. In order to
further assess the selection function applied to the full \gaia
catalogue we compare here the \gaia UCD catalogue with previous
compilations of UCD candidates based on \gaia data. We note that the
main difference resides in the fact that this catalogue is based on
the observed \gaia RP spectra and hence can be expected to be cleaner
but also potentially more incomplete due to the quality criteria
imposed on the spectra.

\subsubsection{The Gaia Ultracool Dwarf Sample (GUCDS)}

\begin{figure}[ht]
	\centering
	\includegraphics[scale=0.38]{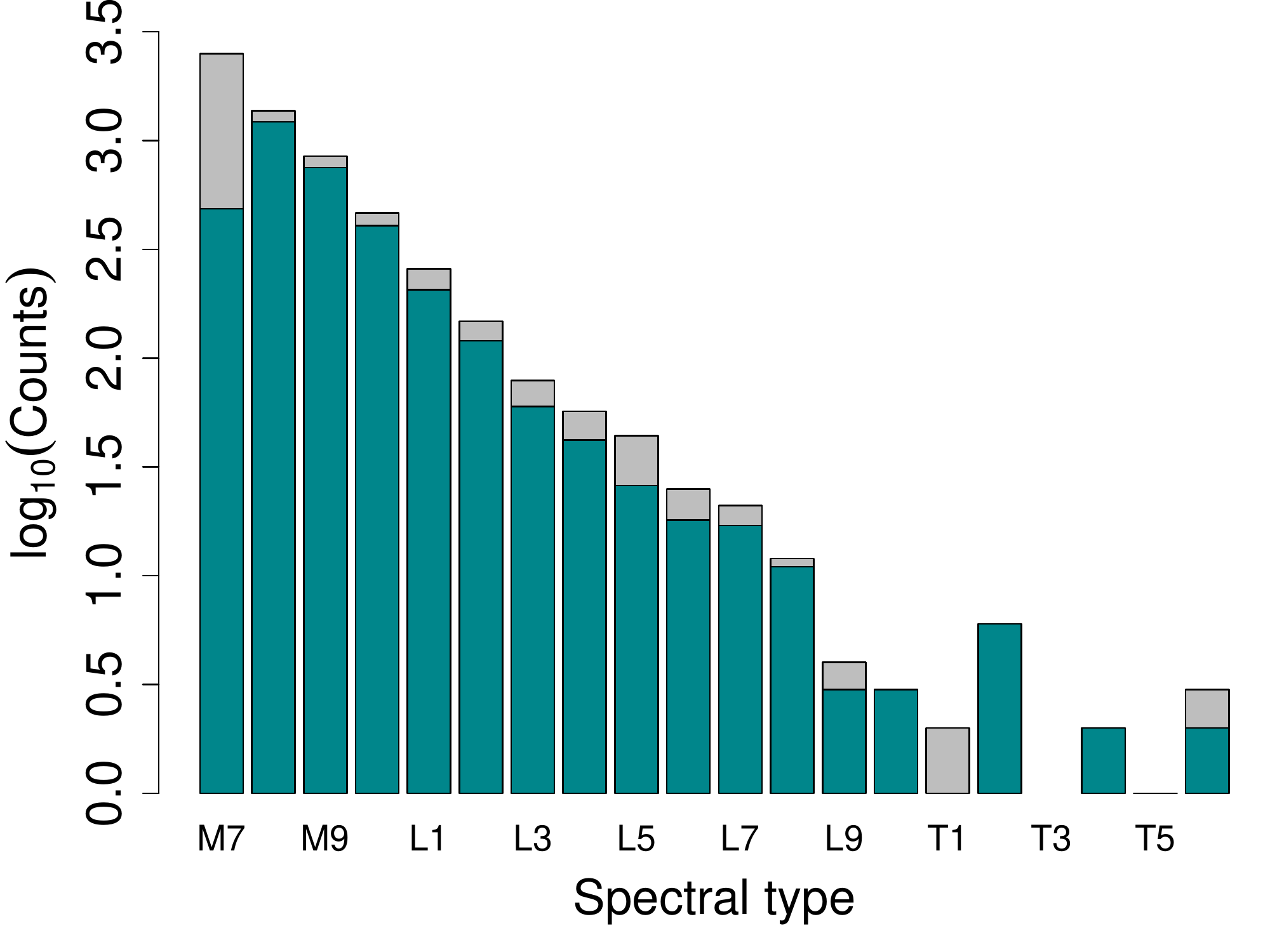}
	\caption{Distribution of the spectral types of GUCDS sources with \gaia cross matches (grey) and of those included in the \gaia UCD catalogue (turquoise). The vertical axis is the decadic logarithm of the counts.}
	\label{fig:gucds-missing}
\end{figure}

The \gaia Ultracool Dwarf Sample \citep[GUCDS][]{2017MNRAS.469..401S}
is a continuous effort to match existing UCD catalogues in the
literature with the \gaia catalogue. We use the version produced in
May 2022 that contains 20108 entries corresponding to UCDs and any
companions identified and spectroscopically confirmed in the
literature. This list of objects is cross-matched using a cone search
with a large 5" radius to the \gaia DR3. This is a relaxed cross-match
that can lead to many mismatches, so for each entry we estimate its
\gaia $G$ magnitude from the 2MASS $J$ magnitude and published
spectral type, if this is more than 2 magnitudes away from the \gaia
DR3 value we reject the match. We then check any outliers in various
colour-magnitude diagrams and we manually remove the
mis-identifications. There are 5856 entries in the GUCDS with spectral
types later than or equal to M7 \citep[corresponding to a temperature
  of 2656 K according to][]{2009ApJ...702..154S} and a \gaia DR3
identifier. Of these, only 5201 have the three \gaia measurements that
are required to be included in the \espucd input list ($\varpi$, $G$
and $G_{RP}$), and 5180 fulfil the input list selection criteria
($\varpi > 1.7$ mas and $(G-G_{RP}) > 1.0$ mag). Finally, 4206 of the
5180 are included in the \gaia DR3 UCD catalogue. Hence, there are 974
sources identified in the literature as UCDs but excluded from the
\gaia UCD catalogue because they did not fulfil one or several of the
selection criteria established for publication. For illustration
purposes, 144 of them did not fulfil the RP flux percentiles
criteria\footnote{In the first stage, \espucd candidates are retained
only if they have a sufficient fraction of the total RP flux at very
red wavelengths. Let $q_N$ denote the pixel position where the $N$-th
percentile of the RP spectrum flux distribution (accumulated from low to
high wavelengths) is attained. Then, the selection sub-module requires
that $q_{33} > 60$, $q_{50} > 71$ and $q_{67} > 83$ which reduces the
47 million input sources to 8.3 millions. For reference, pixel
positions 60, 71 and 83 correspond to central wavelengths of
approximately 776.9, 818.6 and 858.6 nm respectively.}, 362 have
temperature estimates \teff > 2700 K and 384 fail the criterion on the
Euclidean distance to the set of training set templates. The rest are
excluded due to the astrometric quality criterion
($\log_{10}(\sigma_{\varpi}) < -0.8 + 1.3 \cdot \log_{10}(\varpi)$) or
a number of RP transits less than 15. Figure \ref{fig:gucds-missing}
shows the spectral type distribution of the GUCDS sources and of those
in the UCD catalogue. The apparent incompleteness of the leftmost bin
is due to the sharp cut in estimated \teff that only covers part of
the bin temperature range.

Hence, incompleteness of our catalogue is due mostly to unavailability
of the measurements required for selection by \espucd (11\%),
indications of temperatures above the UCD limit (6\%) or quality
criteria (8\%).

\subsubsection{The DR2 catalogue of UCD candidates by \cite{2018A&A...619L...8R}}

We have used the \gaia archive mapping between DR2 and DR3 source
identifiers to track the UCD candidate list by
\cite{2018A&A...619L...8R} in our catalogue. There are 14\,915 sources
in the original list based on DR2 data. Of these, 12\,656 are included
in the DR3 catalogue of UCDs. The remaining 2\,259 were rejected due
to the RP flux percentile filters (210) or because the estimated \teff
> 2\,700 K (1\,928). The remaining 121 fail the criteria for the
number of transits or distance to the templates.

The number of UCDs retrieved in this work is much larger. The main
reason is that \cite{2018A&A...619L...8R} applied very strict filters
on the data based on astrometric and photometric features to define a
list of robust candidates, with a simple selection from their locus in
the CAMD. At that point and without the possibility of confirmation
from RP spectra, the strict quality filters imposed were the only
possibility to avoid the many potential contaminants. Another reason
for the larger number of candidates in DR3 is that the \teff selection
we used includes objects earlier than M7 (the \gaia DR3 UCD limit is
2\,700 K while, according to \cite{2009ApJ...702..154S} M7 corresponds
to 2\,656 K). Nevertheless, if we focus on the region of the CAMD used
by \cite{2018A&A...619L...8R} to select M7 and later dwarfs, the
availability of the RP spectra combined with the use of the Gaussian
Process regression module allows us to retrieve about 60,000 objects
rejected by \cite{2018A&A...619L...8R} including sources in young
associations as described in Sect. \ref{sec:young}.

\subsubsection{The \gaia Catalogue of Nearby stars}

\begin{figure}[ht]
	\centering
	\includegraphics[scale=0.8]{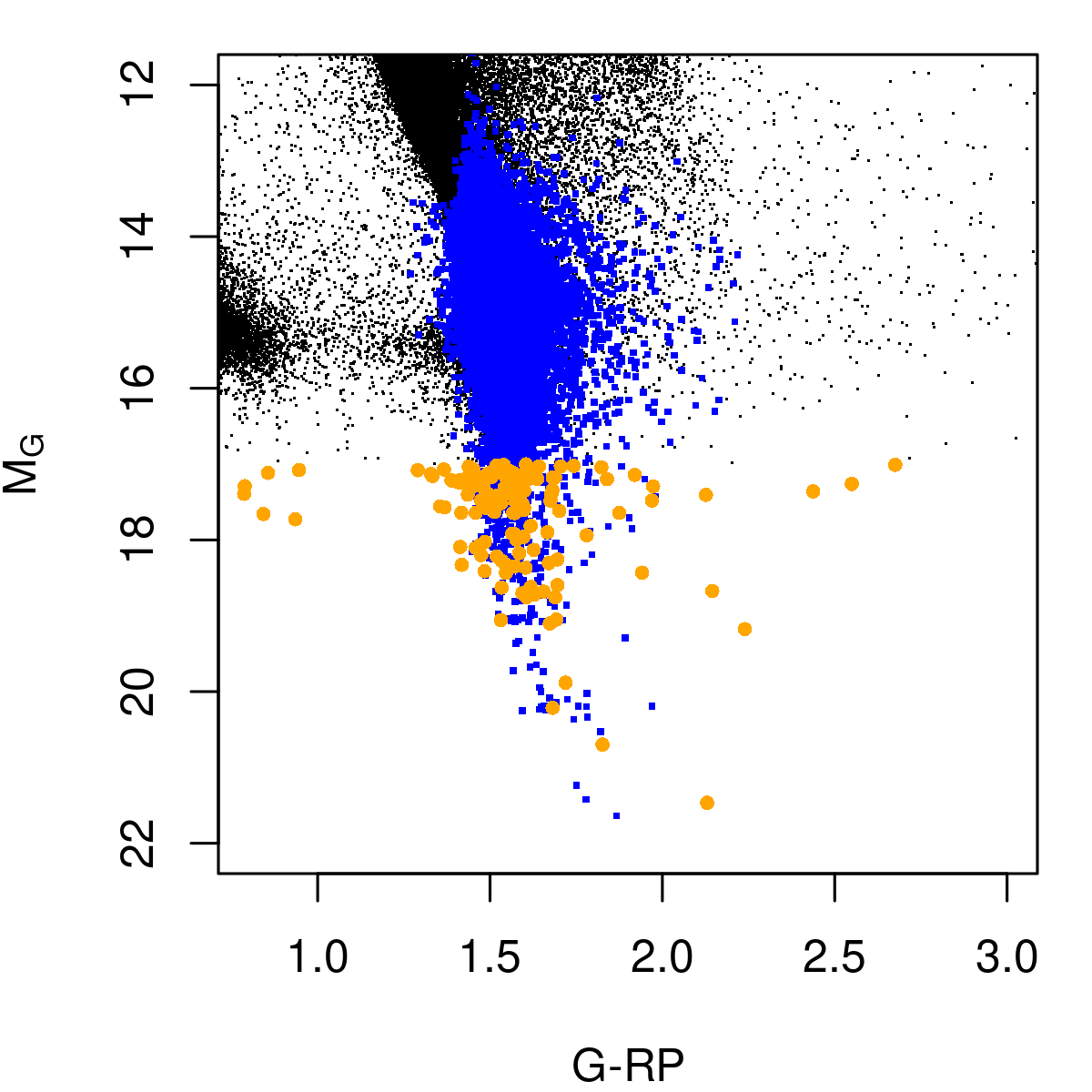}
	\caption{Colour-absolute magnitude diagram of the GCNS sources
          (black) in the \gaia DR3 UCD catalogue (blue). Orange
          circles highlight the GCNS entries with $M_{G} > 17$ mag
          missing from the UCD catalogue.}
	\label{fig:gcns-missing}
\end{figure}

Finally, we compare the UCD content of the \gaia Catalogue of Nearby
stars \citep[GCNS][]{2021A&A...649A...6G}. In this case, a direct
comparison of the samples is difficult because the GCNS is not
restricted to UCDs. For illustration purposes, we consider the
subsample of sources with $G + 5\cdot\log(\varpi/1000)+5$ > 17
mag. There are 155 GCNS sources in that subsample missing from the
\gaia DR3 UCD catalogue. Of these, 122 fulfil the \espucd input list
criteria (again, $\varpi > 1.7$ mas and $(G-G_{RP}) > 1.0$ mag) but
are rejected on the basis of the quality of the RP
spectra\footnote{\teff estimates for sources with total negative
fluxes in the normalised RP spectrum greater than -0.1, with a median
RP curvature $\tau \geq 2.0\cdot 10^{-5}$, or with fluxes at the
reddest bin greater than 0.015 were not selected for publication.}
(103 cases). Also, 19 sources are missing because the mean RP spectrum
was derived from less than 15 transits. Figure \ref{fig:gcns-missing}
shows the two catalogues in the \gaia CAMD, with the GCNS represented
with black dots, those also included in the \gaia DR3 UCD catalogue as
blue small circles and the missing sources fainter than $M_G=17$ mag
as orange circles. The cool end of the White Dwarf sequence is visible
as black dots with $G-G_{RP}$ colour indices bluer that approximately
1.

\section{RP Spectra} \label{sec:rpspectra}

The \gaia DR3 includes for the first time the BP and RP low-resolution
spectra described in~\citet{EDR3-DPACP-118}. These are internally
calibrated spectra affected by the instrumental response, the LSF
(Line Spread Function) and the wavelength dependent dispersion
relation. Only differential effects are tackled by the internal
calibration: variations of the instrument across the different
observing conditions (time, CCD, FoV, window class and gate
configurations). In fact, the \gaia DR3 does not include the
internally calibrated spectra themselves but the coefficients of their
representations in a set of basis functions as described in
\cite{EDR3-DPACP-118} and \cite{EDR3-DPACP-120}. In this work we will
analyse and concentrate on the \gaia internally calibrated spectra
published as part of the \gaia DR3 in the form of coefficients. Since
UCDs are intrinsically faint and very red objects, their BP spectra
only contain noise in all but the brightest and hottest candidates and
even there, only at the reddest wavelengths. Hence, we only discuss
here the internally calibrated RP spectra, abbreviated hereafter as RP
spectra. In this Section we will briefly mention the externally
calibrated spectra. These are available from the \gaia archive only
for sources with $G$ magnitudes brighter than 15 mag (48 UCDs) but can
be generated from the coefficients used to represent the internally
calibrated spectra using the {\sl GaiaXPy} software\footnote{Available
from \url{https://gaia-dpci.github.io/GaiaXPy-website/}}. The
externally calibrated spectra of UCDs present problems discussed in
\cite{EDR3-DPACP-120} and illustrated in
Fig. \ref{fig:gtc_comparison_calibrated} below.

\begin{figure}[ht]
	\centering
	\includegraphics[width=\linewidth]{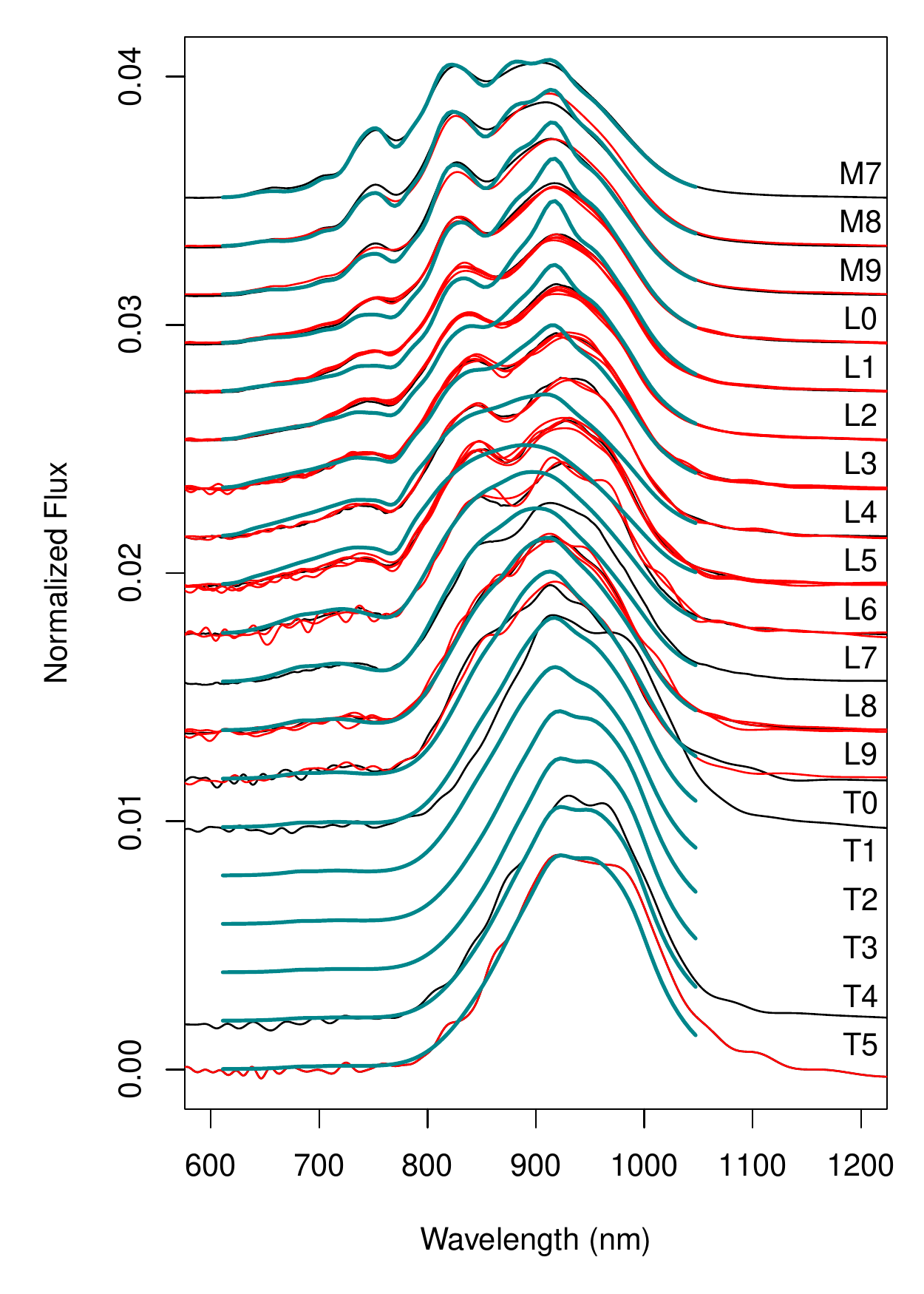}
	\caption{RP spectra for spectral types in the UCD regime,
          from M7 to T6.  The red lines are spectral type comparison
          objects, black lines represent the median RP spectrum of all
          UCD candidates in each spectral type and the turquoise lines
          represent the MIOG simulations based on BT-Settl synthetic
          spectra.  The conversion between effective temperatures and
          spectral types for the black and turquoise lines is done
          using the calibrations of
          \cite{2009ApJ...702..154S}. \label{fig:RPspectra} }
\end{figure}

Figure~\ref{fig:RPspectra} shows in turquoise, simulations of BT-Settl
synthetic spectra obtained using the Mean Instrument Object
Generator~\citep[MIOG, briefly described in][]{DR3-DPACP-157}.  The
spectral types were assigned from the BT-Settl \teff using
the~\cite{2009ApJ...702..154S} calibration.  The black lines
correspond to the median RP spectrum in each spectral type (assigned
again using the same calibration and the \espucd temperatures) and
they aim to exemplify the appearance of a `typical' object; the red
lines show comparison objects with the same spectral types.  The
comparison objects were selected from the spectral classifications
in~\citet{2019ApJS..240...19K} which had RP spectra in our validation
subset and were visually similar.  Three of these objects (VB 10, LP
271-25 and SIPS J1058-1548 with spectral types M8, M9 and L3
respectively) a are listed in~\citet{1999ApJ...519..802K} too whilst
two objects (2MASS J05591914-1404488 and 2MASS J15031961+2525196 with
spectral types T5 and T6 respectively) appear also listed
in~\citet{2003ApJ...594..510B}.  All of these works represent long
accepted optical spectroscopic standards of UCDs.  An example of an
object that was visually rejected is Kelu-1 that - probably because of
its binarity - appeared too red.  This comparison sample is given in
Table~\ref{tab:standards_used}.  The differences between the BT-Settl
models and the literature standards for each spectral type or the
median RP spectra between M9 and L8 are evident and they do not simply
correspond to effective temperature offsets that could be explained by
a different spectral type-temperature calibration.

\begin{table*}
\captionsetup{width=\linewidth}
\caption{
\label{tab:standards_used}
List of comparison UCDs used to calibrate the \espucd module empirical training set in effective temperature.
Astrometry is from \gaia DR3 and the \teff values are those produced by \espucd and published 
as part of the Data Release.
}
\begin{tabular}{r rr r ll r}
\hline
\myalign{c}{
    \gaia DR3}  & \myalign{c}{$\alpha$} & \myalign{c}{$\delta$} & \myalign{c}{$\varpi$} & \myalign{c}{Object}   & \myalign{c}{Spectral}   & \myalign{c}{Teff} \\
     \myalign{c}{Source ID} & \myalign{c}{(hms)}     & \myalign{c}{(dms)}    & \myalign{c}{(mas)}     & \myalign{c}{Name}     & \myalign{c}{Type}       & \myalign{c}{[K]} \\
\hline
    4293315765165489536 & 19 16 57 & +5 08 39.7 & $169.0\pm0.1$ & VB 10$\hyperlink{stdrefs}{^{1}}$ & M8$\hyperlink{stdrefs}{^{2}}$ & $2404\pm8$\\
    1287312100751643776 & 14 28 43 & +33 10 27.9 & $91.2\pm0.1$ & LP  271-25$\hyperlink{stdrefs}{^{3}}$ & M9$\hyperlink{stdrefs}{^{4}}$ & $2238\pm9$\\
    5761985432616501376 & 8 53 36 & -3 29 35.4 & $115.5\pm0.1$ & LP  666-9$\hyperlink{stdrefs}{^{3}}$ & M9$\hyperlink{stdrefs}{^{4}}$ & $2272\pm26$\\
    4595127343251508992 & 17 31 30 & +27 21 19.2 & $83.7\pm0.1$ & LSPM J1731+2721$\hyperlink{stdrefs}{^{5}}$ & L0$\hyperlink{stdrefs}{^{6}}$ & $2233\pm24$\\
    31235033696866688 & 3 14 03 & +16 03 04.6 & $72.6\pm0.2$ & 2MASS J03140344+1603056$\hyperlink{stdrefs}{^{7}}$ & L0$\hyperlink{stdrefs}{^{6}}$ & $2201\pm40$\\
    3701479918946381184 & 12 21 28 & +2 57 19.1 & $53.8\pm0.2$ & 2MASS J12212770+0257198$\hyperlink{stdrefs}{^{7}}$ & L0$\hyperlink{stdrefs}{^{8}}$ & $2210\pm41$\\
    3457493517036545280 & 6 02 31 & +39 10 50.5 & $85.8\pm0.1$ & LSR J0602+3910$\hyperlink{stdrefs}{^{9}}$ & L1$\hyperlink{stdrefs}{^{10}}$ & $2044\pm24$\\
    3802665122192531712 & 10 45 23 & -1 49 57.9 & $58.8\pm0.2$ & 2MASS J10452400-0149576$\hyperlink{stdrefs}{^{11}}$ & L1$\hyperlink{stdrefs}{^{12}}$ & $2073\pm71$\\
    3808159454810609280 & 10 48 42 & +1 11 54.5 & $66.6\pm0.2$ & LSPM J1048+0111$\hyperlink{stdrefs}{^{11}}$ & L1$\hyperlink{stdrefs}{^{8}}$ & $2077\pm38$\\
    1649407285800074240 & 16 58 03 & +70 26 56.7 & $54.1\pm0.1$ & LSPM J1658+7027$\hyperlink{stdrefs}{^{13}}$ & L1$\hyperlink{stdrefs}{^{13}}$ & $2069\pm39$\\
    3460806448649173504 & 11 55 40 & -37 27 48.2 & $84.7\pm0.1$ & 2MASS J11553952-3727350$\hyperlink{stdrefs}{^{12}}$ & L2$\hyperlink{stdrefs}{^{12}}$ & $1978\pm31$\\
    4878035808244168832 & 4 45 54 & -30 48 27.4 & $61.9\pm0.1$ & 2MASS J04455387-3048204$\hyperlink{stdrefs}{^{14}}$ & L2$\hyperlink{stdrefs}{^{15}}$ & $2017\pm47$\\
    5723739672264914176 & 8 28 34 & -13 09 19.4 & $85.6\pm0.1$ & SSSPM J0829-1309$\hyperlink{stdrefs}{^{16}}$ & L2$\hyperlink{stdrefs}{^{16}}$ & $1981\pm51$\\
    851053031037729408 & 10 51 19 & +56 13 03.6 & $63.9\pm0.1$ & 2MASS J10511900+5613086$\hyperlink{stdrefs}{^{7}}$ & L2$\hyperlink{stdrefs}{^{6}}$ & $2025\pm81$\\
    5733429157137237760 & 8 47 29 & -15 32 40.6 & $57.5\pm0.2$ & SIPS J0847-1532$\hyperlink{stdrefs}{^{14}}$ & L2$\hyperlink{stdrefs}{^{8}}$ & $2040\pm50$\\
    4910850870213836928 & 1 28 26 & -55 45 32.5 & $53.9\pm0.2$ & SIPS J0128-5545$\hyperlink{stdrefs}{^{17}}$ & L2$\hyperlink{stdrefs}{^{18}}$ & $1993\pm65$\\
    1182574753387703680 & 15 06 53 & +13 21 05.9 & $85.4\pm0.2$ & 2MASSW J1506544+132106$\hyperlink{stdrefs}{^{19}}$ & L3$\hyperlink{stdrefs}{^{13}}$ & $1787\pm56$\\
    167202325215063168 & 4 01 37 & +28 49 51.1 & $80.4\pm0.2$ & 2MASS J04013766+2849529$\hyperlink{stdrefs}{^{20}}$ & L3$\hyperlink{stdrefs}{^{20}}$ & $1872\pm59$\\
    1329942262499164544 & 16 15 44 & +35 58 51.1 & $50.2\pm0.3$ & 2MASSW J1615441+355900$\hyperlink{stdrefs}{^{21}}$ & L3$\hyperlink{stdrefs}{^{21}}$ & $1791\pm216$\\
    3238449635184620672 & 5 00 21 & +3 30 44.5 & $75.6\pm0.3$ & 2MASS J05002100+0330501$\hyperlink{stdrefs}{^{7}}$ & L3$\hyperlink{stdrefs}{^{22}}$ & $1735\pm119$\\
    3562717226488303360 & 10 58 48 & -15 48 16.8 & $55.1\pm0.3$ & SIPS J1058-1548$\hyperlink{stdrefs}{^{23}}$ & L3$\hyperlink{stdrefs}{^{24}}$ & $1834\pm109$\\
    6118581861234228352 & 14 25 28 & -36 50 30.8 & $84.4\pm0.3$ & 2MASS J14252798-3650229$\hyperlink{stdrefs}{^{25}}$ & L4$\hyperlink{stdrefs}{^{22}}$ & $1819\pm52$\\
    5908794218026022144 & 17 53 45 & -66 00 01.1 & $63.6\pm0.3$ & SIPS J1753-6559$\hyperlink{stdrefs}{^{7}}$ & L4$\hyperlink{stdrefs}{^{18}}$ & $1703\pm147$\\
    6306068659857135232 & 15 07 48 & -16 27 54.5 & $134.9\pm0.3$ & 2MASSW J1507476-162738$\hyperlink{stdrefs}{^{19}}$ & L5$\hyperlink{stdrefs}{^{21}}$ & $1552\pm102$\\
    2467182154313027712 & 1 44 36 & -7 16 17.5 & $78.5\pm0.5$ & 2MASS J01443536-0716142$\hyperlink{stdrefs}{^{26}}$ & L5$\hyperlink{stdrefs}{^{8}}$ & $1603\pm90$\\
    3698979462002285824 & 12 03 57 & +0 15 45.6 & $66.3\pm0.5$ & 2MASS J12035812+0015500$\hyperlink{stdrefs}{^{27}}$ & L5$\hyperlink{stdrefs}{^{28}}$ & $1642\pm219$\\
    3597096309389074816 & 12 13 03 & -4 32 44.3 & $59.1\pm0.6$ & 2MASS J12130336-0432437$\hyperlink{stdrefs}{^{14}}$ & L5$\hyperlink{stdrefs}{^{28}}$ & $1580\pm152$\\
    4220379661283166720 & 20 02 51 & -5 21 54.4 & $56.7\pm1.4$ & 2MASSI J2002507-052152$\hyperlink{stdrefs}{^{29}}$ & L6$\hyperlink{stdrefs}{^{8}}$ & $1547\pm187$\\
    4371611781971072768 & 17 50 24 & -0 16 11.8 & $108.6\pm0.2$ & 2MASS J17502484-0016151$\hyperlink{stdrefs}{^{30}}$ & L6$\hyperlink{stdrefs}{^{31}}$ & $1542\pm71$\\
    1954170404122975232 & 21 48 17 & +40 04 06.7 & $123.7\pm0.4$ & 2MASSW J2148162+400359$\hyperlink{stdrefs}{^{32}}$ & L6$\hyperlink{stdrefs}{^{8}}$ & $1511\pm160$\\
    4752399493622045696 & 2 55 05 & -47 01 00.2 & $205.4\pm0.2$ & DENIS J025503.3-470049$\hyperlink{stdrefs}{^{33}}$ & L8$\hyperlink{stdrefs}{^{8}}$ & $1365\pm38$\\
    5052876333365036928 & 2 57 27 & -31 05 46.7 & $102.7\pm0.5$ & 2MASS J02572581-3105523$\hyperlink{stdrefs}{^{15}}$ & L8$\hyperlink{stdrefs}{^{8}}$ & $1354\pm71$\\
    1037131492704550656 & 8 57 58 & +57 08 45.2 & $72.7\pm0.7$ & 2MASS J08575849+5708514$\hyperlink{stdrefs}{^{11}}$ & L8$\hyperlink{stdrefs}{^{11}}$ & $1361\pm316$\\
    3426333598021539840 & 6 07 38 & +24 29 51.7 & $138.1\pm0.5$ & 2MASS J06073908+2429574$\hyperlink{stdrefs}{^{34}}$ & L9$\hyperlink{stdrefs}{^{20}}$ & $1355\pm105$\\
    2997171394834174976 & 5 59 20 & -14 04 54.6 & $95.3\pm0.7$ & 2MASS J05591914-1404488$\hyperlink{stdrefs}{^{35}}$ & T5$\hyperlink{stdrefs}{^{36}}$ & $1147\pm86$\\
    1267906854386665088 & 15 03 20 & +25 25 28.7 & $155.8\pm0.8$ & 2MASS J15031961+2525196$\hyperlink{stdrefs}{^{37}}$ & T6$\hyperlink{stdrefs}{^{37}}$ & $1132\pm102$\\ 
\hline
\end{tabular}
\caption*{\hypertarget{stdrefs}{References}: 1.~\cite{1955LFT...C......0L}, 2.~\cite{1991ApJS...77..417K}, 3.~\cite{1979LHS...C......0L}, 4.~\cite{2005PASP..117..676R}, 5.~\cite{2005AJ....129.1483L}, 6.~\cite{2008AJ....136.1290R}, 7.~\cite{2006AJ....132..891R}, 8.~\cite{2014AJ....147...34S}, 9.~\cite{2002AJ....124.1190L}, 10.~\cite{2003ApJ...586L.149S}, 11.~\cite{2002AJ....123.3409H}, 12.~\cite{2002ApJ...575..484G}, 13.~\cite{2000AJ....120.1085G}, 14.~\cite{2003AJ....126.2421C}, 15.~\cite{2007AJ....133.2258S}, 16.~\cite{2002MNRAS.336L..49S}, 17.~\cite{2007A&A...468..163D}, 18.~\cite{2013AJ....146..161M}, 19.~\cite{2000AJ....119..369R}, 20.~\cite{2013ApJ...776..126C}, 21.~\cite{2000AJ....120..447K}, 22.~\cite{2015ApJS..219...33G}, 23.~\cite{1997A&A...327L..25D}, 24.~\cite{1999ApJ...519..802K}, 25.~\cite{2004A&A...416L..17K}, 26.~\cite{2002ApJ...580L..77H}, 27.~\cite{2000AJ....119..928F}, 28.~\cite{2014ApJ...794..143B}, 29.~\cite{2007AJ....133..439C}, 30.~\cite{2007MNRAS.374..445K}, 31.~\cite{2010ApJ...710.1142B}, 32.~\cite{2008ApJ...686..528L}, 33.~\cite{1999AJ....118.2466M}, 34.~\cite{2012ApJ...746....3C}, 35.~\cite{2000AJ....120.1100B}, 36.~\cite{2002ApJ...564..466G}, 37.~\cite{2003AJ....125..850B}}
\end{table*}

In Sect.~\ref{sec:young}, we study UCDs in the catalogue that we
identify as young and therefore potentially useful in defining low
gravity diagnostics based on their RP spectra.  The detection of
subdwarfs in the catalogue will be addressed in a subsequent
paper~(Cooper et al., in preparation).

At the low resolution typical of the RP spectra \citep[50--30 in
  $\lambda/\Delta\lambda$][]{EDR3-DPACP-120}, individual features
cannot be discerned since multiple nearby spectral features, both
lines and bands, are blended and merged.  The systematic changes and
dependencies of the RP spectra with astrophysical parameters such as
\teff, \logg or metallicity are not immediately evident due to this
blending of spectral features.  Also, because the effects in different
features appear as opposing factors which can cancel out or partially
compensate each other.  Figure~\ref{fig:RPspectra} shows how these
merged absorption features differ with spectral type.  For example,
the majority of RP spectra of L dwarfs have a peak near 800~nm, the
strength of which (and its red-ward trough) is affected by the pressure
broadening on the K\,\textsc{i} resonance doublet~\citep[which
  strengthens with spectral type,][]{1998MNRAS.301.1031T}; by the
weakening of Na\,\textsc{i} (again with spectral type) and by a
weakening of TiO (vanishing in the early L types but still present in
late M).

\begin{figure}
    \centering
    \includegraphics[width=\linewidth]{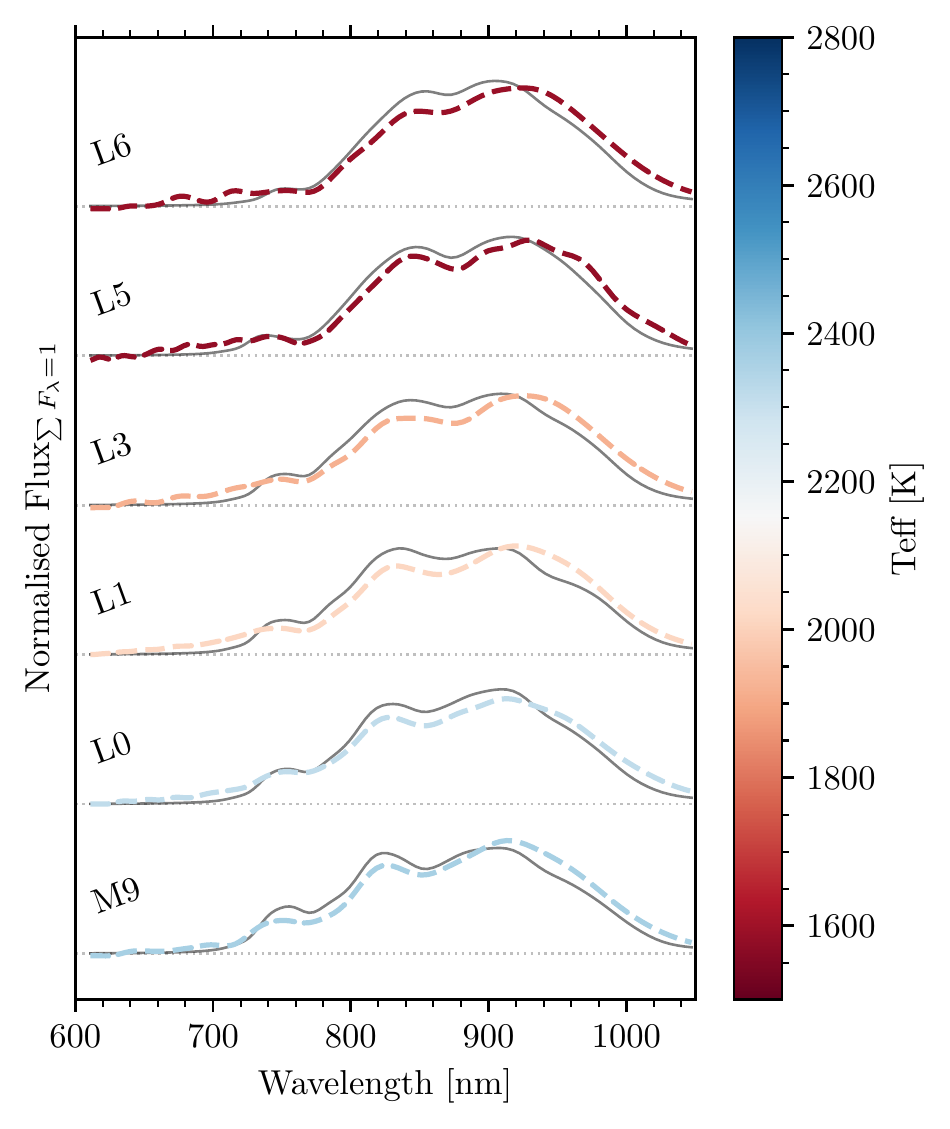}
    \caption{In dark grey, ground-based optical spectra from the GTC
      (Cooper et al., in prep).  These spectra have been simulated
      from the original spectra by passing through MIOG\@.  The
      objects shortnames are: J1717+6526 - L6, J1213-0432 - L5,
      J0453-1751 - L3, J1745-1640 - L1, J0935-2934 - L0, J0938+0443 -
      M9.  Over-plotted are the corresponding observed RP spectra of
      the same objects coloured by effective temperature and labelled
      by spectral type.  All fluxes are normalised by the area and
      linearly offset.  }
    \label{fig:gtc_comparison}
\end{figure}

We use ground-based optical spectra to better understand the
morphological features seen in \gaia RP spectra.  This is demonstrated
in Figure~\ref{fig:gtc_comparison}, which shows the \gaia RP spectra
of six UCDs and the MIOG simulated RP spectra based on mid-resolution
spectra from the GTC/OSIRIS instrument for the same set of
sources. This sequence is also illustrated in
Figure~\ref{fig:gtc_comparison_calibrated} which shows the externally
calibrated (c.f.  \texttt{gaiaxpy.calibrate}) RP
spectra~\citep{EDR3-DPACP-120}.  It shows spurious oscillations and
significant discrepancies with respect to the ground-based spectra
(particularly evident in the L5 case).  The appearance of these
oscillations is discussed in~\citet{EDR3-DPACP-120} and is not yet
fully understood.  However, the apparent amplification of these
wiggles at longer wavelengths is due to the fact that the externally
calibrated spectral energy distributions are normalised by the inverse 
of the response model, which in the RP case drops quickly to very small 
values beyond 900\,nm.  These externally calibrated spectra were not 
used as input to Apsis or \espucd.  \espucd and the rest of Apsis modules 
used only internally calibrated spectra for the prediction of astrophysical
parameters.

\begin{figure}
    \centering
    \includegraphics[width=\linewidth]{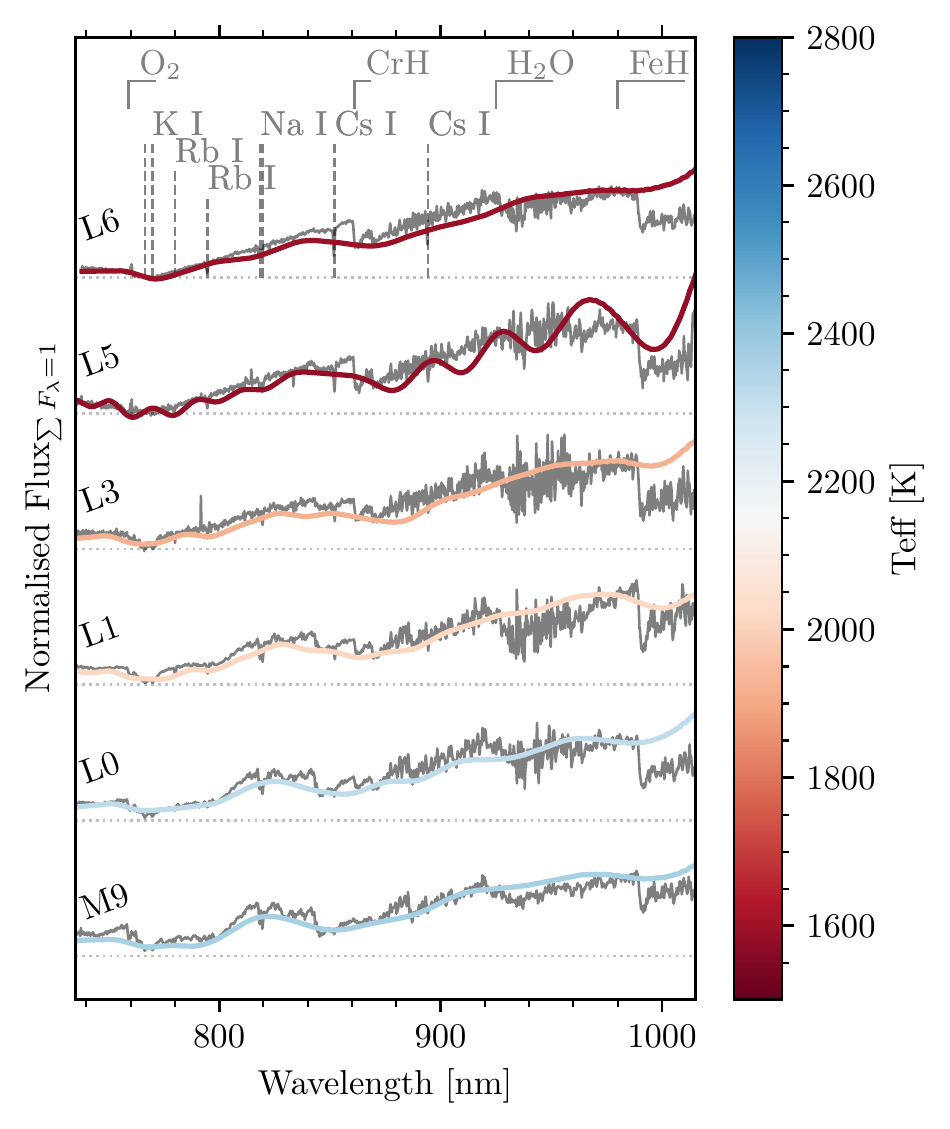}
    \caption{As Figure~\ref{fig:gtc_comparison} but for externally
      calibrated RP spectra.  The calibrated spectra were constructed
      using the \texttt{gaiaxpy.calibrate} function.  The GTC spectra
      as shown here have not been passed through MIOG and represent
      the actual spectra with resolution $\approx2500$.  A selection
      of features typical of late M - mid L dwarfs are shown above.  }
    \label{fig:gtc_comparison_calibrated}
\end{figure}

\section{Bayesian distances and the Luminosity Function}\label{sec:LF}

The \gaia DR3 catalogue of UCDs is truncated as a result of the
various selection filters described in previous Sects. It reaches out
to maximum barycentric distances that depend on the intrinsic
brightness of the UCD and the selection/quality criteria applied
result in further incompletenesses that are difficult to formulate in
terms of the UCD properties. The analysis of the selection function is
out of the scope of this work; it will be addressed in a subsequent
paper and the lessons learnt will be incorporated to the \espucd
processing for \gaia DR4. In this Sect. we elaborate on a simple model
that aims to infer UCD population properties as a first step towards a
full probabilistic treatment that takes into account the selection
biases inherent to the production of the catalogue. In particular, we
attempt to infer true distances from the \gaia observations, the
relationship between absolute $G$ magnitudes and the $(G-G_{RP})$ at
the faint end of the Main Sequence, and the empirical (that is,
affected by truncation) luminosity distribution.

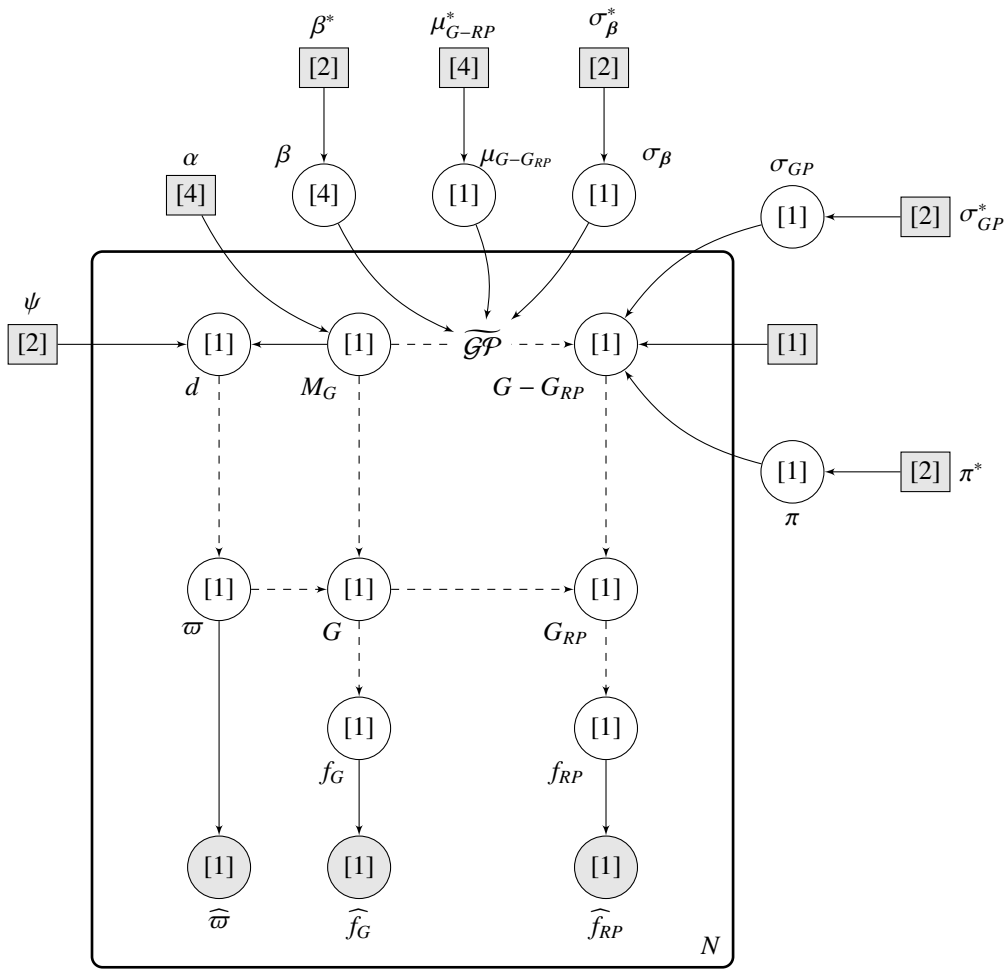
\begin{figure*}
    \centering
%
%
%
\begin{tikzpicture}[>=stealth', node distance = 1cm, every node/.style={rectangle,fill=white}]

\node[draw,  circle, label={[name=Gabs_i_lab,xshift=-15pt,yshift=3pt]below:$M_G$}] (Gabs) at (0,0) {$[1]$};
\node[draw, fill = gray!20, label={[name= alpha_lab]above:$\alpha$}, yshift=20pt, xshift=-25pt] (alpha) [above left = of Gabs] {$[4]$};
\node[draw, circle, label={[name=G_lab,xshift=-10pt,yshift=3pt]below:$G$}] (G) [below = of Gabs, , yshift=-40pt] {$[1]$};
\node[draw, circle, label={[name=flux_G_lab,xshift=-10pt,yshift=3pt]below:$f_G$}] (F_G) [below = of G] {$[1]$};
\node[draw, circle, fill = gray!20, label={[name=flux_G_lab,xshift=0pt,yshift=0pt]below:$\widehat{f}_G$}] (F_G_obs) [below = of F_G] {$[1]$};

\node[draw, circle, label={[name= beta_lab,xshift=-15pt,yshift=-5pt]above:$\beta$}] (beta) [right = of alpha] {$[4]$};
\node[draw, circle, label={[name= mu_RP_lab,xshift=20pt,yshift=-5pt]above:$\mu_{G-G_{RP}}$}] (mu_GP) [right = of beta] {$[1]$};
\node[draw, circle, label={[name= mu_RP_lab,xshift=20pt,yshift=-5pt]above:$\sigma_{\bm{\beta}}$}] (sigma_GP) [right = of mu_GP] {$[1]$};

\node[draw, fill = gray!20, label={[name= beta_lab]above:$\beta^*$}, ] (beta_hyper) [above = of beta] {$[2]$};
\node[draw, fill = gray!20, label={[name= mu_GP_lab]above:$\mu^*_{G-RP}$}] (mu_GP_hyper) [above = of mu_GP] {$[4]$};
\node[draw, fill = gray!20, label={[name= mu_GP_lab]above:$\sigma^*_{\bm{\beta}}$}] (sigma_GP_hyper) [above = of sigma_GP] {$[2]$};

\node[draw, circle, label={[name= alpha_lab, xshift=-25pt,yshift=3pt]below:$G-G_{RP}$}] (G_RP) [right = of Gabs, xshift=40pt] {$[1]$};

\node[draw, circle, label={[name= flux_rp_lab, xshift=-15pt,yshift=3pt]below:$G_{RP}$}] (RP) [below = of G_RP, yshift=-40pt] {$[1]$};
\node[draw, circle, label={[name= flux_rp_lab, xshift=-15pt,yshift=3pt]below:$f_{{RP}}$}] (F_RP) [below = of RP] {$[1]$};
\node[draw, circle, fill = gray!20, label={[name= F_RP_obs_lab]below:$\widehat{f}_{{RP}}$}] (F_RP_obs) [below = of F_RP] {$[1]$};


\node[draw, fill = gray!20, label={[name=gamma_lab_hyper]}] (gamma) [right = of G_RP, xshift=20pt] {$[1]$};
\node[draw, circle, label={[name=sigma_hyper_GP]above:$\sigma_{GP}$}] (sigma_GP2) [above = of gamma] {$[1]$};
\node[draw, fill = gray!20, label={[name=sigma_hyper_hyper_GP]right:$\sigma^*_{GP}$}] (sigma_hyper_GP2) [right = of sigma_GP2] {$[2]$};

\node[draw, circle, label={[name=prop_hyper_GP]below:$\pi$}] (prop_GP) [below = of gamma] {$[1]$};
\node[draw, fill = gray!20, label={[name=prop_hyper_hyper_GP]right:$\pi^*$}] (prop_hyper_GP2) [right = of prop_GP] {$[2]$};



\node[draw,  circle, label={[name=d_i_lab,xshift=-10pt,yshift=3pt]below:$d$}] (d) [left= of Gabs] {$[1]$};
\node[draw, fill = gray!20, label={[name= psi_lab]above:$\psi$}, xshift=-20pt] (psi) [left = of d] {$[2]$};
\node[draw, circle, label={[name= plx_lab_i,xshift=-10pt,yshift=3pt]below:$\varpi$}] (varpi) [below = of d, yshift=-40pt] {$[1]$};
\node[draw, circle, fill=gray!20, label={[name= plx_lab_i,xshift=0pt,yshift=0pt]below:$\widehat{\varpi}$}] (varpi_obs) [left = of F_G_obs] {$[1]$};

\path[line] (alpha) to [bend right=20] (Gabs);
\path[line] (psi) -> (d);
\path[line, dashed] (d) -> (varpi);
\path[line] (varpi) -> (varpi_obs);

\path[line, dashed] (varpi) -> (G);

\path[line, dashed] (Gabs) -> (G);
\path[line] (Gabs) -> (d);

\path[line, dashed] (G) -> (F_G);
\path[line] (F_G) -> (F_G_obs);

\path[line, dashed] (Gabs) edge node[name=la] {$\widetilde{\mathcal{GP}}$} (G_RP);


\path[line, dashed] (G_RP) -> (RP);
\path[line, dashed] (G) -> (RP);
\path[line] (gamma) -> (G_RP);

\path[line] (sigma_GP2) to [bend right=20] (G_RP);
\path[line] (prop_GP) to [bend left=20] (G_RP);
\path[line] (sigma_hyper_GP2) -> (sigma_GP2);
\path[line] (prop_hyper_GP2) -> (prop_GP);

\path[line, dashed] (RP) -> (F_RP);
\path[line] (F_RP) -> (F_RP_obs);





\path[line] (beta) to [bend right=20] (la);
\path[line] (mu_GP) to [bend left=15] (la);

\path[line] (sigma_GP) to [bend left=10] (la);

\path[line] (beta_hyper) -> (beta);
\path[line] (mu_GP_hyper) -> (mu_GP);
\path[line] (sigma_GP_hyper) -> (sigma_GP);

\begin{pgfonlayer}{background}
\node (rect) at (Gabs) [xshift = 20pt, yshift = -100pt, draw, rounded corners, thick, minimum width=240pt, 
minimum height=270pt, line width=1pt] (rect1) {};
\end{pgfonlayer}

\node[anchor=south east,inner sep=5pt, fill = none] at (rect1.south east) {$N$};

\end{tikzpicture}
    \caption{Hierarchical model in plate notation. Squares indicate
      fixed hyperparameters and circles indicate random
      variables. Filled-in shapes indicate fixed values that are not
      subject of inference. Dashed lines indicate a functional
      non-probabilistic relationship. The dimension of the parameter
      vectors are indicated inside the shapes inside brackets and the
      fixed hyperparameters are listed in Table~\ref{tab:prior}. For
      instance, $\sigma^*_{\bm{\beta}} = (0, 0.5)$ are the parameters
      of the Lognormal prior for $\sigma_{\bm{\beta}}$.}
    \label{fig:model_plate_notation}
\end{figure*}

We have defined a hierarchical Bayesian model that attempts to capture
the probabilistic relationship between unobserved physical quantities
(distances, absolute magnitudes, etc.) and the observations
$\mathcal{D}$. In Bayesian inference the posterior probability of the
model parameters $\bm{\theta}$ (the object of our inference) is
related to the prior probability and the likelihood according to
Bayes' theorem:
\begin{equation}
    p(\bm{\theta} | D, \mathcal{H}) \propto p(D|\bm{\theta}, \mathcal{H}) \cdot p(\bm{\theta} | \mathcal{H}),
    \label{eq:bayes}
\end{equation}
\noindent
where $p(\bm{\theta} | D, \mathcal{H})$ is the posterior probability
of the model parameters $\bm{\theta}$, $p(\bm{\theta} | \mathcal{H})$
the prior probability distribution, and $p(\boldsymbol{\mathcal{D}}|
\bm{\theta}, \mathcal{H})$ the likelihood. The notation $\mathcal{H}$
refers to all the physical assumptions that define the hierarchical
model. In our case $\bm{\theta}$ is a high-dimensional parameter
vector and for the sake of clarity, we will distinguish two components
of $\bm{\theta}$ involved in different parts of the model's
hierarchy. We subdivide the set of all parameters as
$\bm{\theta}=(\bm{\theta}_{\mathcal GP}, \bm{\theta}_{star})$, where
$\bm{\theta}_{star} = (\textbf{d}, \textbf{M}_G, \textbf{G}_{RP})$
represents the vectors of true values of the distance ($\textbf{d}$),
absolute magnitude ($\textbf{M}_G$) and $\textbf{G}_{RP}$ magnitude
for the set of $N$ stars used for inference. The relation between the
true absolute magnitude $M_G$ and the true colour index $G-G_{RP}$ is
modelled with an approximation of a Gaussian Process \citep[][see
  Fig.~\ref{fig:model_plate_notation} and the explanation
  below]{Higdon2020, GPs4ML} and $\bm{\theta}_{\mathcal
  GP}=(\mu_{G-G_{RP}}, \bm{\beta}, \sigma_{\bm{\beta}}, \gamma,
\delta)$ is the subset of parameters that defines that
approximation. Taking into account this notation, the prior
distribution in equation \ref{eq:bayes} is such that
\begin{equation*}
    P(\bm{\theta} | \mathcal{H})=
    P((\bm{\theta}_{\mathcal GP} | \mathcal{H}) \cdot
    P(\bm{\theta}_{star} | \bm{\theta}_{\mathcal GP}, \mathcal{H})
\end{equation*}

Fig.~\ref{fig:model_plate_notation} shows graphically the
probabilistic hierarchical model and Table~\ref{tab:prior} lists the
(hyper-)parameters prior probability distributions specified in the
following paragraphs where we also explain the conditional relations
underlying the model.

\setlength{\tabcolsep}{12pt}
\begin{table}[h]
    \centering{
      \caption[]{Prior distributions for the parameters in the
        inference model.  Due to its hierarchical nature, some
        distributions depend on others. Specific details and
        definition of the PERT distribution are given in Appendix      \ref{appendix:PERT}}.
    \label{tab:prior}
    \begin{tabular}{@{}ccc@{}}
    \hline\hline
    Parameter & Prior & Units \\ \hline
    $\mu_{G-RP}$ & ${\rm PERT} (1, 1.5, 2, 5)$ & mag \\
     $\sigma_{\bm{\beta}}$ & ${\rm Lognormal} (0, 0.5)$ & mag \\
    $\bm{\beta}$ & $\mathcal{N}_{4} (0, \sigma_{\bm{\beta}})$ & mag \\
    $M_{G, i}$ & ${\rm PERT} (13.7, 14.49, 21.5, 21.73)$ & mag \\
    $d_i$ & ${\rm CPD} (M_{G,i}, d_{\max}, d_{\rm extra})$ & pc \\
    $\sigma_{GP}$ & ${\rm Lognormal} (-3, 0.1)$ & mag \\
    $\pi$ & $\mathcal{U} (0, 0.15)$ & - \\
    $G_i - G_{RP,i}$ & $(1 - \pi) \cdot \mathcal{N} (\widetilde{\mu}, \sigma_{GP}) + \pi \cdot \mathcal{N} (\widetilde{\mu}, 1.5)$ & mag \\
    \hline 
    \end{tabular}}%
   \end{table}

At the top level of the graph, the prior for the absolute magnitude
$M_G$ is a ${\rm PERT} (13.7, 14.49, 21.5, 21.73)$ distribution (see
Appendix \ref{appendix:PERT} for a definition of the PERT distribution
and parameters). The PERT $peak$ and $temperature$ parameters (14.49
and 21.73 respectively) are the maximum likelihood estimates obtained
after fixing the $low$ and $high$ parameters to 13.7 and 21.5 mag
respectively (values that include and extend the range of absolute
magnitudes obtained by naive inversion of the parallaxes). The
parameter name $temperature$ can be misleading in this context but we
maintain it for consistency with the literature. This distribution
peaks at the same absolute magnitude as the kernel density estimate of
the distribution of observed absolute magnitudes but has non-vanishing
prior probability densities in the range $(13.7, 21.5)$ mag. From the
true value of $M_G$ we obtain the probability distribution for true
distance $d$ using a custom probability distribution (CPD) defined as
\begin{equation}
    f(x) = 
    \begin{cases}
    \cfrac{x^2}{C} &\quad x < d_{\max}\\
    \cfrac{d^2_{\max} \cdot \exp \left(\cfrac{d_{\max}-x}{H} \right)}{C} &\quad x \geq d_{\max}
    \end{cases}
\label{eq:dist-distribution}
\end{equation}
\noindent where $d_{\max}$ and $C$ are as follows
\begin{eqnarray*}
d_{\max} &=& 10^{-(M_G-20.48-5)/5}, \\
d_{\rm extra} &=& 10^{-(M_G-20.7 -5)/5}, \\
H&=&\cfrac{d_{\max}-d_{\rm extra}}{\log(0.01)}, \\
C&=&\frac{d^3_{\max}}{3} - d^2_{\max} \cdot H \cdot \left(\exp \left( \frac{d_{\max} - d_{\rm extra}}{H} \right) -1 \right).
\end{eqnarray*}

Equation \ref{eq:dist-distribution} represents the expected
distribution of distances for a uniform volume density of UCDs with an
exponential decay at the maximum distance $d_{\max}$ defined by the
absolute magnitude $M_G$ and the \gaia limiting magnitude for
completeness assumed here to be 20.48 mag
\citep{2021A&A...649A...6G}. The exponential decay is introduced to
incorporate the fact that there is no hard cut in magnitudes that can
be detected by \gaia and is parametrized by a length scale $H$ that
defines the range of distances $d_{\rm extra}$ beyond $d_{\max}$ over
which the decay takes place. The value of $d_{\rm extra}$ is defined
for convenience based on a nominal limiting magnitude of 20.7 mag
\citep{2016A&A...595A...2G}. Several simplifications are adopted here
that will be lifted in subsequent investigations that will properly
incorporate the selection function. For example, the vertical
stratification of the Milky Way disc is a more realistic prescription
of the volume density than the uniform distribution \citep[see for
  example][]{2021A&A...649A...6G}. Additionally, neither 20.48 nor
20.7 mag are the \gaia limiting magnitudes as these depend on the
celestial coordinates under consideration via the scanning law.

The true $G_{RP}$ magnitude of every star given the absolute magnitude
follows a probability distribution given by a Gaussian Process. This
Gaussian Process models the faint end of the Main Sequence in the CAMD
constructed with $M_G$ and $(G-G_{RP})$. For the sake of computational
efficiency, we use an approximate Gaussian Process
$\widetilde{\mathcal{GP}}$ (see appendix~\ref{appendix:GP}) that
depends on a set of parameters $\mu_{G-G_{RP}}$, $\bm{\beta}$ and
$\sigma_{\bm{\beta}}$. Their prior probability densities are defined
as ${\rm PERT} (1, 1.5, 2, 5)$, $\mathcal{N}_{4}(0, 1)$ and ${\rm
  Lognormal} (0, 0.5)$, respectively. Finally, we assume that the
approximate Gaussian Process describes the faint end of the Main
Sequence, but also that for a given absolute magnitude outliers can
exist that result in a skewed distribution of $G-G_{RP}$ colour
indices with respect to the GP prescription. We model the presence of
outliers using a Gaussian Mixture distribution:
\begin{equation}
(G_i-G_{RP, i}) \sim (1 - \pi) \cdot \mathcal{N} (\widetilde{\mu}, \sigma_{GP}) + \pi \cdot \mathcal{N} (\widetilde{\mu}, 1.5),
\end{equation}
\noindent where $\widetilde{\mu_i} = \widetilde{\mathcal{GP}}(M_{G,i}, \mu_{G-RP}, \bm{\beta})$.
Hence, the distribution of colour indices has its mode given by the
Gaussian Process with an added scatter term parameterised by
$\sigma_{GP}$. Outliers are modelled using a wide second Gaussian
distribution with the same mean and a standard deviation fixed to
$1.5$ mag. The prior probability density for $\sigma_{GP}$ is defined
as a ${\rm Lognormal} (-3,0.1)$. The prior distribution for the
mixture proportion follows a Uniform distribution between 0 and 0.15.

The model parameters $d$, $M_G$ and $G_{RP}$, all represent true
values. From them, we can derive true values of the parallax $\varpi$
and of the fluxes $f_G$ and $f_{RP}$ as
\begin{eqnarray*}
\varpi &=& 1000 / d,\\
G &=& M_{G} - 5 \cdot \log_{10}(\varpi/1000) -5, \\
f_{G} &=& 10^{(G - 25.6874) / -2.5},\\
f_{RP} &=& 10^{(G_{RP} - 24.7479) / -2.5}.
\end{eqnarray*}

In the likelihood, we assume that the \gaia observations for each UCD
are independent and $P(\boldsymbol{\mathcal{D}} | \bm{\theta},
\mathcal{H})$ can be factorised as
$$P(\boldsymbol{\mathcal{D}} | \bm{\theta}, \mathcal{H}) = 
\prod_{i=1}^N p(\mathcal{D}_i | \bm{\theta}, \mathcal{H}),$$
where the likelihood for each UCD, $p(\mathcal{D}_i | \bm{\theta},
\mathcal{H})$, is defined as the product of three Normal distributions
centred at the true values and with standard deviations given by the
catalogue uncertainties:
\begin{equation}
p(\mathcal{D}_i | \bm{\theta}, \mathcal{H}) = 
\mathcal{N}(\widehat{\varpi} \, | \, \varpi, \widehat{\sigma}_{\varpi}) \cdot
\mathcal{N}(\widehat{f}_{G} \, | \, f_{G}, \widehat{\sigma}_{f_{G}}) \cdot
\mathcal{N}(\widehat{f}_{RP} \, | \, f_{RP}, \widehat{\sigma}_{f_{RP}}).
\end{equation}

\begin{figure}
    \centering
    \includegraphics[width=\linewidth]{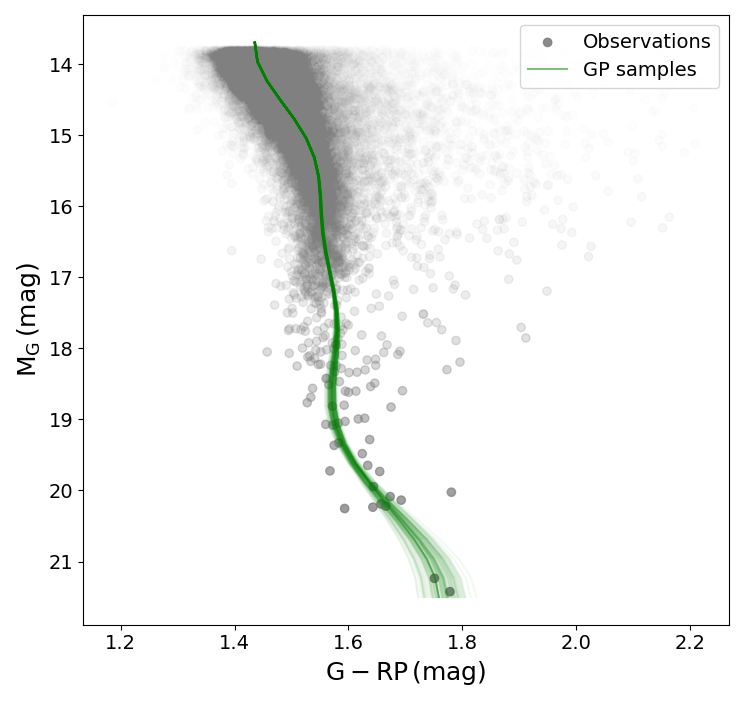}
    \caption{CAMD with posterior samples of the Gaussian Process (green) and the observations (black). The transparency and the symbol
    sizes were chosen to enhance visibility of the main densities.}
    \label{fig:posterior_CMD}
\end{figure}

The data comprise the measured parallaxes and fluxes in the $G$ and
$RP$ bands of all sources in the \gaia UCD catalogue with quality
categories 0 or 1 and their associated uncertainties:
$$\boldsymbol{\mathcal{D}}=
\{\widehat{\varpi}_i, \widehat{\sigma}_{\varpi,i},
\widehat{f}_{G, i}, \widehat{\sigma}_{f_{G},i},
\widehat{f}_{RP, i}, \widehat{\sigma}_{f_{RP},i} \}_{i=1}^{N},$$
\noindent where $\widehat{f}_{G, i}$ and $\widehat{f}_{RP, i}$ are the
measured fluxes in the $G$ and $RP$ bands respectively, and the index
$i$ runs between 1 and the total number of stars in the sample. The
notation uses the circumflex symbol to distinguish observations
(assumed affected by measurement noise) from true values. The
observations amounts to a total number of stars $N$=67\,428. From this
set we select only sources with $G+5\cdot\log_{10}(1000/\varpi)+5 >
13.8$ to avoid including the very young sources discussed in
Sect. \ref{sec:young}. A fraction of very young sources will still be
present in the data but these will be dealt with under the category of
outliers (as described later in the Sect.). The total number of
sources used in the inference is then 43\,795.

As usual in this kind of complex problems, it is not possible to
obtain a closed-form expression for the posterior probability
distribution from the prior and the likelihood, and we will describe
the posterior distribution of the parameters using samples obtained
with Markov Chain Monte Carlo (MCMC) sampling techniques. Figure
\ref{fig:posterior_CMD} shows the data used for the inference and 100
posterior samples of the Gaussian Process (green). We have used the
\espucd effective temperature estimates to define a relationship
between $M_{G}$ and an average \teff using the Gaussian Process mean,
and weighting the contribution of each source to the effective
temperature. Table \ref{tab:mg-teff} lists the average \teff values
for several $M_{G}$ values along the Gaussian Process sequence.

\begin{table}[ht]
    \centering
    \begin{tabular}{cc}
\hline\hline
        $M_{G}$ (mag) & $T_{\rm eff}$ (K)\\
\hline
    14     &  2650.5\\
    15     &  2477\\
    16     &  2317\\
    17     &  2050\\
    18     &  1836\\
    19     &  1553\\
    20     &  1374\\
    21     &  1145\\
\hline
    \end{tabular}
    \caption{Weighted average of the temperatures estimated by the \espucd module for several absolute magnitudes along the mean of the Gaussian Process posterior.}
    \label{tab:mg-teff}
\end{table}

\begin{figure}
    \centering
    \includegraphics[width=\linewidth]{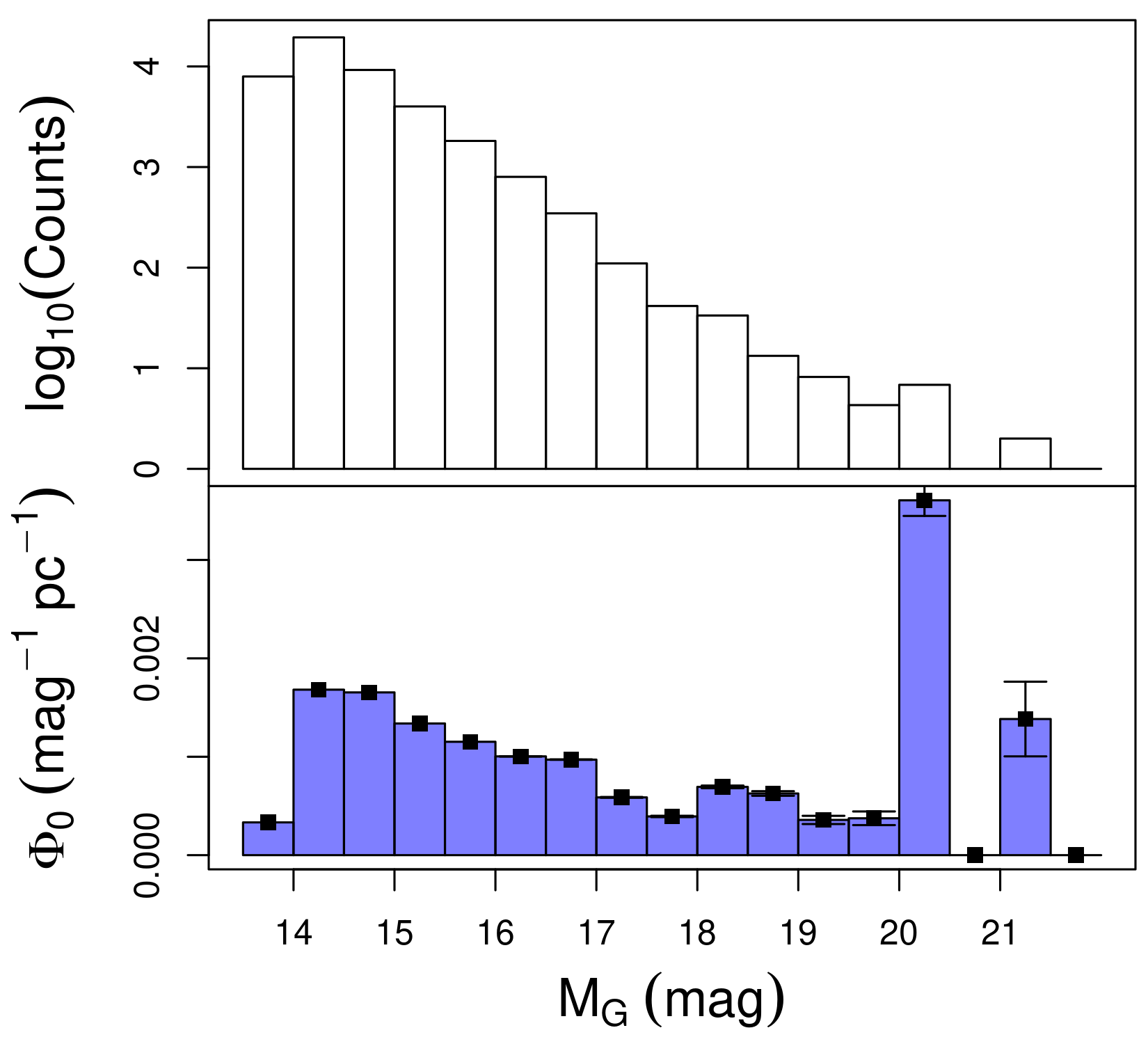}
    \caption{(Top panel) Histogram of the absolute magnitudes inferred
      by the hierarchical Bayesian model. (Bottom panel) Luminosity
      function (in units of mag$^{-1}$ pc$^{-3}$) derived as described
      in the text.}
    \label{fig:luminfuncs}
\end{figure}

Figure \ref{fig:luminfuncs} shows in logarithmic scale the empirical
luminosity function derived from the absolute magnitudes inferred by
the model (top panel). It has been derived using the 600 samples from
the posterior and then dividing the counts in each bin of 0.5 mag by
600. Since we cut the initial sample at $M_G= 13.8$ mag, the first bin
of the histogram is only partially observed and can be
ignored. Obviously this empirical distribution function is affected by
all the selection biases derived from the filters described in
Sect. \ref{sec:candidatesel} including the decreasing 3D volume
explored as the absolute magnitude bins get fainter.

The true luminosity function can in principle be recovered if we can
formulate all the filters applied as selection functions. These
selection filters (listed below) are described in
\linksec{sec_cu8par_apsis/ssec_cu8par_apsis_espucd.html}{the \espucd
  Section of the official documentation}, in \cite{DR3-DPACP-157}, and
in \cite{DR3-DPACP-160}. \cite{2021AJ....162..142R} provide a
simplified example of this type of reconstruction using \gaia data and
focusing on the domain of White Dwarfs. In our case, the first factor
of the selection function comes from the filters that define the
sources to which ESP-UCD is applied. These are the $\approx$ 47
million sources with $G-G_{RP} > 1$ and $\varpi > 1.7$ mas. While
these filters may remove very young and bright sources that could be
detected beyond this limit or subdwarfs redder than 1.0 mag in
$G-G_{RP}$, we assume that their numbers (if not zero) are small
enough to neglect their effect in the reconstruction of the luminosity
function. However, we assume that only 89\% of the UCDs have $\varpi$,
$G$ and $G_{RP}$ measurements available (see the GUCDS analysis in
Sect. \ref{sec:comparisonCats}).

The next selection filter is applied by the \espucd module based on
the RP spectrum pixels where the 33, 50 and 67 flux percentiles are
attained. Since the filter definitions are inclusive definitions of
the hot boundary of the UCD regime, in principle we only expect
selection effects at that boundary. The data set used for inference
with the Bayesian model is cut at $M_G=13.8$ mag and we do not expect
significant selection effects. This is illustrated in
Figs. \ref{fig:camds-dense} and \ref{fig:hmac-banyan-1} that show that
$M_{G}=13.8$ leaves out all the very young sources discussed in
Sect. \ref{sec:young} and the brightest end of the parallel sequence
of equal mass binaries.

Finally, the main selection filters applied are related to the quality
of the RP data. This includes the removal of sources with less than 15
RP transits, high RP median curvature values (above $2.0\cdot
10^{-5}$), large Euclidean distances to the set of templates $d_{TS}$
and/or poor astrometric measurements (the \espucd module removed
sources with $\log_{10}(\sigma_{\varpi}) > -0.8 + 1.3 \cdot
\log_{10}(\varpi)$). While the requirement on the number of RP
transits depends on the celestial coordinates of each source in a way
that is complex but quantifiable, the selection function related to
the RP quality requirements are difficult to estimate. The quality
classes used in this Sect. (0 and 1) were defined in terms of the
decadic logarithm of the Euclidean distances $d_{TS}$ to the spectral
type standards and of the relative RP flux uncertainties
$\sigma_{RP}/f_{RP}$. As a first order approximation to the selection
function, we fit mixtures of Gaussian components to the distributions
of sources in the 2D spaces $\log_{10}(d_{TS})$ --$\sigma_{RP}/f_{RP}$
and $\log_{10}(\varpi)$--$\log_{10}(\sigma_{\varpi})$. We use the
Gaussian components to estimate the number of sources lost in the
application of the filters. We further assume that the distribution of
$M_{G}$ of these sources is the same as that of the UCDs with quality
class 2 (the worst quality class of the three). From them, we estimate
a $M_G$-dependent selection function. The lower panel of
Fig. \ref{fig:luminfuncs} shows the resulting point estimates of the
luminosity function derived as in \cite{2021AJ....162..142R} assuming
an exponentially decaying spatial density with scale height $H=365$
\citep{2021A&A...649A...6G}, Poisson distributions, and a flat prior
for the number density $\Phi_0(M_G)$. The last bin of the luminosity
function is affected by the low number of sources but the Poisson
uncertainty makes it compatible with the trend that can be deduced
from the rest of the function. The bin from $M_{G}=20$ to $20.5$ mag,
however, seems inconsistent with the rest of the bins and is due to
the sequence of UCDs in the CAMD running horizontally at those
absolute magnitudes (see Fig. \ref{fig:camds-dense}). This Fig. is in
broad agreement with other distributions in the literature \citep[for
  example,][]{2021ApJS..253....7K,2019ApJ...883..205B} in the range of
absolute magnitudes from 14 to 20 mag.

As mentioned above, this is the result of a simplified treatment of
the selection function applied to the raw data and a detailed
comparison with existing luminosity functions in the literature is
postponed until a more accurate specification of the selection
function is available.

\section{UCDs in binary systems}
\label{sec:binaries}

In this Section we explore the existence of co-moving pairs consisting
of sources in the UCD catalogue and a primary component. We search the
\gaia archive for potential primaries that fulfil the following
criteria (where the subindices $p$ and $UCD$ denote the primary and
the UCD components of the pair):
\begin{enumerate}
    \item the projected distance (in the tangential plane of the sky
      and calculated using the naive inversion of the parallax and the
      sine of the angular separation) from the UCD candidate is less
      than 0.1 pc
    \item its parallax ($\varpi_{p}$) is greater than 1 mas
    \item $\varpi_{p}$ is in the interval defined by $
      \varpi_{UCD} \pm 3\cdot
      (\sigma_{\varpi_{p}}+\sigma_{\varpi_{UCD}})$
    \item $\mu_{RA;p}$ is in the interval defined by $ \mu_{RA;UCD}
      \pm 3\cdot (\sigma_{\mu_{RA};p}+\sigma_{\mu_{RA};UCD})$
    \item $\mu_{Dec;p}$ is in the interval defined by $ \mu_{Dec;UCD}
      \pm 3\cdot (\sigma_{\mu_Dec;p}+\sigma_{\mu_{Dec};UCD})$
    \item the absolute magnitude of the primary (assuming negligible
      extinction and inferred by inverting the parallax) is brighter
      than that of the UCD.
\end{enumerate}

\begin{figure}[ht]
	\centering
	\includegraphics[scale=0.85]{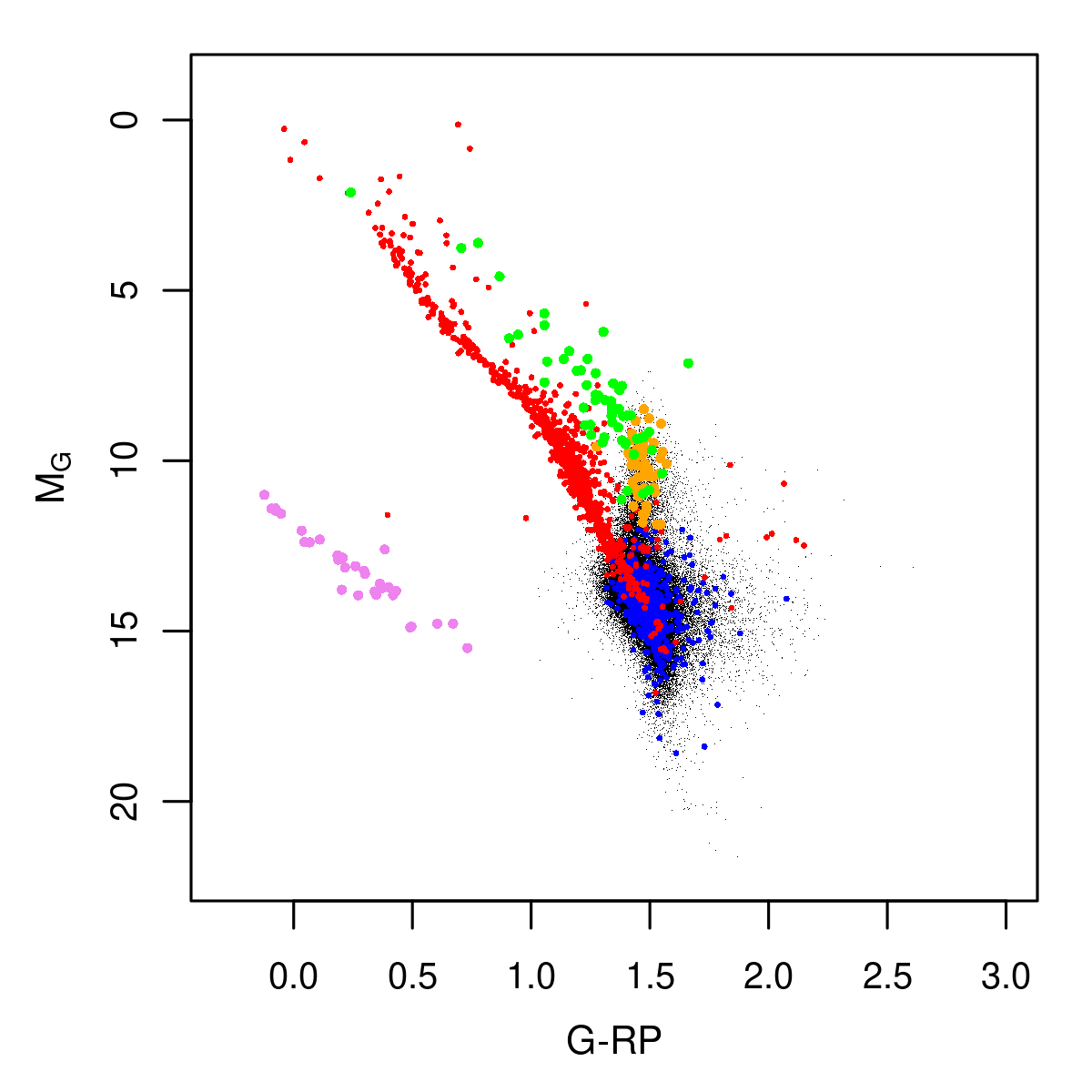}
	\caption{Absolute CMD for binary candidates. Black dots
          represent all UCD candidates, blue and red circles represent
          UCD secondaries and candidate primaries respectively with
          $\varpi/\sigma_{\varpi} > 15$. Magenta circles correspond to
          sources identified as white dwarfs using a linear boundary
          $M_G > 10+6\cdot (G-G_{RP})$. Orange circles tag UCD
          candidates in stellar associations and green circles their
          corresponding primaries.}
	\label{fig:CAMD-primaries}
\end{figure}

Adding the uncertainties in quadrature would imply a more restrictive
threshold which at this stage was not necessary. This leads to a list
of 28\,704 candidates primaries (that is, 28\,704 potential pairs
containing at least one UCD).  Figure \ref{fig:CAMD-primaries} and
subsequent figures in this Section concentrate on the 880 systems
where both components have SNR ratios $\varpi/\sigma_{\varpi} >
15$. This selection removes the large fraction of unlikely primary
candidates with large parallax uncertainties. These large
uncertainties make them pass the selection criteria even though their
parallaxes are very different from the UCD parallaxes. Thirty-two of
the 880 candidate primaries are included in the UCD catalogue and thus
represent cases of UCD pairs and we find two triple systems
candidates.

For the candidate primaries in the White Dwarf (WD) sequence we do not
find differences in the distribution of the separations with respect
to Main Sequence primaries as could be expected if UCDs in wide
binaries do not survive long because they are less gravitationally
bound. In Figure \ref{fig:CAMD-primaries} we have also marked in
orange UCDs with $M_{G}$ > 12 and in green their corresponding
primaries. The overluminous absolute magnitudes of these primaries are
in agreement with the assumed indication of youth ascribed to the UCD
components (see Sect. \ref{sec:young})

\begin{figure}[ht]
	\centering
	\includegraphics[scale=0.6]{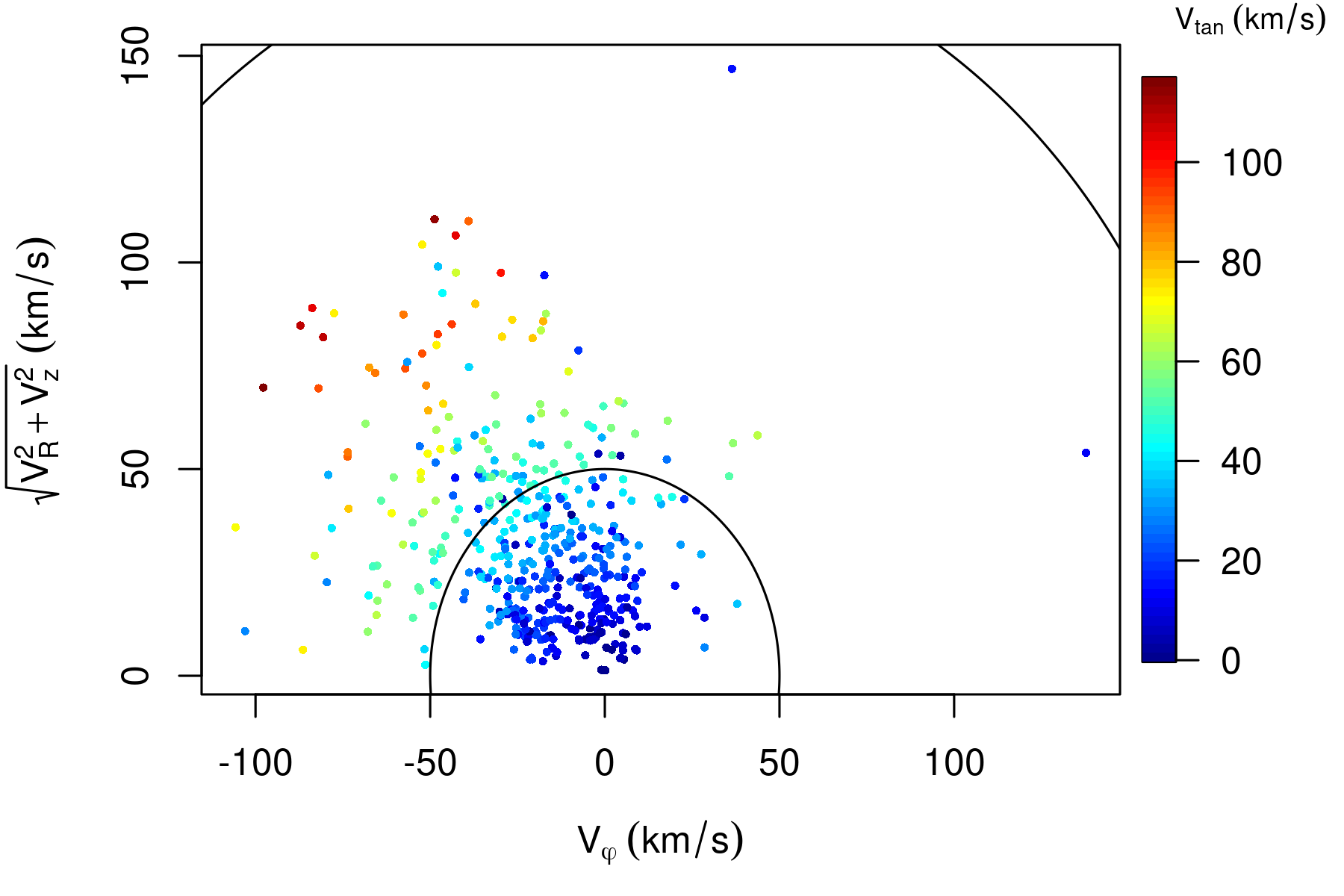}
	\caption{Toomre diagram of the primaries with UCD
          companions. The colour code reflects the tangential
          velocities. Common boundaries for the thin/thick disk and
          thick disk/halo components are included as half circles at
          50 and 180 km/s respectively.}
	\label{fig:toomre}
\end{figure}

We also searched the \gdrthree archive for radial velocities and
astrophysical parameters of the primaries. We find radial velocities
available for 465 of the 880 candidate primaries and results from the
FLAME module for 586 of them \citep[see][for a detailed description of
  the Apsis modules that produced stellar astrophysical parameters as
  part of the \gaia processing, including \flame and
  \gspphot]{DR3-DPACP-157,DR3-DPACP-160}. Figure \ref{fig:toomre}
shows the position of the primaries in the Toomre diagram. It shows
$\sqrt{V_R^2+V_z^2}$ and $V_\phi$ of the sources with radial velocity
measurement, where $(V_R, V_\phi, V_z)$ are the velocity components of
the stars in the Galactocentric cylindrical coordinate system, with
$R$ pointing from the Galactic centre to the Sun, $z$ along the axis
perpendicular to the Galactic plane, and $\phi$ along the azimuthal
direction in the Milky Way disc plane (defined such that $V_\phi$ is
positive for prograde stars in the disc). The calculation of $(V_R,
V_\phi, V_z)$ follows the same assumptions as adopted for the
selection of OBA stars in \cite{DR3-DPACP-123}. In particular, we
assume the local circular velocity from the \texttt{MWPotential2014}
Milky Way model \citep{2015ApJS..216...29B}, which is $219$~\kms at
the distance of the Sun from the Galactic centre
\citep[$8277$~pc,][]{2022A&A...657L..12G}. The height of the Sun above
the disc plane is assumed to be $20.8$~pc \citep{2019MNRAS.482.1417B}
and the peculiar motion of the Sun is assumed to be $(U,V,W)=(11.1,
12.24, 7.25)$~\kms \citep{2010MNRAS.403.1829S}. The Toomre diagram
seems to indicate that the UCDs in binary systems (with radial
velocities of the primary measured by \gaia) are a mixture of the thin
and thick disc components with no conspicuous member of the halo.

\begin{figure}[ht]
	\centering
	\includegraphics[scale=0.85]{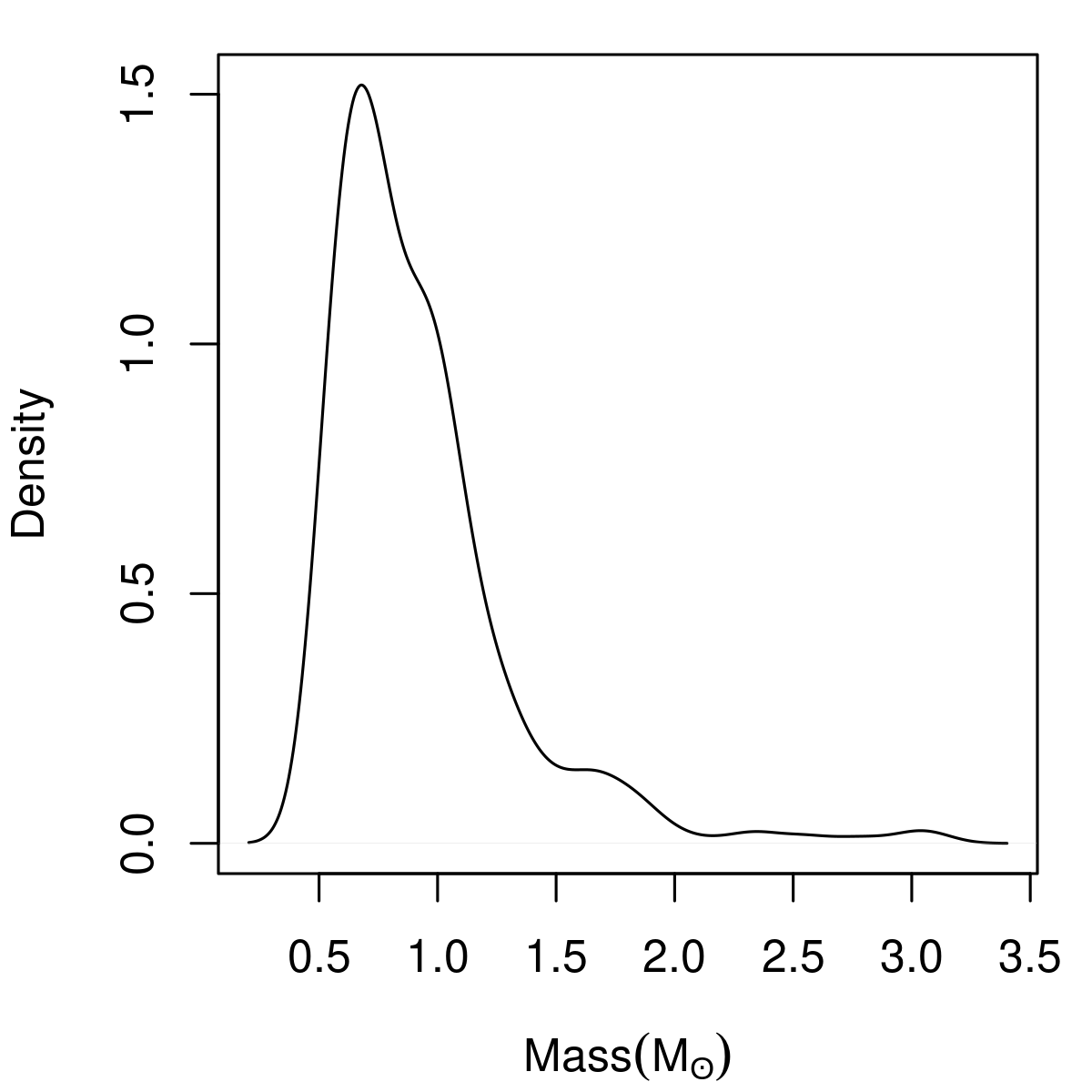}
	\caption{Kernel density estimate of the mass distribution for
          primaries with good astrometric measurements
          ($\varpi/\sigma_{\varpi}>15$).}
	\label{fig:prim-masses}
\end{figure}

\begin{figure}[ht]
	\centering
	\includegraphics[scale=0.6]{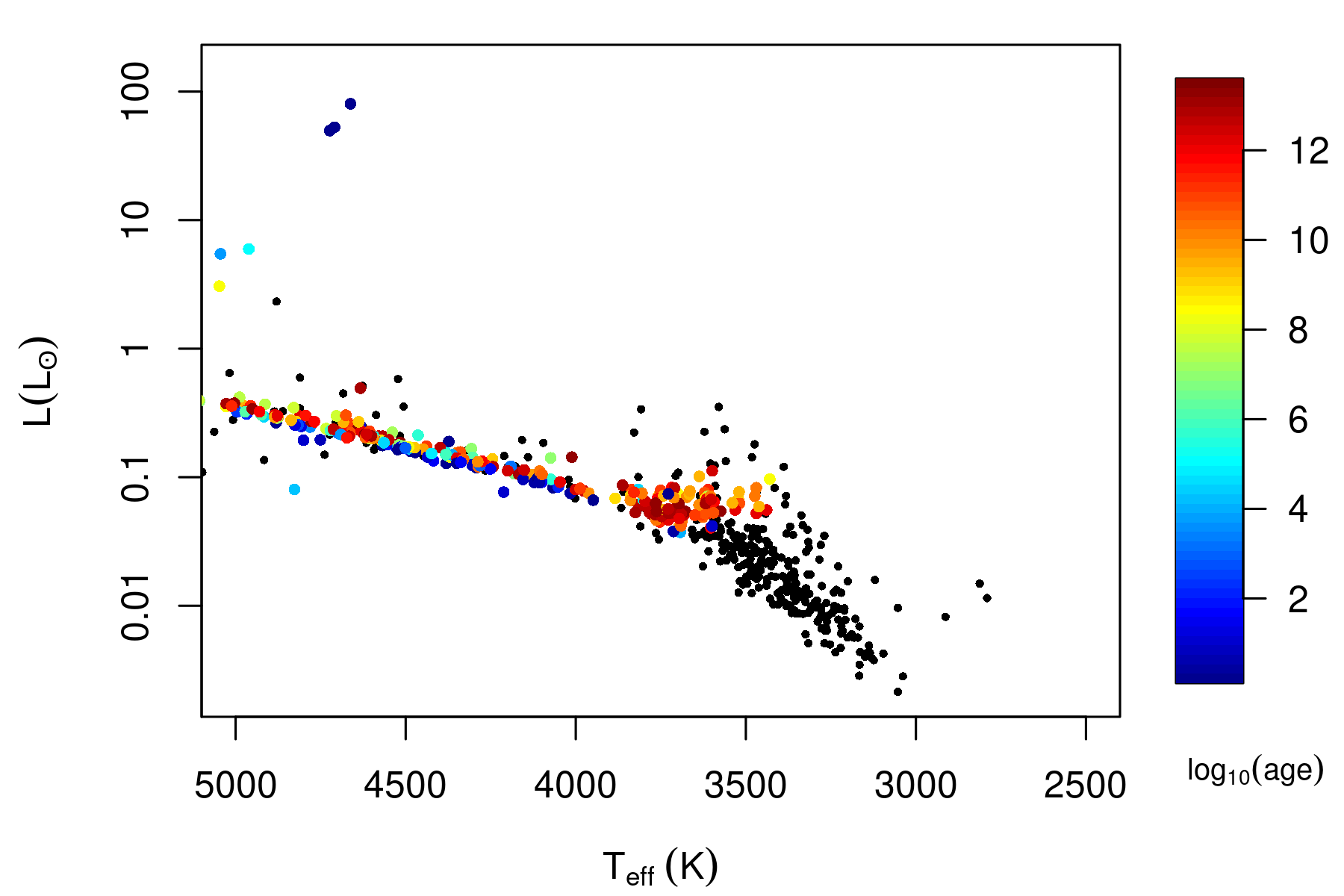}
	\caption{Scatter plot of effective temperatures from the
          \gspphot module ($x$ axis) and luminosities from \flame ($y$
          axis) for primaries in systems where both components have
          $\varpi/\sigma_{\varpi}>15$. Coloured circles correspond to
          sources with available age estimates from \flame. The colour
          code represents the decadic logarithm of the age.}
	\label{fig:prim-ages}
\end{figure}

FLAME produces amongst other, luminosities, ages, masses and radii
based on astrophysical parameters derived by the \gspphot module for
$\approx$ 280 million sources. However, FLAME ages and masses are
derived for a set of stellar models that do not include pre-Main
Sequence stages. They cover from the main-sequence to the tip of the
red giant branch, for masses between 0.5 and 10 solar masses, and for
a solar-metallicity prior. Figure \ref{fig:prim-masses} shows the
distribution of the masses of the (candidate) primaries with a peak at
$\sim 0.8 M_{\odot}$ larger than expected for the solar neighbourhood
\citep[around M3 or 0.3 $M_{\odot}$ according
  to][]{2021A&A...649A...6G,2018ApJ...861L..11J}. This bias arises
mainly from the requirement of availability of \flame masses (with a
minimum mass of 0.5 $M_{\odot}$) but also from the parallax SNR cut at
$\varpi/\sigma_{\varpi} > 15$. Figure \ref{fig:prim-ages} shows the
\teff-luminosity scatter plot colour coded by the decadic logarithm of
the age when available and only for primaries with UCD companions
characterised by $M_G > 12$ mag.

\section{UCDs in star forming regions, clusters and moving groups.}
\label{sec:young}

\begin{figure*}[th]
	\centering
        \includegraphics[width=1.00\textwidth]{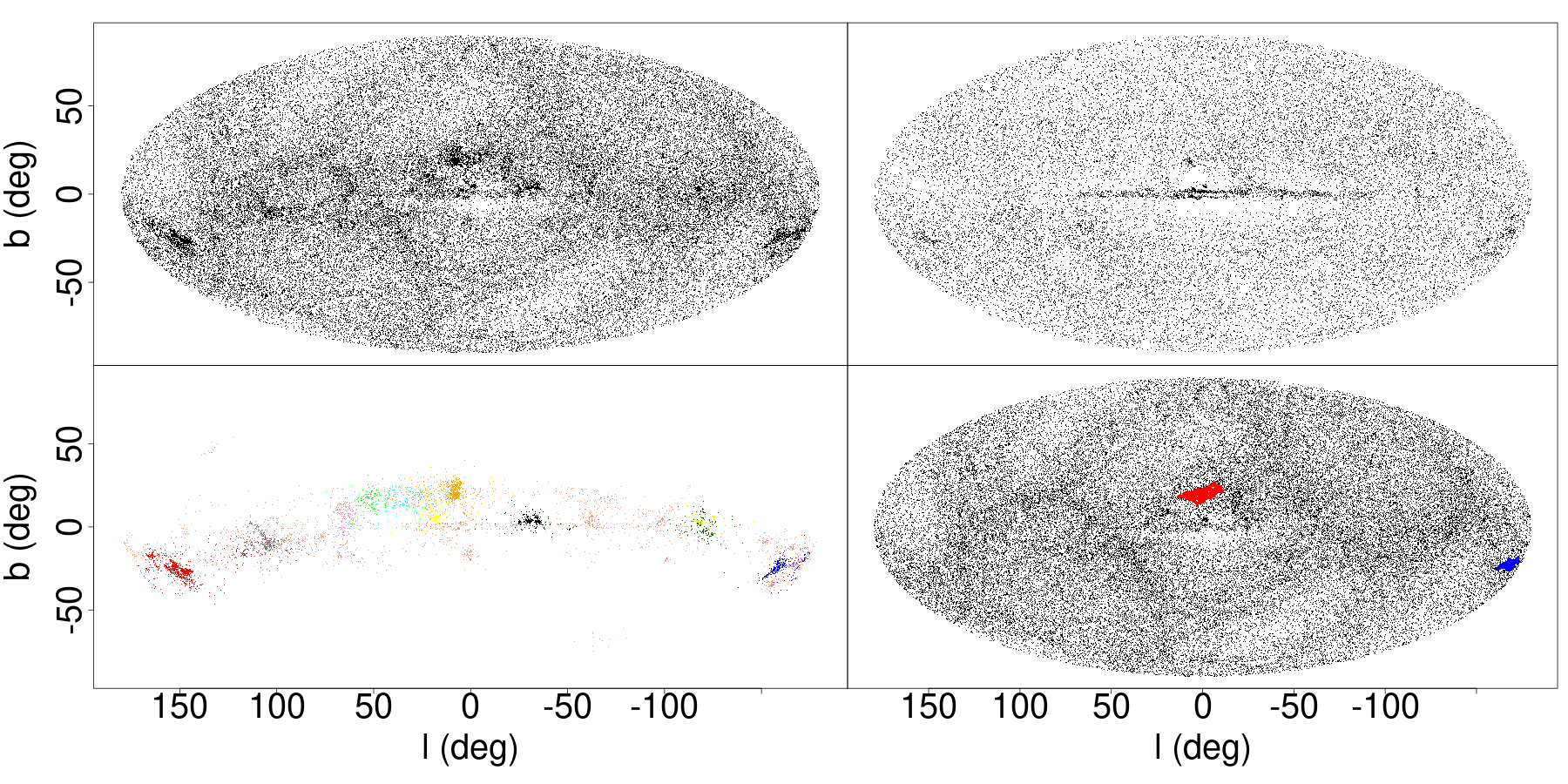}
	        \caption{Distribution in Galactic coordinates of the
                  sources in quality classes 0 and 1 (upper left) and
                  2 (upper right) of the \gaia UCD catalogue using the
                  Aitoff projection. Lower left panel: clusters at
                  level 4 of the hierarchy with more than 10 members
                  identified by HMAC in the set of sources in quality
                  classes 0 and 1. There are 88 such clusters, the
                  largest 13 of which are identified with prominent
                  colours and the rest are plotted in salmon for
                  clarity. Lower right panel: the rest of the sources
                  in clusters with 10 members or less. We mark in red
                  and blue the two most prominent overdensities.}
	        \label{fig:lb}
\end{figure*}

The distribution on the sky of sources in the \gaia UCD catalogue has
clear overdensities apparent in the top row of Figure \ref{fig:lb}. It
shows the distribution of sources in the two best \espucd categories
(qualities 0 and 1; left panel) and quality 2 sources (right). The
former shows overdensities that can be easily identified with open
clusters and star forming regions while the latter is dominated by an
overdensity aligned with the Galactic disc that we interpret as
residual contamination from bad astrometric solutions due to crowding
and reddened sources.

In this Section we explain how we determine membership of the UCD
candidates to several groups, show the position of the candidate
members in several CAMDs and illustrate the differences in the RP
spectra as a function of the age (taken from the literature). The
sizes and nature of the groups found can vary greatly. We will be
using loosely the terms cluster, association or group without implying
a specific range of sizes or complexities in terms of members.

We identify the above mentioned overdensities by using a clustering
technique applied to the set of sources in quality classes 0 and 1 of
the \gaia UCD catalogue in the five-dimensional space of Galactic
coordinates, tangential velocities and parallaxes. The clustering
analysis in this 5D space uses the Hierarchical Mode Association
Clustering (HMAC) algorithm
\citep{651d1d8e5a5a4871b38e424902d0098c}. HMAC defines groups as sets
of points associated to each mode (maxima) of the density distribution
in the input space. It does not explicitly estimate the density but it
makes use of a kernel inside an iterative loop that associates sources
with modes/groups. Each kernel defines a set of modes, with narrower
kernels resulting in many modes (that may include noise) and wider
kernels reflecting only the larger structures. By using several
kernels of increasing size we attain a hierarchical stratification of
clustering groupings. The term cluster in this context will designate
one of the groups of UCDs identified by the HMAC algorithm as
associated to the same mode of the density in the 5D space. The
analysis of the substructures revealed by the different levels of the
HMAC hierarchy is beyond the scope of this paper. It involves the
analysis of the ages, kinematics and star formation histories of these
regions as demonstrated by recent studies of some of the regions
identified here \citep[see for
  example][]{2019A&A...626A..17C,2021ApJ...917...23K,2022MNRAS.517..161K}. In
this Sect. we will not take into account the subtleties due to the
spatial and temporal complexities of the groups found by HMAC and will
give a broad picture of the differences as seen by \gaia.

\begin{figure}[th]
	\centering
        \includegraphics[scale=0.5]{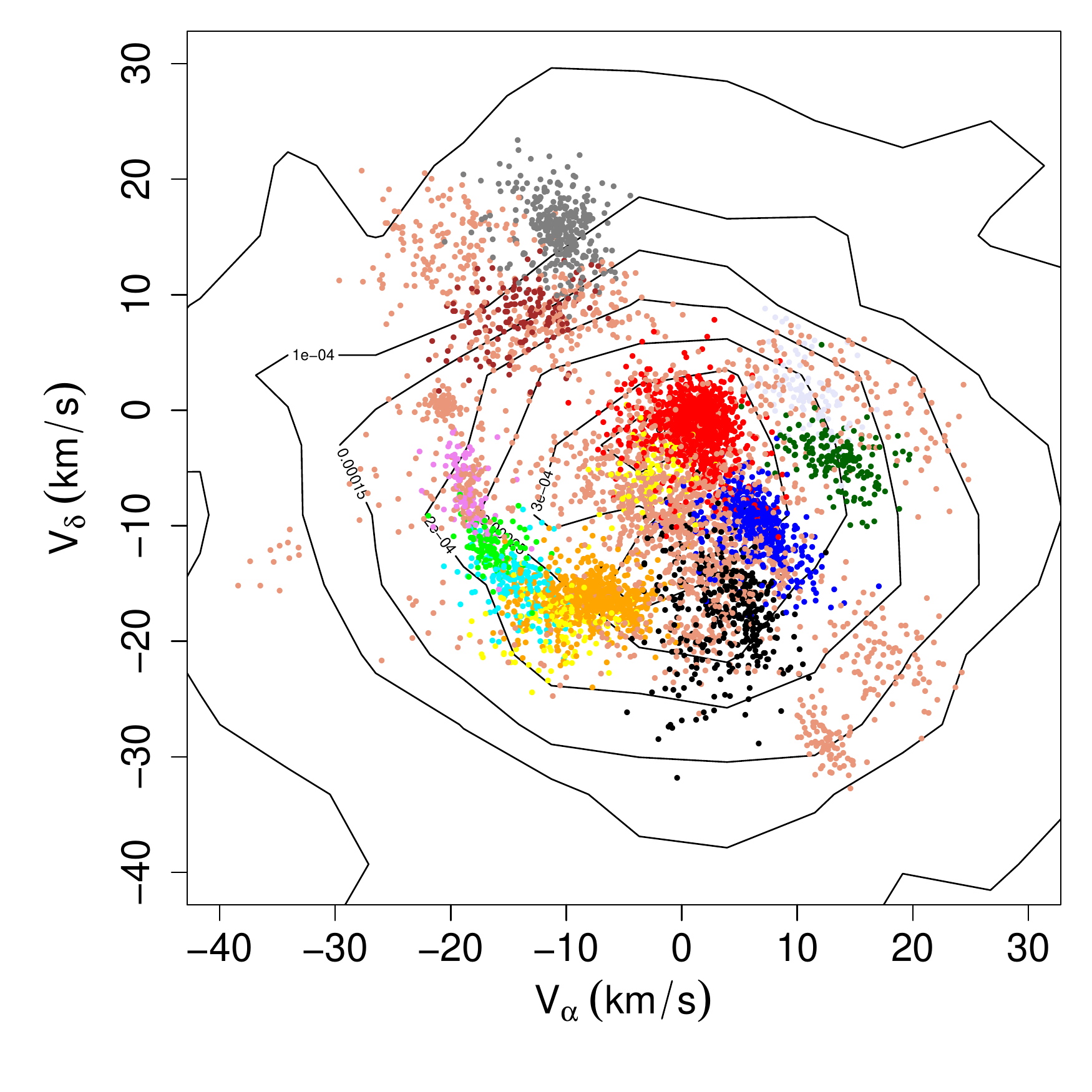}
	        \caption{Distribution in the space of tangential velocities of the sources in clusters with more than 10 members. The colour code is the same as used in the lower right panel of Figure \ref{fig:lb}. The underlying contour lines depict the distribution of the sources in clusters with 10 members or less as estimated using a kernel with $\sigma=3$.}
	        \label{fig:tanvel}
\end{figure}

We derive equatorial tangential velocities (in km/s) using the
measured parallax and a conversion factor of 4.74. In a preprocessing
stage, the data were centred at the median value of the distribution
of each variable in the full set of sources in quality classes 0 and 1
and divided by the median absolute deviation (MAD) to avoid the
clustering being dominated by the variables with a larger range of
values. In our case, we defined isotropic kernels of sizes 0.02 (level
1), 0.04, 0.06, 0.08 and 0.1 (level 5). The bottom panel of Figure
\ref{fig:lb} shows clusters with more than 10 members identified at
level 4 (left) and the distribution of sources not attached to any
cluster (right). The latter shows hints of residual
overdensities. Some of them are due to the scanning law (that results
in more transits, better quality measurements and consequently more
sources passing the quality criteria in certain regions of the sky)
while two of the others (shown in red and blue) are discussed later
below. Figure \ref{fig:tanvel} shows the various clusters depicted in
the lower left panel of Fig. \ref{fig:lb} in the space of tangential
velocities (using the same colour code).

The significant overdensities identified by HMAC as distinct groups
can easily be identified with well known star forming regions and
clusters. These large groups show clear substructures discussed in the
specialised literature. In Section \ref{sec:banyan} we discuss
clusters identified using the BANYAN $\Sigma$ software tool
\citep{2018ApJ...856...23G} and in Section \ref{sec:hmac-only} we
analyse HMAC clusters without members identified in any of the BANYAN
groups.

But before discussing the group identification we next analyse the
residual overdensities not identified as clusters by HMAC. While some
are easily recognised as due to the scanning law, there are two
prominent regions which cannot be explained as related to it. The
positions of at least a significant fraction of the sources in these
two overdensities on the celestial sphere and several CAMDs is
consistent with membership to the Upper Scorpio, $\rho$ Ophiucus and
Taurus star forming regions but their velocities are not concentrated
and do not correspond to those typical of these regions. Figures
\ref{fig:resClusterstanvel} and \ref{fig:resClustersParallaxes} in
Appendix \ref{appendix:clusters} show the distribution of these
sources in the space of tangential velocities and parallaxes. We find
no concentration at the positions expected for these star forming
regions. Figure \ref{fig:resClustersCAMD} provides a potential
explanation for these overdensities. It shows a $M_G$ vs. $(G-J)$ CAMD
including the \gaia Catalogue of Nearby Stars (GCNS, black dots with
transparency), the UCD catalogue discussed in this work (salmon dots
with transparency), the GUCDS (violet dots), the sources in the two
overdensities (red and blue dots) and the corresponding photometry
corrected for extinction (green and orange dots) using the Planck
GNILC map \citep{2016A&A...596A.109P}. It shows a remarkable
coincidence with the GCNS Main Sequence indicating that the
overdensities are mainly background sources of spectral type earlier
than the UCD limit and whose RP spectra appear as those of UCDs due to
the associated reddening.

\subsection{Membership according to BANYAN}
\label{sec:banyan}

In this Section we discuss the determination of membership to star
clusters and moving groups within 150\ pc from the Sun as derived by
the BANYAN $\Sigma$ software tool \citep{2018ApJ...856...23G}. We use
the \gaia DR3 UCD candidates (their celestial positions, proper
motions and parallaxes) as input to BANYAN $\Sigma$ and find 2840
sources with membership probabilities higher than 0.5, 80\% of them
higher than 0.8. In the following we use the lower threshold of 0.5
despite indications that this may include contamination by sources
from the field \citep{2016ApJ...833...96L}. BANYAN $\Sigma$ is based
on multivariate Gaussian modelling of the groups which in some cases
can be a simplification of the true distribution of sources in the
space of measurements. HMAC on the contrary is a non-parametric
clustering technique that can identify groups of arbitrary shapes and
we prefer to be conservative in the definition of the BANYAN
probability threshold for the sake of completeness. The cluster
assignments and membership probabilities are included in the online
Table \ref{tab:hmac+banyan-cluster-tags}. BANYAN $\Sigma$ only
produces membership probabilities to a set of predefined, well known
clusters of stars. As opposed to HMAC, it does not detect clusters or
groups of sources with similar properties.

\begin{table}[t]
\setlength{\tabcolsep}{3pt}
    \caption{\label{tab:hmac+banyan-cluster-tags} First ten lines of
      the table containing the clusters assignments of the HMAC and
      BANYAN groups including the BANYAN membership probability. NA is
      used as a code to denote not available in cases where one UCD
      was assigned to a group by one of the techniques but not by the
      other. The full table contains 7630 entries.}
    \begin{tabular}{
    cccc}
        \hline\hline
Source ID & HMAC & BANYAN & BANYAN \\
 & cluster & group & probability \\
\hline
2572901021957789568&NA&CARN&0.98\\
1016186483391641216&NA&CARN&0.79\\
5556620540565785600&NA&ARG&0.95\\
1954170404122975232&NA&CARN&0.73\\
3411692668689199744&NA&ARG&0.96\\
5541111516746730752&NA&ABDMG&0.89\\
3597096309389074816&NA&CARN&0.99\\
6031367499416648192&NA&ARG&0.72\\
5908794218026022144&NA&ARG&1\\
6118581861234228352&NA&ABDMG&1\\

    \hline
    \end{tabular}
\end{table}

\begin{figure*}[th]
	\centering
        \includegraphics[scale=0.4]{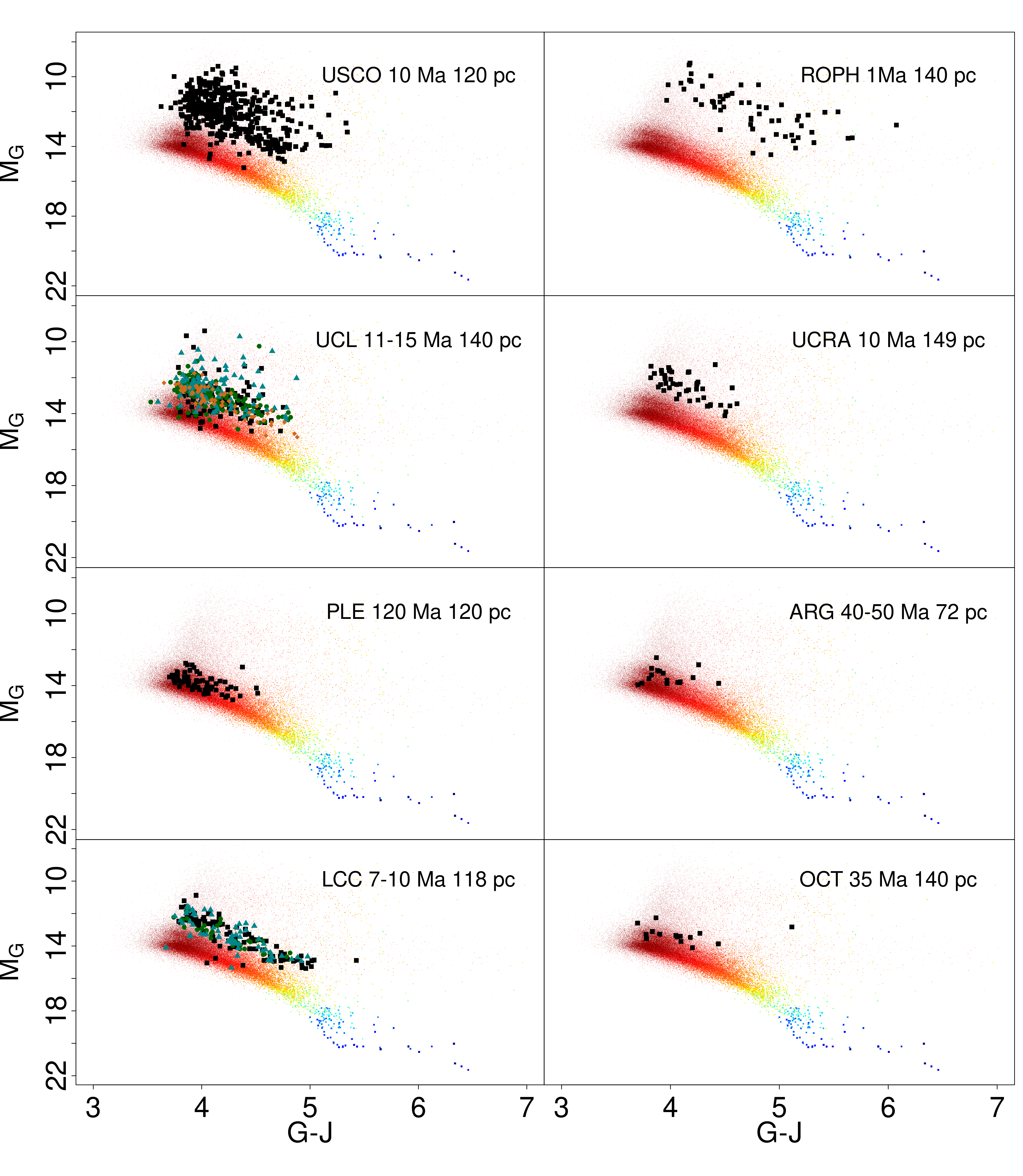}
	\caption{Position in the CAMD of sources identified as members
          of one of the BANYAN associations/clusters and assigned to
          (up to three) HMAC clusters with more than 50 members. We
          use the same colour scale as in Figure \ref{fig:camds-dense}
          for the full set of UCD candidates in the catalogue and
          black filled squares, turquoise filled triangles and brown
          filled diamonds are used to identify the HMAC groups over
          which BANYAN members can be spread if they correspond to
          more than one (see Table \ref{tab:clusters-table}).}
	\label{fig:hmac-banyan-1}
\end{figure*}

Table \ref{tab:clusters-table} shows the number of sources in common
between the BANYAN $\Sigma$ groups and HMAC clusters, and Figure
\ref{fig:hmac-banyan-1} shows the position of these sources in a CAMD
for a few selected groups.

\subsection{Stellar groups without members of the BANYAN clusters}
\label{sec:hmac-only}

Apart from the 19 HMAC clusters with members in common with one or
several BANYAN groups, there are 38 additional HMAC clusters with more
than 20 members with no common member with BANYAN groups.  Figures
\ref{fig:hmac-only-610}-\ref{fig:hmac-only-8984} show the celestial
coordinates (leftmost plot), CAMD (mid-left), tangential velocities
(mid-right) and distance (estimated by naive parallax inversion;
rightmost) distributions of the 14 most numerous groups identified by
the HMAC algorithm but without identifications in any of the BANYAN
groups. Each Figure represents one HMAC cluster at level 4 while the
colours represent subclusters with more that five members identified
at lower levels of the hierarchy (that is, for narrower kernels). The
black circles denote sources in subclusters with 5 members or
less. The substructures respond mainly to variations in the space of
tangential velocities and not to the space of celestial coordinates
where they sometimes mix without clear separations. The tentative
identifications provided in the captions are only orientative and do
not aim to reflect the spatial complexity of these stellar
associations.

Figure \ref{fig:hmac-only-restGT20} depicts the remaining 24 HMAC
clusters with more than 20 members in a similar way as that used in
Figures \ref{fig:hmac-only-610}-\ref{fig:hmac-only-8984} except for
the rightmost panel that represents the kernel density estimates of
the distance estimated by naive inversion of the parallax.

\subsection{RP spectra as a function of age}

\begin{figure*}[t]
	\centering
        \includegraphics[scale=0.45]{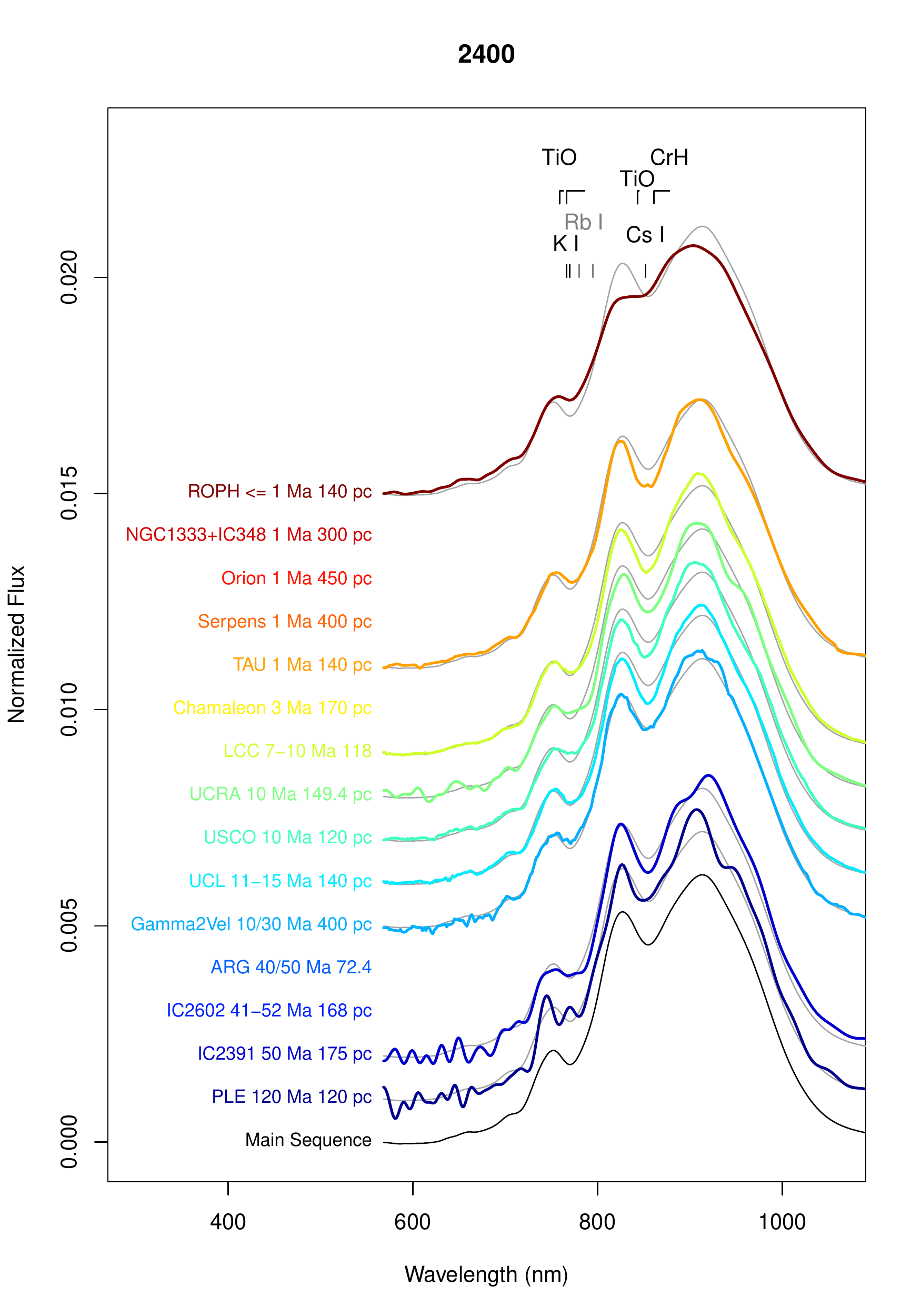}
        \includegraphics[scale=0.45]{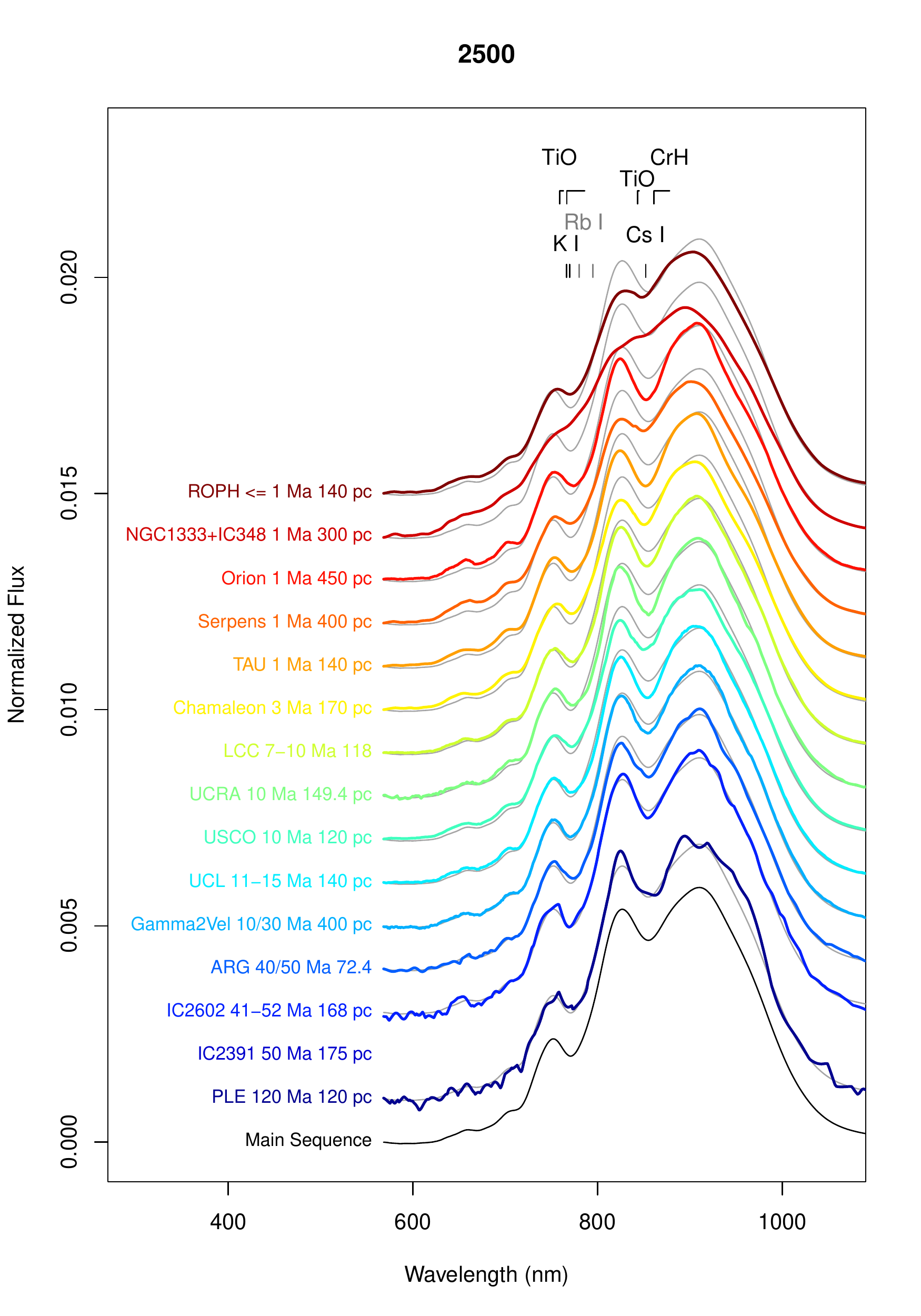}
        \includegraphics[scale=0.45]{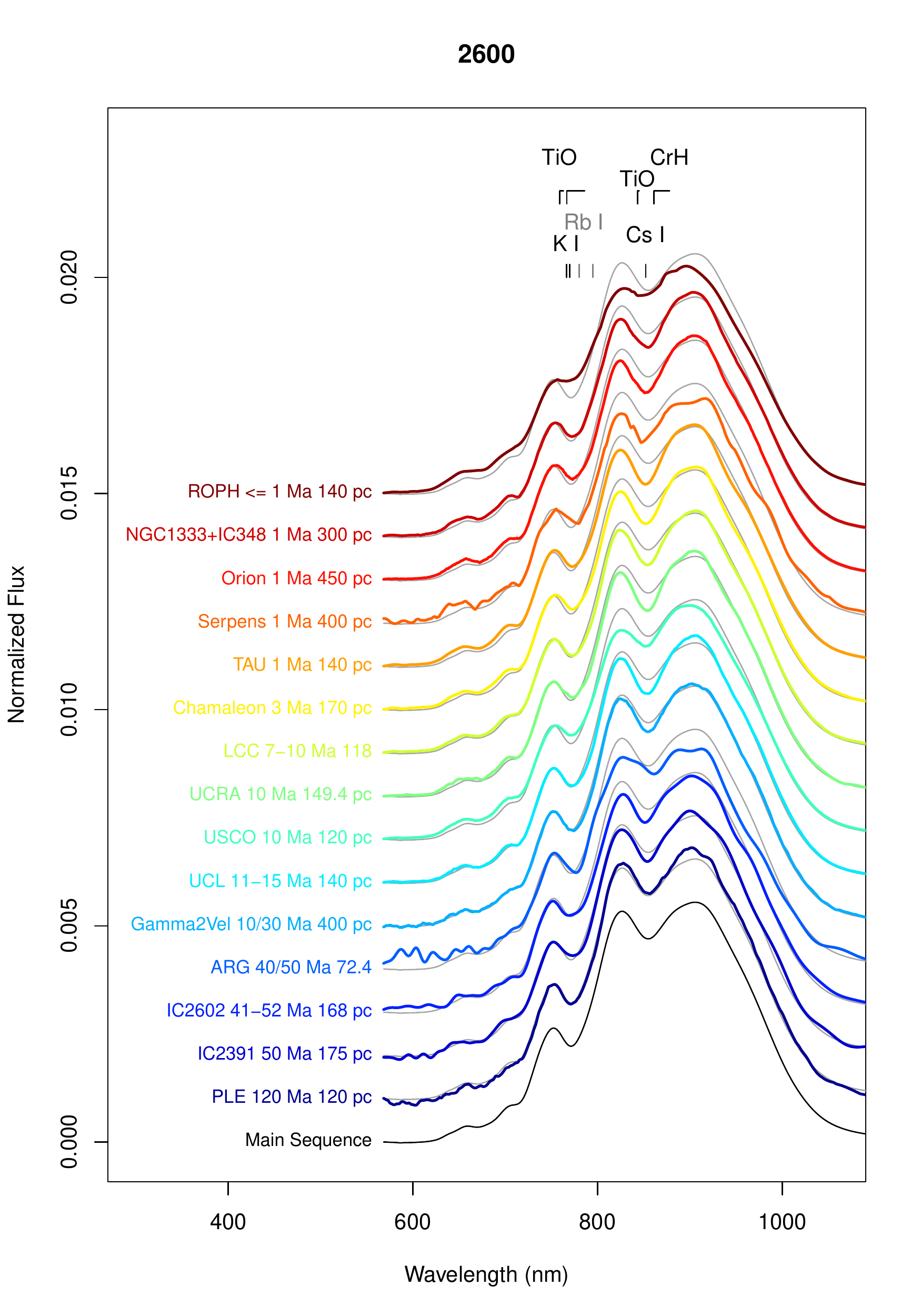}
        \includegraphics[scale=0.45]{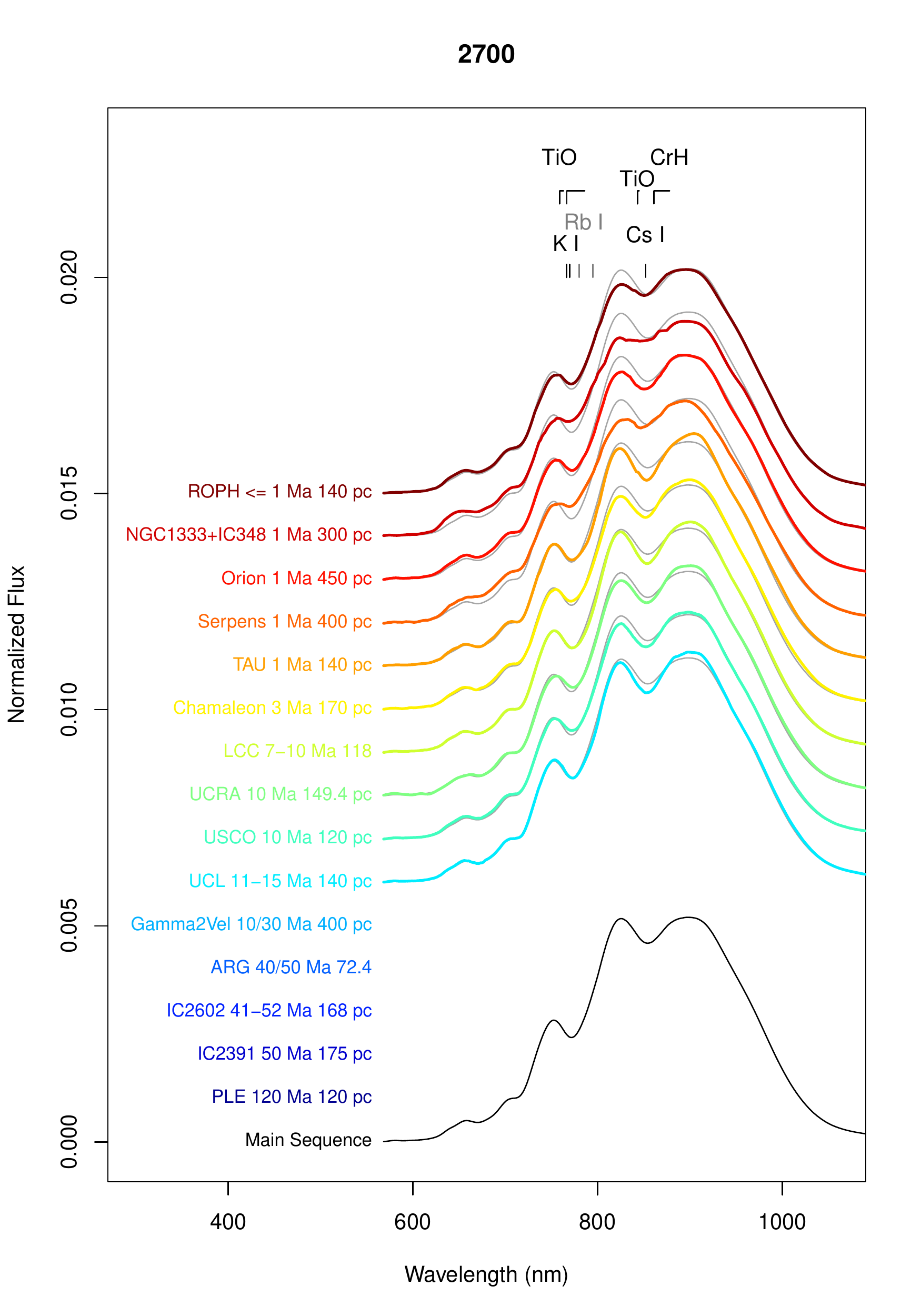}
        \caption{Median RP spectra calculated for sources in the range
          between 2350 and 2450 K (top left), 2450 and 2550 K (top
          right), 2550 and 2650 K (bottom left), and 2650 and 2700 K
          (bottom right) and assigned to the most prominent BANYAN and
          HMAC groups. The spectra are ordered in age from the top
          (youngest) to the bottom (oldest). The median RP spectrum
          for sources outside the overdensities in each temperature
          bin is labelled as Main Sequence and is shown in black at
          the bottom and in light grey superimposed on each group.}
	\label{fig:rps-clusters}
\end{figure*}

Figure \ref{fig:rps-clusters} shows the median RP spectrum in some of
the clusters identified in the previous Sects. Again, the distances
and ages used are only included as a guide to order the spectra. The
medians are calculated in bins of increasing effective temperature
between 2350 and 2450\,K (top left), between 2450 and 2550\,K (top
right), between 2550 and 2650\,K (bottom left), and between 2650 and
2700\,K (bottom right). When interpreting these Figs., bear in mind
that the regression module estimates temperatures using an empirical
training set that does not include young or non-solar metallicity
sources. Hence, the effective temperatures assigned to the sources in
these stellar associations may be biased. Each of the Figs. includes
the median calculated for sources outside the overdensities with a
black line labelled Main Sequence at the bottom of each Fig., and with
a light grey line superimposed on each association to facilitate the
comparison. We interpret these sources as representing evolved
examples from the Zero-Age Main Sequence and so, well represented in
the training set. It is impossible to separate the contribution of the
various absorption lines and bands at the low resolution of the RP
spectra. \cite{2007A&A...473..245R} provide us with high-resolution
spectra and line/band identifications in that wavelength range and for
spectral types slightly cooler that those corresponding to the
temperatures represented in Fig. \ref{fig:rps-clusters}. From them, we
hypothesise that the two absorption features visible in these RP
spectra are mainly due to TiO (band heads at 758.9 nm, 766.6 \nm,
843.2 \nm, and 885.9 nm), VO (785.1 nm and 852.1 nm), CrH (861.1 nm)
and the alkali spectral lines of K~{\sc i} (766.5 and 7698 nm),
Rb~{\sc i} (780.0 nm and 794.7 nm) and Cs~{\sc i} (852.1 nm). In all
four Figs. (but more prominently in the top right panel) we see that
the absorption bands get deeper as the association age becomes
older. As mentioned above, the HMAC cluster identifications involves
complex groups with several substructures and a range of ages. Orion
for example contains star forming regions with ages estimated between
1 and 10 Ma. In these cases, the median RP spectrum plotted in the
Figs. represents a weighted average of slightly different spectra.

These figures seem to indicate that by 10 Ma the RP spectrum can
hardly be distinguished from that of the evolved ones in the Main
Sequence and therefore, low gravity detection can only be accomplished
for very young associations. Plans for DR4 include the
parameterisation of these changes in the RP spectrum and the inclusion
of youth indication flags in the archive for these young UCD
candidates.

\section{Variability of UCDs}
\label{sec:vari}

\begin{figure}[ht]
  \centering
  \includegraphics[width=\linewidth]{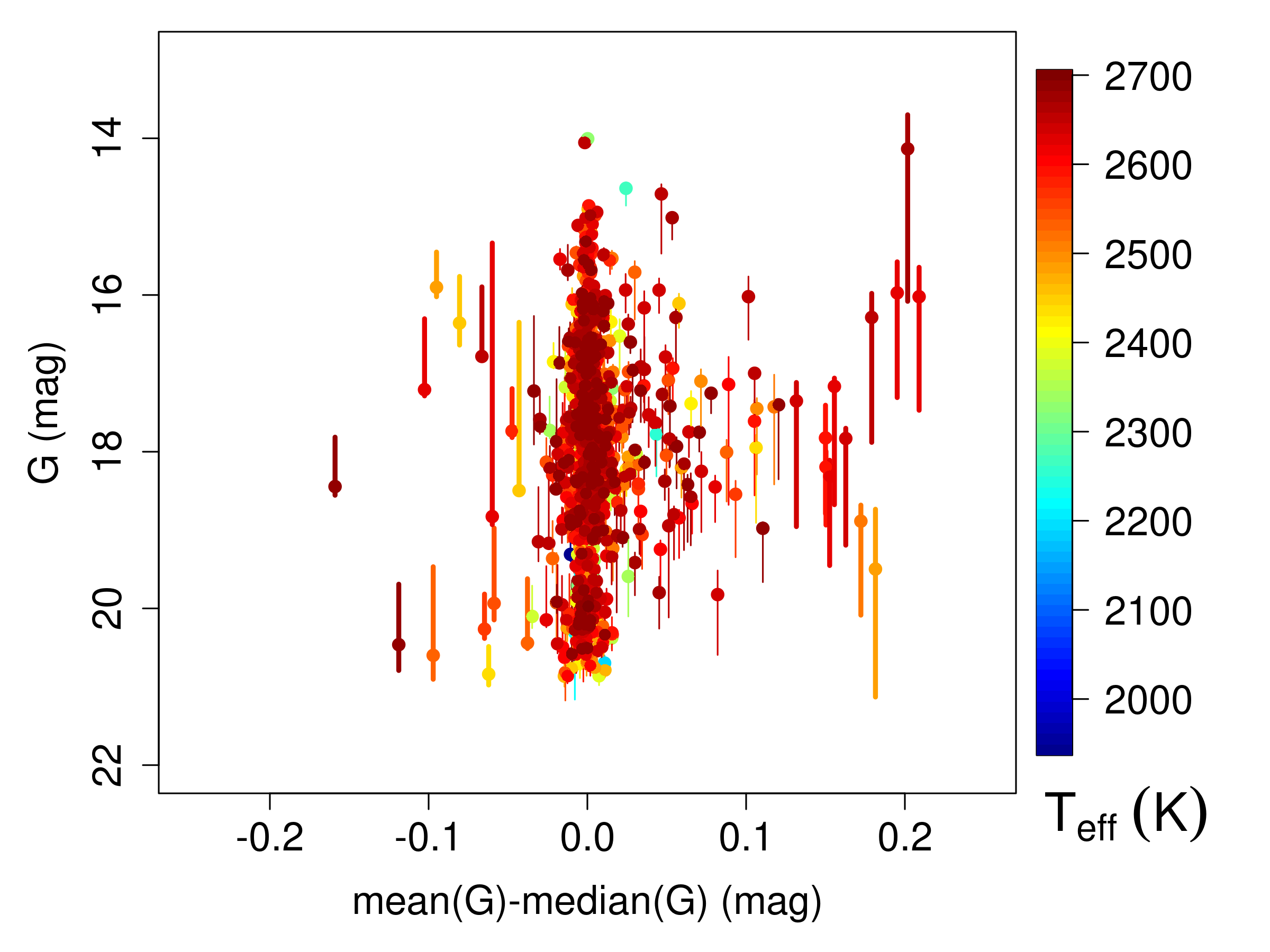}
  \caption{Ranges of \gmag\ values in the time series of UCD
    candidates included in the variability tables of
    \gaia\ DR3. The filled circle in each segment marks the time
    series median value.}
  \label{fig:vari-range}
\end{figure}

\begin{figure}[ht]
  \centering
  \includegraphics[width=\linewidth]{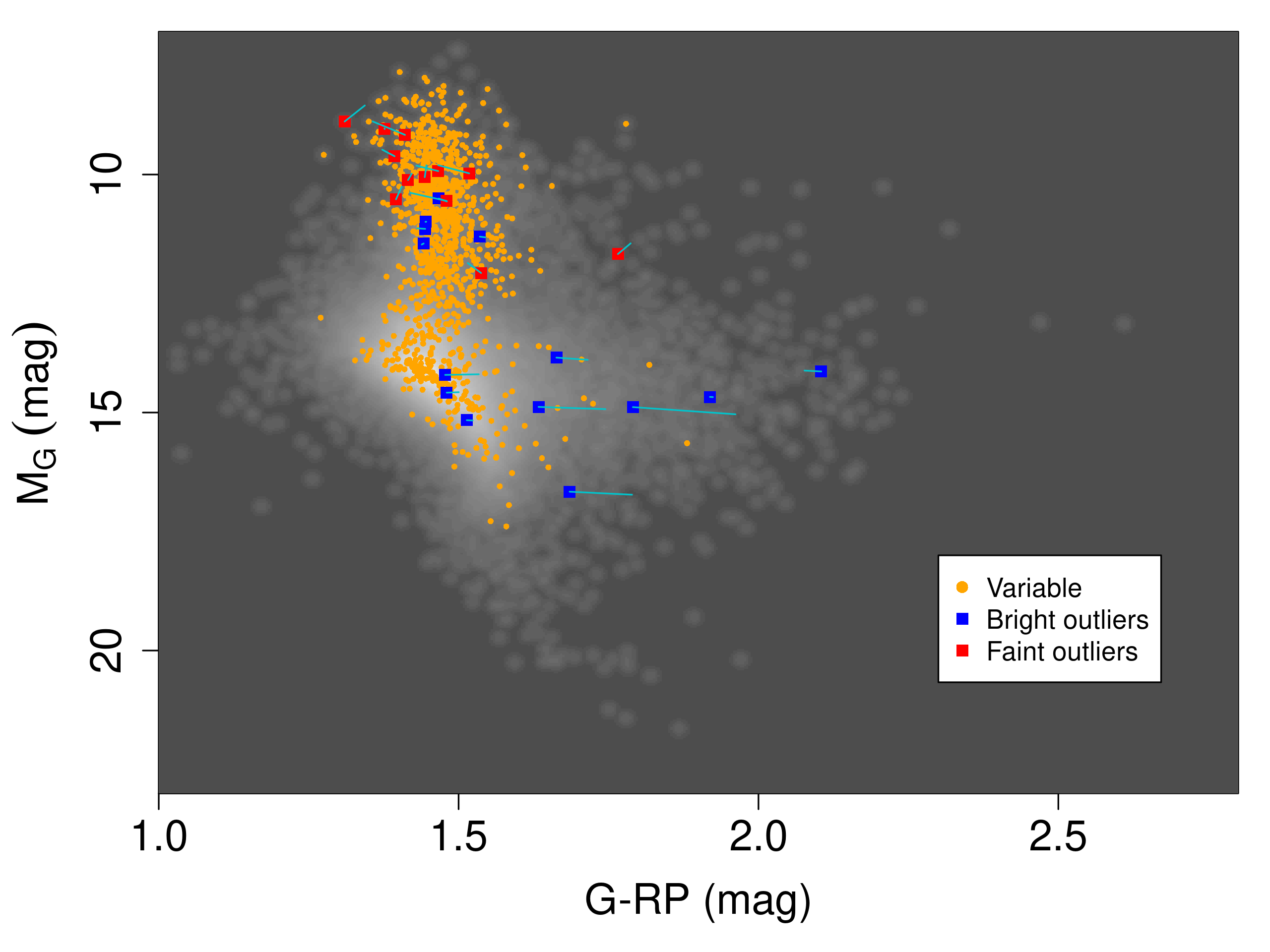}
  \caption{Kernel density estimate of the distribution of UCD
    candidates in the \gaia\ CAMD (grey scale). Candidates with
    entries in the vari\_summary table are marked as orange circles;
    blue and red squares mark the extreme cases of bright and faint
    outliers shown in Fig. \ref{fig:vari-range}, respectively. The
    segments illustrate the displacement in the CAMD between the main
    archive values of the photometry and the medians available in the
    vari\_summary table.}
  \label{fig:vari-camd}
\end{figure}

\gaia DR3 includes the results of the processing and analysis of the
time series of individual sources. We have searched the associated
archive variability tables for entries in common with the UCD
catalogue discussed here. Figure \ref{fig:vari-range} shows the median
$G$ magnitude of the 1109 sources in common as a filled circle and its
range of values in the time series (segments) as a function of
difference between the mean and the median of the time series. The
colour code reflects the \teff value inferred by the \espucd
module. We have highlighted sources at the two extremes of the
distribution of this difference (mean-median) with thicker lines. It
shows that the main bulk of UCD candidates with entries in the
vari\_summary table concentrates around the origin of the $x$ axis
implying that the time series does not show outliers. But it also
shows UCD candidates with very asymmetric distributions. On the left
hand side of the plot we encounter sources with bright outliers and on
the right hand side, sources with faint outliers. Figure
\ref{fig:vari-camd} shows the position of the UCD candidates that have
entries in the vari\_summary table (orange) in the \gaia\ CAMD,
superimposed on a kernel-based estimate of the density of UCD
candidates. Red squares represent the variable UCDs with faint
outliers and the blue asterisks, UCDs with bright outliers. The line
segments represent the displacement between the values of the $G$ and
$G_{RP}$ magnitudes in the main \gaia\ catalogue (used in all
Sects. of this paper) and the median values from the time series. The
CAMD already shows that most of the variable UCD candidates are placed
in the region of young, pre-Main Sequence sources. In fact, 728 of the
1\,109 UCD candidates with variability data are identified by HMAC as
belonging to overdensities, and 353 by BANYAN $\Sigma$ as members of
the BANYAN clusters. The most numerous HMAC groups are Upper Sco (250
sources), Orion (162), Serpens (53) and the Perseus association
(48). Those of the BANYAN class set are Upper Sco (203 sources), Upper
Centaurus Lupus (57) and Taurus (45).

We initially interpreted the sources with bright outliers as
candidates to sources with flares, and the latter as candidates to
eclipsing binaries. However, the UCD candidates with outliers
significantly fainter that the median (the examples to the right which
we had interpreted initially as potential eclipsing binaries) are all
classified in the vari\_classifier\_result as Young Stellar Objects
(YSOs) except for one source that is classified as Long Period
Variable albeit with a probability of 0.3. This is in agreement with
their position in the CAMD.

The sources to the left of Fig. \ref{fig:vari-range} were initially
assumed to be candidates for flaring UCDs. We searched the identifiers
in the Variability tables of the \gaia DR3 tables and in SIMBAD. The
latter returned seven cross matches with object type identifications:
two high proper motion sources, one low-mass object, two rotating
variables and two young stellar objects. The former returned 5
classifications as YSOs (with probabilities greater than 0.85 except
in one case); three classifications in the RR~Lyrae class (with
probabilities below 0.15); and two Long Period Variables
(probabilities of 0.06).

Variability of UCDs has been studied in the past \citep[see e.g.][and
  references
  therein]{2014A&A...566A.111W,2017Sci...357..683A,2017AstRv..13....1B,2018haex.bookE..94A}. Some
variability might be associated to binarity but intrinsic atmospheric
changes should be present too. Note, however, that young, low mass
stars present other phenomena, such as the presence of dips in their
light curve possibly due to stellar disk occultations
\citep{2015AJ....149..130S}. This explanation cannot be ruled out
since most of our variable sources are associated to stellar
associations. However, the irregular sampling of the \gaia time series
and their unavailability in DR3 precludes a deeper analysis in the
context of this work.

\section{Summary}
\label{sec:conclusions}

In this work we present a first overview of the UCD content in the
\gaia DR3 archive. More specifically, we present the typical (median)
RP spectra in spectral type bins and compare them to UCD standards and
ground-based high-resolution spectra; we present CAMDs including \gaia
and external photometry; we compare the catalogue with previous lists
of UCD candidates based on \gaia data; we construct a simple
hierarchical probabilistic model as a first step towards inferring the
spatial and luminosity distribution of UCDs when a deeper knowledge of
the catalogue selection function is available; we provide a list of
candidate companions in wide binary systems including UCDs from the
catalogue; we analyse the overdensities of UCDs in the celestial
sphere and identify them (at a coarse grain level) with known star
forming regions and stellar associations; and, finally, we briefly
review the variability properties of the UCD candidates. These
identifications will serve as a basis for the inclusion of youth
indicators in future versions of the catalogue.

This global overview will help the community to further explore some
of the aspects of the catalogue highlighted here and inspire
subsequent analyses with complementary observations.

\begin{acknowledgements}

This work has made use of results from the European Space Agency (ESA)
space mission {\it Gaia}, the data from which were processed by the
{\it Gaia Data Processing and Analysis Consortium} (DPAC).  The {\it
  Gaia} mission website is \url{http://www.cosmos.esa.int/gaia}.

This work was supported by the MCIN (Spanish Ministry of Science and
Innovation) through grant PID2020-112949GB-I00; the MINECO (Spanish
Ministry of Economy) through grants AyA2017-84089,
ESP2016-80079-C2-1-R, ESP2014-55996-C2-1-R, and RTI2018-095076-B-C22
(MINECO/FEDER, UE). This research has been funded by the Spanish State
Research Agency (AEI) Projects No.PID2019-107061GB-C61 and
No. MDM-2017-0737 Unidad de Excelencia “María de Maeztu”- Centro de
Astrobiología (CSIC/INTA).

Ground based spectra is from observations made with the Gran
Telescopio Canarias (GTC), installed in the Spanish Observatorio del
Roque de los Muchachos of the Instituto de Astrofísica de Canarias, on
the island of La Palma.  The GTC data was obtained with the instrument
OSIRIS, built by a Consortium led by the Instituto de Astrofísica de
Canarias in collaboration with the Instituto de Astronomía of the
Universidad Autónoma de México.  OSIRIS was funded by GRANTECAN and
the National Plan of Astronomy and Astrophysics of the Spanish
Government.  The programme codes were GTC54-15A0 \& GTC8-15ITP.

This work has made use of the Python package GaiaXPy, developed and
maintained by members of the Gaia Data Processing and Analysis
Consortium (DPAC), and in particular, Coordination Unit 5 (CU5), and
the Data Processing Centre located at the Institute of Astronomy,
Cambridge, UK (DPCI).

We derived extiction corrections using the Python package dustmap
\citep{2018JOSS....3..695M}.

\end{acknowledgements}

\bibliographystyle{aa}
\bibliography{main} 

\begin{thebibliography}{108}
\expandafter\ifx\csname natexlab\endcsname\relax\def\natexlab#1{#1}\fi

\bibitem[{Abadi {et~al.}(2015)Abadi, Agarwal, Barham, Brevdo, Chen, Citro,
  Corrado, Davis, Dean, Devin, Ghemawat, Goodfellow, Harp, Irving, Isard, Jia,
  Jozefowicz, Kaiser, Kudlur, Levenberg, Man\'{e}, Monga, Moore, Murray, Olah,
  Schuster, Shlens, Steiner, Sutskever, Talwar, Tucker, Vanhoucke, Vasudevan,
  Vi\'{e}gas, Vinyals, Warden, Wattenberg, Wicke, Yu, \&
  Zheng}]{tensorflow2015-whitepaper}
Abadi, M., Agarwal, A., Barham, P., {et~al.} 2015, {TensorFlow}: Large-Scale
  Machine Learning on Heterogeneous Systems, software available from
  tensorflow.org

\bibitem[{{Almendros-Abad} {et~al.}(2022){Almendros-Abad}, {Mu{\v{z}}i{\'c}},
  {Moitinho}, {Krone-Martins}, \& {Kubiak}}]{2022A&A...657A.129A}
{Almendros-Abad}, V., {Mu{\v{z}}i{\'c}}, K., {Moitinho}, A., {Krone-Martins},
  A., \& {Kubiak}, K. 2022, \aap, 657, A129

\bibitem[{{Apai} {et~al.}(2017){Apai}, {Karalidi}, {Marley}, {Yang}, {Flateau},
  {Metchev}, {Cowan}, {Buenzli}, {Burgasser}, {Radigan}, {Artigau}, \&
  {Lowrance}}]{2017Sci...357..683A}
{Apai}, D., {Karalidi}, T., {Marley}, M.~S., {et~al.} 2017, Science, 357, 683

\bibitem[{{Artigau}(2018)}]{2018haex.bookE..94A}
{Artigau}, {\'E}. 2018, in Handbook of Exoplanets, ed. H.~J. {Deeg} \& J.~A.
  {Belmonte}, 94

\bibitem[{{Bailer-Jones}(2015)}]{2015PASP..127..994B}
{Bailer-Jones}, C.~A.~L. 2015, \pasp, 127, 994

\bibitem[{{Baraffe} {et~al.}(2015){Baraffe}, {Homeier}, {Allard}, \&
  {Chabrier}}]{2015A&A...577A..42B}
{Baraffe}, I., {Homeier}, D., {Allard}, F., \& {Chabrier}, G. 2015, \aap, 577,
  A42

\bibitem[{{Bardalez Gagliuffi} {et~al.}(2014){Bardalez Gagliuffi}, {Burgasser},
  {Gelino}, {Looper}, {Nicholls}, {Schmidt}, {Cruz}, {West}, {Gizis}, \&
  {Metchev}}]{2014ApJ...794..143B}
{Bardalez Gagliuffi}, D.~C., {Burgasser}, A.~J., {Gelino}, C.~R., {et~al.}
  2014, \apj, 794, 143

\bibitem[{{Bardalez Gagliuffi} {et~al.}(2019){Bardalez Gagliuffi}, {Burgasser},
  {Schmidt}, {Theissen}, {Gagn{\'e}}, {Gillon}, {Sahlmann}, {Faherty},
  {Gelino}, {Cruz}, {Skrzypek}, \& {Looper}}]{2019ApJ...883..205B}
{Bardalez Gagliuffi}, D.~C., {Burgasser}, A.~J., {Schmidt}, S.~J., {et~al.}
  2019, \apj, 883, 205

\bibitem[{{Bate}(2012)}]{2012MNRAS.419.3115B}
{Bate}, M.~R. 2012, \mnras, 419, 3115

\bibitem[{{Bennett} \& {Bovy}(2019)}]{2019MNRAS.482.1417B}
{Bennett}, M. \& {Bovy}, J. 2019, \mnras, 482, 1417

\bibitem[{{Biller}(2017)}]{2017AstRv..13....1B}
{Biller}, B. 2017, The Astronomical Review, 13, 1

\bibitem[{{Bonnell} {et~al.}(2008){Bonnell}, {Clark}, \&
  {Bate}}]{2008MNRAS.389.1556B}
{Bonnell}, I.~A., {Clark}, P., \& {Bate}, M.~R. 2008, \mnras, 389, 1556

\bibitem[{{Bovy}(2015)}]{2015ApJS..216...29B}
{Bovy}, J. 2015, \apjs, 216, 29

\bibitem[{{Burgasser} {et~al.}(2010){Burgasser}, {Cruz}, {Cushing}, {Gelino},
  {Looper}, {Faherty}, {Kirkpatrick}, \& {Reid}}]{2010ApJ...710.1142B}
{Burgasser}, A.~J., {Cruz}, K.~L., {Cushing}, M., {et~al.} 2010, \apj, 710,
  1142

\bibitem[{{Burgasser} {et~al.}(2002){Burgasser}, {Kirkpatrick}, {Brown},
  {Reid}, {Burrows}, {Liebert}, {Matthews}, {Gizis}, {Dahn}, {Monet}, {Cutri},
  \& {Skrutskie}}]{2002ApJ...564..421B}
{Burgasser}, A.~J., {Kirkpatrick}, J.~D., {Brown}, M.~E., {et~al.} 2002, \apj,
  564, 421

\bibitem[{{Burgasser} {et~al.}(2003{\natexlab{a}}){Burgasser}, {Kirkpatrick},
  {Liebert}, \& {Burrows}}]{2003ApJ...594..510B}
{Burgasser}, A.~J., {Kirkpatrick}, J.~D., {Liebert}, J., \& {Burrows}, A.
  2003{\natexlab{a}}, \apj, 594, 510

\bibitem[{{Burgasser} {et~al.}(2003{\natexlab{b}}){Burgasser}, {Kirkpatrick},
  {McElwain}, {Cutri}, {Burgasser}, \& {Skrutskie}}]{2003AJ....125..850B}
{Burgasser}, A.~J., {Kirkpatrick}, J.~D., {McElwain}, M.~W., {et~al.}
  2003{\natexlab{b}}, \aj, 125, 850

\bibitem[{{Burgasser} {et~al.}(2000){Burgasser}, {Wilson}, {Kirkpatrick},
  {Skrutskie}, {Colonno}, {Enos}, {Smith}, {Henderson}, {Gizis}, {Brown}, \&
  {Houck}}]{2000AJ....120.1100B}
{Burgasser}, A.~J., {Wilson}, J.~C., {Kirkpatrick}, J.~D., {et~al.} 2000, \aj,
  120, 1100

\bibitem[{{Burningham} {et~al.}(2008){Burningham}, {Pinfield}, {Leggett},
  {Tamura}, {Lucas}, {Homeier}, {Day-Jones}, {Jones}, {Clarke}, {Ishii},
  {Kuzuhara}, {Lodieu}, {Zapatero Osorio}, {Venemans}, {Mortlock}, {Barrado Y
  Navascu{\'e}s}, {Martin}, \& {Magazz{\`u}}}]{2008MNRAS.391..320B}
{Burningham}, B., {Pinfield}, D.~J., {Leggett}, S.~K., {et~al.} 2008, \mnras,
  391, 320

\bibitem[{{Cantat-Gaudin} {et~al.}(2019){Cantat-Gaudin}, {Jordi}, {Wright},
  {Armstrong}, {Vallenari}, {Balaguer-N{\'u}{\~n}ez}, {Ramos}, {Bossini},
  {Padoan}, {Pelkonen}, {Mapelli}, \& {Jeffries}}]{2019A&A...626A..17C}
{Cantat-Gaudin}, T., {Jordi}, C., {Wright}, N.~J., {et~al.} 2019, \aap, 626,
  A17

\bibitem[{{Carrasco} {et~al.}(2021){Carrasco}, {Weiler}, {Jordi}, {Fabricius},
  {De Angeli}, {Evans}, {van Leeuwen}, {Riello}, \&
  {Montegriffo}}]{2021A&A...652A..86C}
{Carrasco}, J.~M., {Weiler}, M., {Jordi}, C., {et~al.} 2021, \aap, 652, A86

\bibitem[{{Castro} \& {Gizis}(2012)}]{2012ApJ...746....3C}
{Castro}, P.~J. \& {Gizis}, J.~E. 2012, \apj, 746, 3

\bibitem[{{Castro} {et~al.}(2013){Castro}, {Gizis}, {Harris}, {Mace},
  {Kirkpatrick}, {McLean}, {Pattarakijwanich}, \&
  {Skrutskie}}]{2013ApJ...776..126C}
{Castro}, P.~J., {Gizis}, J.~E., {Harris}, H.~C., {et~al.} 2013, \apj, 776, 126

\bibitem[{{Chabrier} {et~al.}(2014){Chabrier}, {Johansen}, {Janson}, \&
  {Rafikov}}]{2014prpl.conf..619C}
{Chabrier}, G., {Johansen}, A., {Janson}, M., \& {Rafikov}, R. 2014, in
  Protostars and Planets VI, ed. H.~{Beuther}, R.~S. {Klessen}, C.~P.
  {Dullemond}, \& T.~{Henning}, 619

\bibitem[{{Creevey} {et~al.}(2022){Creevey}, {Sordo}, {Pailler}, {Fr{\'e}mat},
  {Heiter}, {Th{\'e}venin}, {Andrae}, {Fouesneau}, {Lobel}, {Bailer-Jones},
  {Garabato}, {Bellas-Velidis}, {Brugaletta}, {Lorca}, {Ordenovic}, {Palicio},
  {Sarro}, {Delchambre}, {Drimmel}, {Rybizki}, {Torralba Elipe}, {Korn},
  {Recio-Blanco}, {Schultheis}, {De Angeli}, {Montegriffo}, {Abreu Aramburu},
  {Accart}, {{\'A}lvarez}, {Bakker}, {Brouillet}, {Burlacu}, {Carballo},
  {Casamiquela}, {Chiavassa}, {Contursi}, {Cooper}, {Dafonte}, {Dapergolas},
  {de Laverny}, {Dharmawardena}, {Edvardsson}, {Le Fustec},
  {Garc{\'\i}a-Lario}, {Garc{\'\i}a-Torres}, {Gomez},
  {Gonz{\'a}lez-Santamar{\'\i}a}, {Hatzidimitriou}, {Jean-Antoine Piccolo},
  {Kontizas}, {Kordopatis}, {Lanzafame}, {Lebreton}, {Licata}, {Lindstr{\o}m},
  {Livanou}, {Magdaleno Romeo}, {Manteiga}, {Marocco}, {Marshall}, {Mary},
  {Nicolas}, {Pallas-Quintela}, {Panem}, {Pichon}, {Poggio}, {Riclet}, {Robin},
  {Santove{\~n}a}, {Silvelo}, {Slezak}, {Smart}, {Soubiran}, {S{\"u}veges},
  {Ulla}, {Utrilla}, {Vallenari}, {Zhao}, {Zorec}, {Barrado}, {Bijaoui},
  {Bouret}, {Blomme}, {Brott}, {Cassisi}, {Kochukhov}, {Martayan}, {Shulyak},
  \& {Silvester}}]{DR3-DPACP-157}
{Creevey}, O.~L., {Sordo}, R., {Pailler}, F., {et~al.} 2022, arXiv e-prints,
  arXiv:2206.05864

\bibitem[{{Cruz} {et~al.}(2007){Cruz}, {Reid}, {Kirkpatrick}, {Burgasser},
  {Liebert}, {Solomon}, {Schmidt}, {Allen}, {Hawley}, \&
  {Covey}}]{2007AJ....133..439C}
{Cruz}, K.~L., {Reid}, I.~N., {Kirkpatrick}, J.~D., {et~al.} 2007, \aj, 133,
  439

\bibitem[{{Cruz} {et~al.}(2003){Cruz}, {Reid}, {Liebert}, {Kirkpatrick}, \&
  {Lowrance}}]{2003AJ....126.2421C}
{Cruz}, K.~L., {Reid}, I.~N., {Liebert}, J., {Kirkpatrick}, J.~D., \&
  {Lowrance}, P.~J. 2003, \aj, 126, 2421

\bibitem[{{De Angeli} {et~al.}(2022){De Angeli}, {Weiler}, {Montegriffo},
  {Evans}, {Riello}, {Andrae}, {Carrasco}, {Busso}, {Burgess}, {Cacciari},
  {Davidson}, {Harrison}, {Hodgkin}, {Jordi}, {Osborne}, {Pancino},
  {Altavilla}, {Barstow}, {Bailer-Jones}, {Bellazzini}, {Brown}, {Castellani},
  {Cowell}, {Delchambre}, {De Luise}, {Diener}, {Fabricius}, {Fouesneau},
  {Fremat}, {Gilmore}, {Giuffrida}, {Hambly}, {Hidalgo}, {Holland},
  {Kostrzewa-Rutkowska}, {van Leeuwen}, {Lobel}, {Marinoni}, {Miller},
  {Pagani}, {Palaversa}, {Piersimoni}, {Pulone}, {Ragaini}, {Rainer},
  {Richards}, {Rixon}, {Ruz-Mieres}, {Sanna}, {Sarro}, {Rowell}, {Sordo},
  {Walton}, \& {Yoldas}}]{EDR3-DPACP-118}
{De Angeli}, F., {Weiler}, M., {Montegriffo}, P., {et~al.} 2022, arXiv
  e-prints, arXiv:2206.06143

\bibitem[{{Deacon} \& {Hambly}(2007)}]{2007A&A...468..163D}
{Deacon}, N.~R. \& {Hambly}, N.~C. 2007, \aap, 468, 163

\bibitem[{{Delchambre} {et~al.}(2022){Delchambre}, {Bailer-Jones},
  {Bellas-Velidis}, {Drimmel}, {Garabato}, {Carballo}, {Hatzidimitriou},
  {Marshall}, {Andrae}, {Dafonte}, {Livanou}, {Fouesneau}, {Licata},
  {Lindstrom}, {Manteiga}, {Robin}, {Silvelo}, {Abreu Aramburu}, {Alvarez},
  {Bakker}, {Bijaoui}, {Brouillet}, {Brugaletta}, {Burlacu}, {Casamiquela},
  {Chaoul}, {Chiavassa}, {Contursi}, {Cooper}, {Creevey}, {Dapergolas}, {de
  Laverny}, {Demouchy}, {Dharmawardena}, {Edvardsson}, {Fremat},
  {Garcia-Lario}, {Garcia-Torres}, {Gavel}, {Gomez}, {Gonzalez-Santamaria},
  {Heiter}, {Jean-Antoine Piccolo}, {Kontizas}, {Kordopatis}, {Korn},
  {Lanzafame}, {Lebreton}, {Lobel}, {Lorca}, {Magdaleno Romeo}, {Marocco},
  {Mary}, {Nicolas}, {Ordenovic}, {Pailler}, {Palicio}, {Pallas-Quintela},
  {Panem}, {Pichon}, {Poggio}, {Recio-Blanco}, {Riclet}, {Rybizki},
  {Santovena}, {Sarro}, {Schultheis}, {Segol}, {Slezak}, {Smart}, {Sordo},
  {Soubiran}, {Suveges}, {Thevenin}, {Torralba Elipe}, {Ulla}, {Utrilla},
  {Vallenari}, {van Dillen}, {Zhao}, \& {Zorec}}]{DR3-DPACP-158}
{Delchambre}, L., {Bailer-Jones}, C.~A.~L., {Bellas-Velidis}, I., {et~al.}
  2022, arXiv e-prints, arXiv:2206.06710

\bibitem[{{Delfosse} {et~al.}(1997){Delfosse}, {Tinney}, {Forveille},
  {Epchtein}, {Bertin}, {Borsenberger}, {Copet}, {de Batz}, {Fouque},
  {Kimeswenger}, {Le Bertre}, {Lacombe}, {Rouan}, \&
  {Tiphene}}]{1997A&A...327L..25D}
{Delfosse}, X., {Tinney}, C.~G., {Forveille}, T., {et~al.} 1997, \aap, 327, L25

\bibitem[{Dillon {et~al.}(2017)Dillon, Langmore, Tran, Brevdo, Vasudevan,
  Moore, Patton, Alemi, Hoffman, \& Saurous}]{TensorflowProbability}
Dillon, J.~V., Langmore, I., Tran, D., {et~al.} 2017, TensorFlow Distributions

\bibitem[{{Fan} {et~al.}(2000){Fan}, {Knapp}, {Strauss}, {Gunn}, {Lupton},
  {Ivezi{\'c}}, {Rockosi}, {Yanny}, {Kent}, {Schneider}, {Kirkpatrick},
  {Annis}, {Bastian}, {Berman}, {Brinkmann}, {Csabai}, {Federwitz}, {Fukugita},
  {Gurbani}, {Hennessy}, {Hindsley}, {Ichikawa}, {Lamb}, {Lindenmeyer},
  {Mantsch}, {McKay}, {Munn}, {Nash}, {Okamura}, {Pauls}, {Pier},
  {Rechenmacher}, {Rivetta}, {Sergey}, {Stoughton}, {Szalay}, {Szokoly},
  {Tucker}, {York}, \& {SDSS Collaboration}}]{2000AJ....119..928F}
{Fan}, X., {Knapp}, G.~R., {Strauss}, M.~A., {et~al.} 2000, \aj, 119, 928

\bibitem[{{Fouesneau} {et~al.}(2022){Fouesneau}, {Fr{\'e}mat}, {Andrae},
  {Korn}, {Soubiran}, {Kordopatis}, {Vallenari}, {Heiter}, {Creevey}, {Sarro},
  {de Laverny}, {Lanzafame}, {Lobel}, {Sordo}, {Rybizki}, {Slezak},
  {{\'A}lvarez}, {Drimmel}, {Garabato}, {Delchambre}, {Bailer-Jones},
  {Hatzidimitriou}, {Lorca}, {Le Fustec}, {Pailler}, {Mary}, {Robin},
  {Utrilla}, {Abreu Aramburu}, {Bakker}, {Bellas-Velidis}, {Bijaoui}, {Blomme},
  {Bouret}, {Brouillet}, {Brugaletta}, {Burlacu}, {Carballo}, {Casamiquela},
  {Chaoul}, {Chiavassa}, {Contursi}, {Cooper}, {Dafonte}, {Demouchy},
  {Dharmawardena}, {Garc{\'\i}a-Lario}, {Garc{\'\i}a-Torres}, {Gomez},
  {Gonz{\'a}lez-Santamar{\'\i}a}, {Jean-Antoine Piccolo}, {Kontizas},
  {Lebreton}, {Licata}, {Lindstr{\o}m}, {Livanou}, {Magdaleno Romeo},
  {Manteiga}, {Marocco}, {Martayan}, {Marshall}, {Nicolas}, {Ordenovic},
  {Palicio}, {Pallas-Quintela}, {Pichon}, {Poggio}, {Recio-Blanco}, {Riclet},
  {Santove{\~n}a}, {Schultheis}, {Segol}, {Silvelo}, {Smart}, {S{\"u}veges},
  {Th{\'e}venin}, {Torralba Elipe}, {Ulla}, {van Dillen}, {Zhao}, \&
  {Zorec}}]{DR3-DPACP-160}
{Fouesneau}, M., {Fr{\'e}mat}, Y., {Andrae}, R., {et~al.} 2022, arXiv e-prints,
  arXiv:2206.05992

\bibitem[{{Gagn{\'e}} {et~al.}(2015){Gagn{\'e}}, {Faherty}, {Cruz},
  {Lafreni{\'e}re}, {Doyon}, {Malo}, {Burgasser}, {Naud}, {Artigau},
  {Bouchard}, {Gizis}, \& {Albert}}]{2015ApJS..219...33G}
{Gagn{\'e}}, J., {Faherty}, J.~K., {Cruz}, K.~L., {et~al.} 2015, \apjs, 219, 33

\bibitem[{{Gagn{\'e}} {et~al.}(2018){Gagn{\'e}}, {Mamajek}, {Malo}, {Riedel},
  {Rodriguez}, {Lafreni{\`e}re}, {Faherty}, {Roy-Loubier}, {Pueyo}, {Robin}, \&
  {Doyon}}]{2018ApJ...856...23G}
{Gagn{\'e}}, J., {Mamajek}, E.~E., {Malo}, L., {et~al.} 2018, \apj, 856, 23

\bibitem[{{Gaia Collaboration} {et~al.}(2018{\natexlab{a}}){Gaia
  Collaboration}, {Babusiaux}, {van Leeuwen}, {Barstow}, {Jordi}, {Vallenari},
  {Bossini}, {Bressan}, {Cantat-Gaudin}, {van Leeuwen}, \&
  et~al.}]{2018A&A...616A..10G}
{Gaia Collaboration}, {Babusiaux}, C., {van Leeuwen}, F., {et~al.}
  2018{\natexlab{a}}, \aap, 616, A10

\bibitem[{{Gaia Collaboration} {et~al.}(2018{\natexlab{b}}){Gaia
  Collaboration}, {Brown}, {Vallenari}, {Prusti}, {de Bruijne}, {Babusiaux},
  {Bailer-Jones}, {Biermann}, {Evans}, {Eyer}, \& et~al.}]{2018A&A...616A...1G}
{Gaia Collaboration}, {Brown}, A.~G.~A., {Vallenari}, A., {et~al.}
  2018{\natexlab{b}}, \aap, 616, A1

\bibitem[{{Gaia Collaboration} {et~al.}(2021{\natexlab{a}}){Gaia
  Collaboration}, {Brown}, {Vallenari}, {Prusti}, {de Bruijne}, {Babusiaux},
  {Biermann}, {Creevey}, {Evans}, {Eyer}, {Hutton}, {Jansen}, {Jordi},
  {Klioner}, {Lammers}, {Lindegren}, {Luri}, {Mignard}, {Panem}, {Pourbaix},
  {Randich}, {Sartoretti}, {Soubiran}, {Walton}, {Arenou}, {Bailer-Jones},
  {Bastian}, {Cropper}, {Drimmel}, {Katz}, {Lattanzi}, {van Leeuwen}, {Bakker},
  {Cacciari}, {Casta{\~n}eda}, {De Angeli}, {Ducourant}, {Fabricius},
  {Fouesneau}, {Fr{\'e}mat}, {Guerra}, {Guerrier}, {Guiraud}, {Jean-Antoine
  Piccolo}, {Masana}, {Messineo}, {Mowlavi}, {Nicolas}, {Nienartowicz},
  {Pailler}, {Panuzzo}, {Riclet}, {Roux}, {Seabroke}, {Sordo}, {Tanga},
  {Th{\'e}venin}, {Gracia-Abril}, {Portell}, {Teyssier}, {Altmann}, {Andrae},
  {Bellas-Velidis}, {Benson}, {Berthier}, {Blomme}, {Brugaletta}, {Burgess},
  {Busso}, {Carry}, {Cellino}, {Cheek}, {Clementini}, {Damerdji}, {Davidson},
  {Delchambre}, {Dell'Oro}, {Fern{\'a}ndez-Hern{\'a}ndez}, {Galluccio},
  {Garc{\'\i}a-Lario}, {Garcia-Reinaldos}, {Gonz{\'a}lez-N{\'u}{\~n}ez},
  {Gosset}, {Haigron}, {Halbwachs}, {Hambly}, {Harrison}, {Hatzidimitriou},
  {Heiter}, {Hern{\'a}ndez}, {Hestroffer}, {Hodgkin}, {Holl}, {Jan{\ss}en},
  {Jevardat de Fombelle}, {Jordan}, {Krone-Martins}, {Lanzafame},
  {L{\"o}ffler}, {Lorca}, {Manteiga}, {Marchal}, {Marrese}, {Moitinho}, {Mora},
  {Muinonen}, {Osborne}, {Pancino}, {Pauwels}, {Petit}, {Recio-Blanco},
  {Richards}, {Riello}, {Rimoldini}, {Robin}, {Roegiers}, {Rybizki}, {Sarro},
  {Siopis}, {Smith}, {Sozzetti}, {Ulla}, {Utrilla}, {van Leeuwen}, {van
  Reeven}, {Abbas}, {Abreu Aramburu}, {Accart}, {Aerts}, {Aguado}, {Ajaj},
  {Altavilla}, {{\'A}lvarez}, {{\'A}lvarez Cid-Fuentes}, {Alves}, {Anderson},
  {Anglada Varela}, {Antoja}, {Audard}, {Baines}, {Baker},
  {Balaguer-N{\'u}{\~n}ez}, {Balbinot}, {Balog}, {Barache}, {Barbato},
  {Barros}, {Barstow}, {Bartolom{\'e}}, {Bassilana}, {Bauchet},
  {Baudesson-Stella}, {Becciani}, {Bellazzini}, {Bernet}, {Bertone}, {Bianchi},
  {Blanco-Cuaresma}, {Boch}, {Bombrun}, {Bossini}, {Bouquillon}, {Bragaglia},
  {Bramante}, {Breedt}, {Bressan}, {Brouillet}, {Bucciarelli}, {Burlacu},
  {Busonero}, {Butkevich}, {Buzzi}, {Caffau}, {Cancelliere}, {C{\'a}novas},
  {Cantat-Gaudin}, {Carballo}, {Carlucci}, {Carnerero}, {Carrasco},
  {Casamiquela}, {Castellani}, {Castro-Ginard}, {Castro Sampol}, {Chaoul},
  {Charlot}, {Chemin}, {Chiavassa}, {Cioni}, {Comoretto}, {Cooper}, {Cornez},
  {Cowell}, {Crifo}, {Crosta}, {Crowley}, {Dafonte}, {Dapergolas}, {David},
  {David}, {de Laverny}, {De Luise}, {De March}, {De Ridder}, {de Souza}, {de
  Teodoro}, {de Torres}, {del Peloso}, {del Pozo}, {Delbo}, {Delgado},
  {Delgado}, {Delisle}, {Di Matteo}, {Diakite}, {Diener}, {Distefano},
  {Dolding}, {Eappachen}, {Edvardsson}, {Enke}, {Esquej}, {Fabre}, {Fabrizio},
  {Faigler}, {Fedorets}, {Fernique}, {Fienga}, {Figueras}, {Fouron},
  {Fragkoudi}, {Fraile}, {Franke}, {Gai}, {Garabato}, {Garcia-Gutierrez},
  {Garc{\'\i}a-Torres}, {Garofalo}, {Gavras}, {Gerlach}, {Geyer}, {Giacobbe},
  {Gilmore}, {Girona}, {Giuffrida}, {Gomel}, {Gomez}, {Gonzalez-Santamaria},
  {Gonz{\'a}lez-Vidal}, {Granvik}, {Guti{\'e}rrez-S{\'a}nchez}, {Guy},
  {Hauser}, {Haywood}, {Helmi}, {Hidalgo}, {Hilger}, {H{\l}adczuk}, {Hobbs},
  {Holland}, {Huckle}, {Jasniewicz}, {Jonker}, {Juaristi Campillo}, {Julbe},
  {Karbevska}, {Kervella}, {Khanna}, {Kochoska}, {Kontizas}, {Kordopatis},
  {Korn}, {Kostrzewa-Rutkowska}, {Kruszy{\'n}ska}, {Lambert}, {Lanza}, {Lasne},
  {Le Campion}, {Le Fustec}, {Lebreton}, {Lebzelter}, {Leccia}, {Leclerc},
  {Lecoeur-Taibi}, {Liao}, {Licata}, {Lindstr{\o}m}, {Lister}, {Livanou},
  {Lobel}, {Madrero Pardo}, {Managau}, {Mann}, {Marchant}, {Marconi}, {Marcos
  Santos}, {Marinoni}, {Marocco}, {Marshall}, {Martin Polo},
  {Mart{\'\i}n-Fleitas}, {Masip}, {Massari}, {Mastrobuono-Battisti}, {Mazeh},
  {McMillan}, {Messina}, {Michalik}, {Millar}, {Mints}, {Molina}, {Molinaro},
  {Moln{\'a}r}, {Montegriffo}, {Mor}, {Morbidelli}, {Morel}, {Morris},
  {Mulone}, {Munoz}, {Muraveva}, {Murphy}, {Musella}, {Noval}, {Ord{\'e}novic},
  {Orr{\`u}}, {Osinde}, {Pagani}, {Pagano}, {Palaversa}, {Palicio}, {Panahi},
  {Pawlak}, {Pe{\~n}alosa Esteller}, {Penttil{\"a}}, {Piersimoni}, {Pineau},
  {Plachy}, {Plum}, {Poggio}, {Poretti}, {Poujoulet}, {Pr{\v{s}}a}, {Pulone},
  {Racero}, {Ragaini}, {Rainer}, {Raiteri}, {Rambaux}, {Ramos}, {Ramos-Lerate},
  {Re Fiorentin}, {Regibo}, {Reyl{\'e}}, {Ripepi}, {Riva}, {Rixon}, {Robichon},
  {Robin}, {Roelens}, {Rohrbasser}, {Romero-G{\'o}mez}, {Rowell}, {Royer},
  {Rybicki}, {Sadowski}, {Sagrist{\`a} Sell{\'e}s}, {Sahlmann}, {Salgado},
  {Salguero}, {Samaras}, {Sanchez Gimenez}, {Sanna}, {Santove{\~n}a},
  {Sarasso}, {Schultheis}, {Sciacca}, {Segol}, {Segovia}, {S{\'e}gransan},
  {Semeux}, {Shahaf}, {Siddiqui}, {Siebert}, {Siltala}, {Slezak}, {Smart},
  {Solano}, {Solitro}, {Souami}, {Souchay}, {Spagna}, {Spoto}, {Steele},
  {Steidelm{\"u}ller}, {Stephenson}, {S{\"u}veges}, {Szabados}, {Szegedi-Elek},
  {Taris}, {Tauran}, {Taylor}, {Teixeira}, {Thuillot}, {Tonello}, {Torra},
  {Torra}, {Turon}, {Unger}, {Vaillant}, {van Dillen}, {Vanel}, {Vecchiato},
  {Viala}, {Vicente}, {Voutsinas}, {Weiler}, {Wevers}, {Wyrzykowski}, {Yoldas},
  {Yvard}, {Zhao}, {Zorec}, {Zucker}, {Zurbach}, \&
  {Zwitter}}]{2021A&A...649A...1G}
{Gaia Collaboration}, {Brown}, A.~G.~A., {Vallenari}, A., {et~al.}
  2021{\natexlab{a}}, \aap, 649, A1

\bibitem[{{Gaia Collaboration} {et~al.}(2016{\natexlab{a}}){Gaia
  Collaboration}, {Brown}, {Vallenari}, {Prusti}, {de Bruijne}, {Mignard},
  {Drimmel}, {Babusiaux}, {Bailer-Jones}, {Bastian}, \&
  et~al.}]{2016A&A...595A...2G}
{Gaia Collaboration}, {Brown}, A.~G.~A., {Vallenari}, A., {et~al.}
  2016{\natexlab{a}}, \aap, 595, A2

\bibitem[{{Gaia Collaboration} {et~al.}(2022){Gaia Collaboration}, {Creevey},
  {Sarro}, {Lobel}, {Pancino}, {Andrae}, {Smart}, {Clementini}, {Heiter},
  {Korn}, {Fouesneau}, {Fr{\'e}mat}, {De Angeli}, {Vallenari}, {Harrison},
  {Th{\'e}venin}, {Reyl{\'e}}, {Sordo}, {Garofalo}, {Brown}, {Eyer}, {Prusti},
  {de Bruijne}, {Arenou}, {Babusiaux}, {Biermann}, {Ducourant}, {Evans},
  {Guerra}, {Hutton}, {Jordi}, {Klioner}, {Lammers}, {Lindegren}, {Luri},
  {Mignard}, {Panem}, {Pourbaix}, {Randich}, {Sartoretti}, {Soubiran}, {Tanga},
  {Walton}, {Bailer-Jones}, {Bastian}, {Drimmel}, {Jansen}, {Katz}, {Lattanzi},
  {van Leeuwen}, {Bakker}, {Cacciari}, {Casta{\~n}eda}, {Fabricius},
  {Galluccio}, {Guerrier}, {Masana}, {Messineo}, {Mowlavi}, {Nicolas},
  {Nienartowicz}, {Pailler}, {Panuzzo}, {Riclet}, {Roux}, {Seabroke},
  {Gracia-Abril}, {Portell}, {Teyssier}, {Altmann}, {Audard}, {Bellas-Velidis},
  {Benson}, {Berthier}, {Blomme}, {Burgess}, {Busonero}, {Busso},
  {C{\'a}novas}, {Carry}, {Cellino}, {Cheek}, {Damerdji}, {Davidson}, {de
  Teodoro}, {Nu{\~n}ez Campos}, {Delchambre}, {Dell'Oro}, {Esquej},
  {Fern{\'a}ndez-Hern{\'a}ndez}, {Fraile}, {Garabato}, {Garc{\'\i}a-Lario},
  {Gosset}, {Haigron}, {Halbwachs}, {Hambly}, {Hern{\'a}ndez}, {Hestroffer},
  {Hodgkin}, {Holl}, {Jan{\ss}en}, {Jevardat de Fombelle}, {Jordan},
  {Krone-Martins}, {Lanzafame}, {L{\"o}ffler}, {Marchal}, {Marrese},
  {Moitinho}, {Muinonen}, {Osborne}, {Pauwels}, {Recio-Blanco}, {Riello},
  {Rimoldini}, {Roegiers}, {Rybizki}, {Siopis}, {Smith}, {Sozzetti}, {Utrilla},
  {van Leeuwen}, {Abbas}, {{\'A}brah{\'a}m}, {Abreu Aramburu}, {Aerts},
  {Aguado}, {Ajaj}, {Aldea-Montero}, {Altavilla}, {{\'A}lvarez}, {Alves},
  {Anders}, {Anderson}, {Anglada Varela}, {Antoja}, {Baines}, {Baker},
  {Balaguer-N{\'u}{\~n}ez}, {Balbinot}, {Balog}, {Barache}, {Barbato},
  {Barros}, {Barstow}, {Bartolom{\'e}}, {Bassilana}, {Bauchet}, {Becciani},
  {Bellazzini}, {Berihuete}, {Bernet}, {Bertone}, {Bianchi}, {Binnenfeld},
  {Blanco-Cuaresma}, {Boch}, {Bombrun}, {Bossini}, {Bouquillon}, {Bragaglia},
  {Bramante}, {Breedt}, {Bressan}, {Brouillet}, {Brugaletta}, {Bucciarelli},
  {Burlacu}, {Butkevich}, {Buzzi}, {Caffau}, {Cancelliere}, {Cantat-Gaudin},
  {Carballo}, {Carlucci}, {Carnerero}, {Carrasco}, {Casamiquela}, {Castellani},
  {Castro-Ginard}, {Chaoul}, {Charlot}, {Chemin}, {Chiaramida}, {Chiavassa},
  {Chornay}, {Comoretto}, {Contursi}, {Cooper}, {Cornez}, {Cowell}, {Crifo},
  {Cropper}, {Crosta}, {Crowley}, {Dafonte}, {Dapergolas}, {David}, {de
  Laverny}, {De Luise}, {De March}, {De Ridder}, {de Souza}, {de Torres}, {del
  Peloso}, {del Pozo}, {Delbo}, {Delgado}, {Delisle}, {Demouchy},
  {Dharmawardena}, {Di Matteo}, {Diakite}, {Diener}, {Distefano}, {Dolding},
  {Enke}, {Fabre}, {Fabrizio}, {Faigler}, {Fedorets}, {Fernique}, {Figueras},
  {Fournier}, {Fouron}, {Fragkoudi}, {Gai}, {Garcia-Gutierrez},
  {Garcia-Reinaldos}, {Garc{\'\i}a-Torres}, {Gavel}, {Gavras}, {Gerlach},
  {Geyer}, {Giacobbe}, {Gilmore}, {Girona}, {Giuffrida}, {Gomel}, {Gomez},
  {Gonz{\'a}lez-N{\'u}{\~n}ez}, {Gonz{\'a}lez-Santamar{\'\i}a},
  {Gonz{\'a}lez-Vidal}, {Granvik}, {Guillout}, {Guiraud},
  {Guti{\'e}rrez-S{\'a}nchez}, {Guy}, {Hatzidimitriou}, {Hauser}, {Haywood},
  {Helmer}, {Helmi}, {Sarmiento}, {Hidalgo}, {H{\l}adczuk}, {Hobbs}, {Holland},
  {Huckle}, {Jardine}, {Jasniewicz}, {Jean-Antoine Piccolo},
  {Jim{\'e}nez-Arranz}, {Juaristi Campillo}, {Julbe}, {Karbevska}, {Kervella},
  {Khanna}, {Kordopatis}, {K{\'o}sp{\'a}l}, {Kostrzewa-Rutkowska},
  {Kruszy{\'n}ska}, {Kun}, {Laizeau}, {Lambert}, {Lanza}, {Lasne}, {Le
  Campion}, {Lebreton}, {Lebzelter}, {Leccia}, {Leclerc}, {Lecoeur-Taibi},
  {Liao}, {Licata}, {Lindstr{\o}m}, {Lister}, {Livanou}, {Lorca}, {Loup},
  {Madrero Pardo}, {Magdaleno Romeo}, {Managau}, {Mann}, {Manteiga},
  {Marchant}, {Marconi}, {Marcos}, {Marcos Santos}, {Mar{\'\i}n Pina},
  {Marinoni}, {Marocco}, {Marshall}, {Polo}, {Mart{\'\i}n-Fleitas}, {Marton},
  {Mary}, {Masip}, {Massari}, {Mastrobuono-Battisti}, {Mazeh}, {McMillan},
  {Messina}, {Michalik}, {Millar}, {Mints}, {Molina}, {Molinaro}, {Moln{\'a}r},
  {Monari}, {Mongui{\'o}}, {Montegriffo}, {Montero}, {Mor}, {Mora},
  {Morbidelli}, {Morel}, {Morris}, {Muraveva}, {Murphy}, {Musella}, {Nagy},
  {Noval}, {Oca{\~n}a}, {Ogden}, {Ordenovic}, {Osinde}, {Pagani}, {Pagano},
  {Palaversa}, {Palicio}, {Pallas-Quintela}, {Panahi}, {Payne-Wardenaar},
  {Pe{\~n}alosa Esteller}, {Penttil{\"a}}, {Pichon}, {Piersimoni}, {Pineau},
  {Plachy}, {Plum}, {Poggio}, {Pr{\v{s}}a}, {Pulone}, {Racero}, {Ragaini},
  {Rainer}, {Raiteri}, {Ramos}, {Ramos-Lerate}, {Re Fiorentin}, {Regibo},
  {Richards}, {Rios Diaz}, {Ripepi}, {Riva}, {Rix}, {Rixon}, {Robichon},
  {Robin}, {Robin}, {Roelens}, {Rogues}, {Rohrbasser}, {Romero-G{\'o}mez},
  {Rowell}, {Royer}, {Ruz Mieres}, {Rybicki}, {Sadowski}, {S{\'a}ez
  N{\'u}{\~n}ez}, {Sagrist{\`a} Sell{\'e}s}, {Sahlmann}, {Salguero}, {Samaras},
  {Sanchez Gimenez}, {Sanna}, {Santove{\~n}a}, {Sarasso}, {Schultheis},
  {Sciacca}, {Segol}, {Segovia}, {S{\'e}gransan}, {Semeux}, {Shahaf},
  {Siddiqui}, {Siebert}, {Siltala}, {Silvelo}, {Slezak}, {Slezak}, {Snaith},
  {Solano}, {Solitro}, {Souami}, {Souchay}, {Spagna}, {Spina}, {Spoto},
  {Steele}, {Steidelm{\"u}ller}, {Stephenson}, {S{\"u}veges}, {Surdej},
  {Szabados}, {Szegedi-Elek}, {Taris}, {Taylor}, {Teixeira}, {Tolomei},
  {Tonello}, {Torra}, {Torra}, {Torralba Elipe}, {Trabucchi}, {Tsounis},
  {Turon}, {Ulla}, {Unger}, {Vaillant}, {van Dillen}, {van Reeven}, {Vanel},
  {Vecchiato}, {Viala}, {Vicente}, {Voutsinas}, {Weiler}, {Wevers},
  {Wyrzykowski}, {Yoldas}, {Yvard}, {Zhao}, {Zorec}, {Zucker}, \&
  {Zwitter}}]{DR3-DPACP-123}
{Gaia Collaboration}, {Creevey}, O.~L., {Sarro}, L.~M., {et~al.} 2022, arXiv
  e-prints, arXiv:2206.05870

\bibitem[{{Gaia Collaboration} {et~al.}(2016{\natexlab{b}}){Gaia
  Collaboration}, {Prusti}, {de Bruijne}, {Brown}, {Vallenari}, {Babusiaux},
  {Bailer-Jones}, {Bastian}, {Biermann}, {Evans}, \&
  et~al.}]{2016A&A...595A...1G}
{Gaia Collaboration}, {Prusti}, T., {de Bruijne}, J.~H.~J., {et~al.}
  2016{\natexlab{b}}, \aap, 595, A1

\bibitem[{{Gaia Collaboration} {et~al.}(2021{\natexlab{b}}){Gaia
  Collaboration}, {Smart}, {Sarro}, {Rybizki}, {Reyl{\'e}}, {Robin}, {Hambly},
  {Abbas}, {Barstow}, {de Bruijne}, {Bucciarelli}, {Carrasco}, {Cooper},
  {Hodgkin}, {Masana}, {Michalik}, {Sahlmann}, {Sozzetti}, {Brown},
  {Vallenari}, {Prusti}, {Babusiaux}, {Biermann}, {Creevey}, {Evans}, {Eyer},
  {Hutton}, {Jansen}, {Jordi}, {Klioner}, {Lammers}, {Lindegren}, {Luri},
  {Mignard}, {Panem}, {Pourbaix}, {Randich}, {Sartoretti}, {Soubiran},
  {Walton}, {Arenou}, {Bailer-Jones}, {Bastian}, {Cropper}, {Drimmel}, {Katz},
  {Lattanzi}, {van Leeuwen}, {Bakker}, {Casta{\~n}eda}, {De Angeli},
  {Ducourant}, {Fabricius}, {Fouesneau}, {Fr{\'e}mat}, {Guerra}, {Guerrier},
  {Guiraud}, {Jean-Antoine Piccolo}, {Messineo}, {Mowlavi}, {Nicolas},
  {Nienartowicz}, {Pailler}, {Panuzzo}, {Riclet}, {Roux}, {Seabroke}, {Sordo},
  {Tanga}, {Th{\'e}venin}, {Gracia-Abril}, {Portell}, {Teyssier}, {Altmann},
  {Andrae}, {Bellas-Velidis}, {Benson}, {Berthier}, {Blomme}, {Brugaletta},
  {Burgess}, {Busso}, {Carry}, {Cellino}, {Cheek}, {Clementini}, {Damerdji},
  {Davidson}, {Delchambre}, {Dell'Oro}, {Fern{\'a}ndez-Hern{\'a}ndez},
  {Galluccio}, {Garc{\'\i}a-Lario}, {Garcia-Reinaldos},
  {Gonz{\'a}lez-N{\'u}{\~n}ez}, {Gosset}, {Haigron}, {Halbwachs}, {Harrison},
  {Hatzidimitriou}, {Heiter}, {Hern{\'a}ndez}, {Hestroffer}, {Holl},
  {Jan{\ss}en}, {Jevardat de Fombelle}, {Jordan}, {Krone-Martins}, {Lanzafame},
  {L{\"o}ffler}, {Lorca}, {Manteiga}, {Marchal}, {Marrese}, {Moitinho}, {Mora},
  {Muinonen}, {Osborne}, {Pancino}, {Pauwels}, {Recio-Blanco}, {Richards},
  {Riello}, {Rimoldini}, {Roegiers}, {Siopis}, {Smith}, {Ulla}, {Utrilla}, {van
  Leeuwen}, {van Reeven}, {Abreu Aramburu}, {Accart}, {Aerts}, {Aguado},
  {Ajaj}, {Altavilla}, {{\'A}lvarez}, {{\'A}lvarez Cid-Fuentes}, {Alves},
  {Anderson}, {Anglada Varela}, {Antoja}, {Audard}, {Baines}, {Baker},
  {Balaguer-N{\'u}{\~n}ez}, {Balbinot}, {Balog}, {Barache}, {Barbato},
  {Barros}, {Bartolom{\'e}}, {Bassilana}, {Bauchet}, {Baudesson-Stella},
  {Becciani}, {Bellazzini}, {Bernet}, {Bertone}, {Bianchi}, {Blanco-Cuaresma},
  {Boch}, {Bombrun}, {Bossini}, {Bouquillon}, {Bragaglia}, {Bramante},
  {Breedt}, {Bressan}, {Brouillet}, {Burlacu}, {Busonero}, {Butkevich},
  {Buzzi}, {Caffau}, {Cancelliere}, {C{\'a}novas}, {Cantat-Gaudin}, {Carballo},
  {Carlucci}, {Carnerero}, {Casamiquela}, {Castellani}, {Castro-Ginard},
  {Castro Sampol}, {Chaoul}, {Charlot}, {Chemin}, {Chiavassa}, {Cioni},
  {Comoretto}, {Cornez}, {Cowell}, {Crifo}, {Crosta}, {Crowley}, {Dafonte},
  {Dapergolas}, {David}, {David}, {de Laverny}, {De Luise}, {De March}, {De
  Ridder}, {de Souza}, {de Teodoro}, {de Torres}, {del Peloso}, {del Pozo},
  {Delgado}, {Delgado}, {Delisle}, {Di Matteo}, {Diakite}, {Diener},
  {Distefano}, {Dolding}, {Eappachen}, {Edvardsson}, {Enke}, {Esquej}, {Fabre},
  {Fabrizio}, {Faigler}, {Fedorets}, {Fernique}, {Fienga}, {Figueras},
  {Fouron}, {Fragkoudi}, {Fraile}, {Franke}, {Gai}, {Garabato},
  {Garcia-Gutierrez}, {Garc{\'\i}a-Torres}, {Garofalo}, {Gavras}, {Gerlach},
  {Geyer}, {Giacobbe}, {Gilmore}, {Girona}, {Giuffrida}, {Gomel}, {Gomez},
  {Gonzalez-Santamaria}, {Gonz{\'a}lez-Vidal}, {Granvik},
  {Guti{\'e}rrez-S{\'a}nchez}, {Guy}, {Hauser}, {Haywood}, {Helmi}, {Hidalgo},
  {Hilger}, {H{\l}adczuk}, {Hobbs}, {Holland}, {Huckle}, {Jasniewicz},
  {Jonker}, {Juaristi Campillo}, {Julbe}, {Karbevska}, {Kervella}, {Khanna},
  {Kochoska}, {Kontizas}, {Kordopatis}, {Korn}, {Kostrzewa-Rutkowska},
  {Kruszy{\'n}ska}, {Lambert}, {Lanza}, {Lasne}, {Le Campion}, {Le Fustec},
  {Lebreton}, {Lebzelter}, {Leccia}, {Leclerc}, {Lecoeur-Taibi}, {Liao},
  {Licata}, {Lindstr{\o}m}, {Lister}, {Livanou}, {Lobel}, {Madrero Pardo},
  {Managau}, {Mann}, {Marchant}, {Marconi}, {Marcos Santos}, {Marinoni},
  {Marocco}, {Marshall}, {Martin Polo}, {Mart{\'\i}n-Fleitas}, {Masip},
  {Massari}, {Mastrobuono-Battisti}, {Mazeh}, {McMillan}, {Messina}, {Millar},
  {Mints}, {Molina}, {Molinaro}, {Moln{\'a}r}, {Montegriffo}, {Mor},
  {Morbidelli}, {Morel}, {Morris}, {Mulone}, {Munoz}, {Muraveva}, {Murphy},
  {Musella}, {Noval}, {Ord{\'e}novic}, {Orr{\`u}}, {Osinde}, {Pagani},
  {Pagano}, {Palaversa}, {Palicio}, {Panahi}, {Pawlak}, {Pe{\~n}alosa
  Esteller}, {Penttil{\"a}}, {Piersimoni}, {Pineau}, {Plachy}, {Plum},
  {Poggio}, {Poretti}, {Poujoulet}, {Pr{\v{s}}a}, {Pulone}, {Racero},
  {Ragaini}, {Rainer}, {Raiteri}, {Rambaux}, {Ramos}, {Ramos-Lerate}, {Re
  Fiorentin}, {Regibo}, {Ripepi}, {Riva}, {Rixon}, {Robichon}, {Robin},
  {Roelens}, {Rohrbasser}, {Romero-G{\'o}mez}, {Rowell}, {Royer}, {Rybicki},
  {Sadowski}, {Sagrist{\`a} Sell{\'e}s}, {Salgado}, {Salguero}, {Samaras},
  {Sanchez Gimenez}, {Sanna}, {Santove{\~n}a}, {Sarasso}, {Schultheis},
  {Sciacca}, {Segol}, {Segovia}, {S{\'e}gransan}, {Semeux}, {Shahaf},
  {Siddiqui}, {Siebert}, {Siltala}, {Slezak}, {Solano}, {Solitro}, {Souami},
  {Souchay}, {Spagna}, {Spoto}, {Steele}, {Steidelm{\"u}ller}, {Stephenson},
  {S{\"u}veges}, {Szabados}, {Szegedi-Elek}, {Taris}, {Tauran}, {Taylor},
  {Teixeira}, {Thuillot}, {Tonello}, {Torra}, {Torra}, {Turon}, {Unger},
  {Vaillant}, {van Dillen}, {Vanel}, {Vecchiato}, {Viala}, {Vicente},
  {Voutsinas}, {Weiler}, {Wevers}, {Wyrzykowski}, {Yoldas}, {Yvard}, {Zhao},
  {Zorec}, {Zucker}, {Zurbach}, \& {Zwitter}}]{2021A&A...649A...6G}
{Gaia Collaboration}, {Smart}, R.~L., {Sarro}, L.~M., {et~al.}
  2021{\natexlab{b}}, \aap, 649, A6

\bibitem[{{Geballe} {et~al.}(2002){Geballe}, {Knapp}, {Leggett}, {Fan},
  {Golimowski}, {Anderson}, {Brinkmann}, {Csabai}, {Gunn}, {Hawley},
  {Hennessy}, {Henry}, {Hill}, {Hindsley}, {Ivezi{\'c}}, {Lupton}, {McDaniel},
  {Munn}, {Narayanan}, {Peng}, {Pier}, {Rockosi}, {Schneider}, {Smith},
  {Strauss}, {Tsvetanov}, {Uomoto}, {York}, \& {Zheng}}]{2002ApJ...564..466G}
{Geballe}, T.~R., {Knapp}, G.~R., {Leggett}, S.~K., {et~al.} 2002, \apj, 564,
  466

\bibitem[{{Gizis}(2002)}]{2002ApJ...575..484G}
{Gizis}, J.~E. 2002, \apj, 575, 484

\bibitem[{{Gizis} {et~al.}(2000){Gizis}, {Monet}, {Reid}, {Kirkpatrick},
  {Liebert}, \& {Williams}}]{2000AJ....120.1085G}
{Gizis}, J.~E., {Monet}, D.~G., {Reid}, I.~N., {et~al.} 2000, \aj, 120, 1085

\bibitem[{{Gravity Collaboration} {et~al.}(2022){Gravity Collaboration},
  {Abuter}, {Aimar}, {Amorim}, {Ball}, {Baub{\"o}ck}, {Berger}, {Bonnet},
  {Bourdarot}, {Brandner}, {Cardoso}, {Cl{\'e}net}, {Dallilar}, {Davies}, {de
  Zeeuw}, {Dexter}, {Drescher}, {Eisenhauer}, {F{\"o}rster Schreiber},
  {Foschi}, {Garcia}, {Gao}, {Gendron}, {Genzel}, {Gillessen}, {Habibi},
  {Haubois}, {Hei{\ss}el}, {Henning}, {Hippler}, {Horrobin}, {Jochum}, {Jocou},
  {Kaufer}, {Kervella}, {Lacour}, {Lapeyr{\`e}re}, {Le Bouquin}, {L{\'e}na},
  {Lutz}, {Ott}, {Paumard}, {Perraut}, {Perrin}, {Pfuhl}, {Rabien},
  {Shangguan}, {Shimizu}, {Scheithauer}, {Stadler}, {Stephens}, {Straub},
  {Straubmeier}, {Sturm}, {Tacconi}, {Tristram}, {Vincent}, {von Fellenberg},
  {Widmann}, {Wieprecht}, {Wiezorrek}, {Woillez}, {Yazici}, \&
  {Young}}]{2022A&A...657L..12G}
{Gravity Collaboration}, {Abuter}, R., {Aimar}, N., {et~al.} 2022, \aap, 657,
  L12

\bibitem[{{Green}(2018)}]{2018JOSS....3..695M}
{Green}, G. 2018, The Journal of Open Source Software, 3, 695

\bibitem[{{Hall}(2002)}]{2002ApJ...580L..77H}
{Hall}, P.~B. 2002, \apjl, 580, L77

\bibitem[{{Hawley} {et~al.}(2002){Hawley}, {Covey}, {Knapp}, {Golimowski},
  {Fan}, {Anderson}, {Gunn}, {Harris}, {Ivezi{\'c}}, {Long}, {Lupton},
  {McGehee}, {Narayanan}, {Peng}, {Schlegel}, {Schneider}, {Spahn}, {Strauss},
  {Szkody}, {Tsvetanov}, {Walkowicz}, {Brinkmann}, {Harvanek}, {Hennessy},
  {Kleinman}, {Krzesinski}, {Long}, {Neilsen}, {Newman}, {Nitta}, {Snedden}, \&
  {York}}]{2002AJ....123.3409H}
{Hawley}, S.~L., {Covey}, K.~R., {Knapp}, G.~R., {et~al.} 2002, \aj, 123, 3409

\bibitem[{{Hennebelle} \& {Chabrier}(2008)}]{2008ApJ...684..395H}
{Hennebelle}, P. \& {Chabrier}, G. 2008, \apj, 684, 395

\bibitem[{Higdon(2002)}]{Higdon2020}
Higdon, D. 2002, in Quantitative Methods for Current Environmental Issues, ed.
  C.~W. Anderson, V.~Barnett, P.~C. Chatwin, \& A.~H. El-Shaarawi (London:
  Springer London), 37--56

\bibitem[{Hoffman {et~al.}(2014)Hoffman, Gelman, {et~al.}}]{hoffman2014no}
Hoffman, M.~D., Gelman, A., {et~al.} 2014, J. Mach. Learn. Res., 15, 1593

\bibitem[{{Jao} {et~al.}(2018){Jao}, {Henry}, {Gies}, \&
  {Hambly}}]{2018ApJ...861L..11J}
{Jao}, W.-C., {Henry}, T.~J., {Gies}, D.~R., \& {Hambly}, N.~C. 2018, \apjl,
  861, L11

\bibitem[{{Kendall} {et~al.}(2004){Kendall}, {Delfosse}, {Mart{\'\i}n}, \&
  {Forveille}}]{2004A&A...416L..17K}
{Kendall}, T.~R., {Delfosse}, X., {Mart{\'\i}n}, E.~L., \& {Forveille}, T.
  2004, \aap, 416, L17

\bibitem[{{Kendall} {et~al.}(2007){Kendall}, {Jones}, {Pinfield}, {Pokorny},
  {Folkes}, {Weights}, {Jenkins}, \& {Mauron}}]{2007MNRAS.374..445K}
{Kendall}, T.~R., {Jones}, H.~R.~A., {Pinfield}, D.~J., {et~al.} 2007, \mnras,
  374, 445

\bibitem[{{Kerr} {et~al.}(2021){Kerr}, {Rizzuto}, {Kraus}, \&
  {Offner}}]{2021ApJ...917...23K}
{Kerr}, R. M.~P., {Rizzuto}, A.~C., {Kraus}, A.~L., \& {Offner}, S. S.~R. 2021,
  \apj, 917, 23

\bibitem[{{Kirkpatrick} {et~al.}(2021){Kirkpatrick}, {Gelino}, {Faherty},
  {Meisner}, {Caselden}, {Schneider}, {Marocco}, {Cayago}, {Smart},
  {Eisenhardt}, {Kuchner}, {Wright}, {Cushing}, {Allers}, {Bardalez Gagliuffi},
  {Burgasser}, {Gagn{\'e}}, {Logsdon}, {Martin}, {Ingalls}, {Lowrance},
  {Abrahams}, {Aganze}, {Gerasimov}, {Gonzales}, {Hsu}, {Kamraj}, {Kiman},
  {Rees}, {Theissen}, {Ammar}, {Andersen}, {Beaulieu}, {Colin}, {Elachi},
  {Goodman}, {Gramaize}, {Hamlet}, {Hong}, {Jonkeren}, {Khalil}, {Martin},
  {Pendrill}, {Pumphrey}, {Rothermich}, {Sainio}, {Stenner}, {Tanner},
  {Th{\'e}venot}, {Voloshin}, {Walla}, {W{\k{e}}dracki}, \& {Backyard Worlds:
  Planet 9 Collaboration}}]{2021ApJS..253....7K}
{Kirkpatrick}, J.~D., {Gelino}, C.~R., {Faherty}, J.~K., {et~al.} 2021, \apjs,
  253, 7

\bibitem[{{Kirkpatrick} {et~al.}(1991){Kirkpatrick}, {Henry}, \&
  {McCarthy}}]{1991ApJS...77..417K}
{Kirkpatrick}, J.~D., {Henry}, T.~J., \& {McCarthy}, Donald~W., J. 1991, \apjs,
  77, 417

\bibitem[{{Kirkpatrick} {et~al.}(2019){Kirkpatrick}, {Martin}, {Smart},
  {Cayago}, {Beichman}, {Marocco}, {Gelino}, {Faherty}, {Cushing}, {Schneider},
  {Mace}, {Tinney}, {Wright}, {Lowrance}, {Ingalls}, {Vrba}, {Munn}, {Dahm}, \&
  {McLean}}]{2019ApJS..240...19K}
{Kirkpatrick}, J.~D., {Martin}, E.~C., {Smart}, R.~L., {et~al.} 2019, \apjs,
  240, 19

\bibitem[{{Kirkpatrick} {et~al.}(1999){Kirkpatrick}, {Reid}, {Liebert},
  {Cutri}, {Nelson}, {Beichman}, {Dahn}, {Monet}, {Gizis}, \&
  {Skrutskie}}]{1999ApJ...519..802K}
{Kirkpatrick}, J.~D., {Reid}, I.~N., {Liebert}, J., {et~al.} 1999, \apj, 519,
  802

\bibitem[{{Kirkpatrick} {et~al.}(2000){Kirkpatrick}, {Reid}, {Liebert},
  {Gizis}, {Burgasser}, {Monet}, {Dahn}, {Nelson}, \&
  {Williams}}]{2000AJ....120..447K}
{Kirkpatrick}, J.~D., {Reid}, I.~N., {Liebert}, J., {et~al.} 2000, \aj, 120,
  447

\bibitem[{{Kounkel} {et~al.}(2022){Kounkel}, {Stassun}, {Covey}, \&
  {Hartmann}}]{2022MNRAS.517..161K}
{Kounkel}, M., {Stassun}, K.~G., {Covey}, K., \& {Hartmann}, L. 2022, \mnras,
  517, 161

\bibitem[{{L{\'e}pine} \& {Shara}(2005)}]{2005AJ....129.1483L}
{L{\'e}pine}, S. \& {Shara}, M.~M. 2005, \aj, 129, 1483

\bibitem[{{L{\'e}pine} {et~al.}(2002){L{\'e}pine}, {Shara}, \&
  {Rich}}]{2002AJ....124.1190L}
{L{\'e}pine}, S., {Shara}, M.~M., \& {Rich}, R.~M. 2002, \aj, 124, 1190

\bibitem[{Li {et~al.}(2007)Li, Ray, \&
  Lindsay}]{651d1d8e5a5a4871b38e424902d0098c}
Li, J., Ray, S., \& Lindsay, B. 2007, Journal of Machine Learning Research, 8,
  1687

\bibitem[{{Liu} {et~al.}(2016){Liu}, {Dupuy}, \&
  {Allers}}]{2016ApJ...833...96L}
{Liu}, M.~C., {Dupuy}, T.~J., \& {Allers}, K.~N. 2016, \apj, 833, 96

\bibitem[{{Looper} {et~al.}(2008){Looper}, {Kirkpatrick}, {Cutri}, {Barman},
  {Burgasser}, {Cushing}, {Roellig}, {McGovern}, {McLean}, {Rice}, {Swift}, \&
  {Schurr}}]{2008ApJ...686..528L}
{Looper}, D.~L., {Kirkpatrick}, J.~D., {Cutri}, R.~M., {et~al.} 2008, \apj,
  686, 528

\bibitem[{{Luri} {et~al.}(2018){Luri}, {Brown}, {Sarro}, {Arenou},
  {Bailer-Jones}, {Castro-Ginard}, {de Bruijne}, {Prusti}, {Babusiaux}, \&
  {Delgado}}]{2018A&A...616A...9L}
{Luri}, X., {Brown}, A.~G.~A., {Sarro}, L.~M., {et~al.} 2018, \aap, 616, A9

\bibitem[{{Luyten}(1955)}]{1955LFT...C......0L}
{Luyten}, W.~J. 1955, Luyten's Five Tenths. (1955, 0

\bibitem[{{Luyten}(1979)}]{1979LHS...C......0L}
{Luyten}, W.~J. 1979, {LHS catalogue. A catalogue of stars with proper motions
  exceeding 0''5 annually}

\bibitem[{{Mainzer} {et~al.}(2011){Mainzer}, {Bauer}, {Grav}, {Masiero},
  {Cutri}, {Dailey}, {Eisenhardt}, {McMillan}, {Wright}, {Walker}, {Jedicke},
  {Spahr}, {Tholen}, {Alles}, {Beck}, {Brandenburg}, {Conrow}, {Evans},
  {Fowler}, {Jarrett}, {Marsh}, {Masci}, {McCallon}, {Wheelock}, {Wittman},
  {Wyatt}, {DeBaun}, {Elliott}, {Elsbury}, {Gautier}, {Gomillion}, {Leisawitz},
  {Maleszewski}, {Micheli}, \& {Wilkins}}]{2011ApJ...731...53M}
{Mainzer}, A., {Bauer}, J., {Grav}, T., {et~al.} 2011, \apj, 731, 53

\bibitem[{{Marocco} {et~al.}(2013){Marocco}, {Andrei}, {Smart}, {Jones},
  {Pinfield}, {Day-Jones}, {Clarke}, {Sozzetti}, {Lucas}, {Bucciarelli}, \&
  {Penna}}]{2013AJ....146..161M}
{Marocco}, F., {Andrei}, A.~H., {Smart}, R.~L., {et~al.} 2013, \aj, 146, 161

\bibitem[{{Mart\'{i}n} {et~al.}(1997){Mart\'{i}n}, {Basri}, {Delfosse}, \&
  {Forveille}}]{1997A&A...327L..29M}
{Mart\'{i}n}, E.~L., {Basri}, G., {Delfosse}, X., \& {Forveille}, T. 1997,
  \aap, 327, L29

\bibitem[{{Mart{\'\i}n} {et~al.}(1999){Mart{\'\i}n}, {Delfosse}, {Basri},
  {Goldman}, {Forveille}, \& {Zapatero Osorio}}]{1999AJ....118.2466M}
{Mart{\'\i}n}, E.~L., {Delfosse}, X., {Basri}, G., {et~al.} 1999, \aj, 118,
  2466

\bibitem[{{Montegriffo} {et~al.}(2022){Montegriffo}, {De Angeli}, {Andrae},
  {Riello}, {Pancino}, {Sanna}, {Bellazzini}, {Evans}, {Carrasco}, {Sordo},
  {Busso}, {Cacciari}, {Jordi}, {van Leeuwen}, {Vallenari}, {Altavilla},
  {Barstow}, {Brown}, {Burgess}, {Castellani}, {Cowell}, {Davidson}, {De
  Luise}, {Delchambre}, {Diener}, {Fabricius}, {Fremat}, {Fouesneau},
  {Gilmore}, {Giuffrida}, {Hambly}, {Harrison}, {Hidalgo}, {Hodgkin},
  {Holland}, {Marinoni}, {Osborne}, {Pagani}, {Palaversa}, {Piersimoni},
  {Pulone}, {Ragaini}, {Rainer}, {Richards}, {Rowell}, {Ruz-Mieres}, {Sarro},
  {Walton}, \& {Yoldas}}]{EDR3-DPACP-120}
{Montegriffo}, P., {De Angeli}, F., {Andrae}, R., {et~al.} 2022, arXiv
  e-prints, arXiv:2206.06205

\bibitem[{{Padoan} \& {Nordlund}(2004)}]{2004ApJ...617..559P}
{Padoan}, P. \& {Nordlund}, {\r{A}}. 2004, \apj, 617, 559

\bibitem[{{Planck Collaboration} {et~al.}(2016){Planck Collaboration},
  {Aghanim}, {Ashdown}, {Aumont}, {Baccigalupi}, {Ballardini}, {Banday},
  {Barreiro}, {Bartolo}, {Basak}, {Benabed}, {Bernard}, {Bersanelli},
  {Bielewicz}, {Bonavera}, {Bond}, {Borrill}, {Bouchet}, {Boulanger},
  {Burigana}, {Calabrese}, {Cardoso}, {Carron}, {Chiang}, {Colombo}, {Comis},
  {Couchot}, {Coulais}, {Crill}, {Curto}, {Cuttaia}, {de Bernardis}, {de
  Zotti}, {Delabrouille}, {Di Valentino}, {Dickinson}, {Diego}, {Dor{\'e}},
  {Douspis}, {Ducout}, {Dupac}, {Dusini}, {Elsner}, {En{\ss}lin}, {Eriksen},
  {Falgarone}, {Fantaye}, {Finelli}, {Forastieri}, {Frailis}, {Fraisse},
  {Franceschi}, {Frolov}, {Galeotta}, {Galli}, {Ganga}, {G{\'e}nova-Santos},
  {Gerbino}, {Ghosh}, {Giraud-H{\'e}raud}, {Gonz{\'a}lez-Nuevo}, {G{\'o}rski},
  {Gruppuso}, {Gudmundsson}, {Hansen}, {Helou}, {Henrot-Versill{\'e}},
  {Herranz}, {Hivon}, {Huang}, {Jaffe}, {Jones}, {Keih{\"a}nen}, {Keskitalo},
  {Kiiveri}, {Kisner}, {Krachmalnicoff}, {Kunz}, {Kurki-Suonio}, {Lamarre},
  {Langer}, {Lasenby}, {Lattanzi}, {Lawrence}, {Le Jeune}, {Levrier}, {Lilje},
  {Lilley}, {Lindholm}, {L{\'o}pez-Caniego}, {Ma}, {Mac{\'\i}as-P{\'e}rez},
  {Maggio}, {Maino}, {Mandolesi}, {Mangilli}, {Maris}, {Martin},
  {Mart{\'\i}nez-Gonz{\'a}lez}, {Matarrese}, {Mauri}, {McEwen}, {Melchiorri},
  {Mennella}, {Migliaccio}, {Miville-Desch{\^e}nes}, {Molinari}, {Moneti},
  {Montier}, {Morgante}, {Moss}, {Natoli}, {Oxborrow}, {Pagano}, {Paoletti},
  {Patanchon}, {Perdereau}, {Perotto}, {Pettorino}, {Piacentini},
  {Plaszczynski}, {Polastri}, {Polenta}, {Puget}, {Rachen}, {Racine},
  {Reinecke}, {Remazeilles}, {Renzi}, {Rocha}, {Rosset}, {Rossetti}, {Roudier},
  {Rubi{\~n}o-Mart{\'\i}n}, {Ruiz-Granados}, {Salvati}, {Sandri}, {Savelainen},
  {Scott}, {Sirignano}, {Sirri}, {Soler}, {Spencer}, {Suur-Uski}, {Tauber},
  {Tavagnacco}, {Tenti}, {Toffolatti}, {Tomasi}, {Tristram}, {Trombetti},
  {Valiviita}, {Van Tent}, {Vielva}, {Villa}, {Vittorio}, {Wandelt}, {Wehus},
  {Zacchei}, \& {Zonca}}]{2016A&A...596A.109P}
{Planck Collaboration}, {Aghanim}, N., {Ashdown}, M., {et~al.} 2016, \aap, 596,
  A109

\bibitem[{Rasmussen \& Williams(2006)}]{GPs4ML}
Rasmussen, C. \& Williams, C. 2006, Gaussian Processes for Machine Learning,
  Adaptive Computation and Machine Learning (Cambridge, MA, USA: MIT Press),
  248

\bibitem[{{Reid} {et~al.}(2008){Reid}, {Cruz}, {Kirkpatrick}, {Allen},
  {Mungall}, {Liebert}, {Lowrance}, \& {Sweet}}]{2008AJ....136.1290R}
{Reid}, I.~N., {Cruz}, K.~L., {Kirkpatrick}, J.~D., {et~al.} 2008, \aj, 136,
  1290

\bibitem[{{Reid} \& {Gizis}(2005)}]{2005PASP..117..676R}
{Reid}, I.~N. \& {Gizis}, J.~E. 2005, \pasp, 117, 676

\bibitem[{{Reid} {et~al.}(2000){Reid}, {Kirkpatrick}, {Gizis}, {Dahn}, {Monet},
  {Williams}, {Liebert}, \& {Burgasser}}]{2000AJ....119..369R}
{Reid}, I.~N., {Kirkpatrick}, J.~D., {Gizis}, J.~E., {et~al.} 2000, \aj, 119,
  369

\bibitem[{{Reid} {et~al.}(2006){Reid}, {Lewitus}, {Allen}, {Cruz}, \&
  {Burgasser}}]{2006AJ....132..891R}
{Reid}, I.~N., {Lewitus}, E., {Allen}, P.~R., {Cruz}, K.~L., \& {Burgasser},
  A.~J. 2006, \aj, 132, 891

\bibitem[{{Reiners} {et~al.}(2007){Reiners}, {Homeier}, {Hauschildt}, \&
  {Allard}}]{2007A&A...473..245R}
{Reiners}, A., {Homeier}, D., {Hauschildt}, P.~H., \& {Allard}, F. 2007, \aap,
  473, 245

\bibitem[{{Reipurth} \& {Clarke}(2001)}]{2001AJ....122..432R}
{Reipurth}, B. \& {Clarke}, C. 2001, \aj, 122, 432

\bibitem[{{Reyl{\'e}}(2018)}]{2018A&A...619L...8R}
{Reyl{\'e}}, C. 2018, \aap, 619, L8

\bibitem[{{Riaz} {et~al.}(2018){Riaz}, {Vanaverbeke}, \&
  {Schleicher}}]{2018MNRAS.478.5460R}
{Riaz}, R., {Vanaverbeke}, S., \& {Schleicher}, D.~R.~G. 2018, \mnras, 478,
  5460

\bibitem[{{Riello} {et~al.}(2021){Riello}, {De Angeli}, {Evans}, {Montegriffo},
  {Carrasco}, {Busso}, {Palaversa}, {Burgess}, {Diener}, {Davidson}, {Rowell},
  {Fabricius}, {Jordi}, {Bellazzini}, {Pancino}, {Harrison}, {Cacciari}, {van
  Leeuwen}, {Hambly}, {Hodgkin}, {Osborne}, {Altavilla}, {Barstow}, {Brown},
  {Castellani}, {Cowell}, {De Luise}, {Gilmore}, {Giuffrida}, {Hidalgo},
  {Holland}, {Marinoni}, {Pagani}, {Piersimoni}, {Pulone}, {Ragaini}, {Rainer},
  {Richards}, {Sanna}, {Walton}, {Weiler}, \& {Yoldas}}]{EDR3-DPACP-117}
{Riello}, M., {De Angeli}, F., {Evans}, D.~W., {et~al.} 2021, \aap, 649, A3

\bibitem[{{Rix} {et~al.}(2021){Rix}, {Hogg}, {Boubert}, {Brown}, {Casey},
  {Drimmel}, {Everall}, {Fouesneau}, \& {Price-Whelan}}]{2021AJ....162..142R}
{Rix}, H.-W., {Hogg}, D.~W., {Boubert}, D., {et~al.} 2021, \aj, 162, 142

\bibitem[{{Salim} {et~al.}(2003){Salim}, {L{\'e}pine}, {Rich}, \&
  {Shara}}]{2003ApJ...586L.149S}
{Salim}, S., {L{\'e}pine}, S., {Rich}, R.~M., \& {Shara}, M.~M. 2003, \apjl,
  586, L149

\bibitem[{{Sarro} {et~al.}(2013){Sarro}, {Berihuete}, {Carri{\'o}n}, {Barrado},
  {Cruz}, \& {Isasi}}]{2013A&A...550A..44S}
{Sarro}, L.~M., {Berihuete}, A., {Carri{\'o}n}, C., {et~al.} 2013, \aap, 550,
  A44

\bibitem[{{Saumon} {et~al.}(2000){Saumon}, {Geballe}, {Leggett}, {Marley},
  {Freedman}, {Lodders}, {Fegley}, \& {Sengupta}}]{2000ApJ...541..374S}
{Saumon}, D., {Geballe}, T.~R., {Leggett}, S.~K., {et~al.} 2000, \apj, 541, 374

\bibitem[{{Schmidt} {et~al.}(2007){Schmidt}, {Cruz}, {Bongiorno}, {Liebert}, \&
  {Reid}}]{2007AJ....133.2258S}
{Schmidt}, S.~J., {Cruz}, K.~L., {Bongiorno}, B.~J., {Liebert}, J., \& {Reid},
  I.~N. 2007, \aj, 133, 2258

\bibitem[{{Schneider} {et~al.}(2014){Schneider}, {Cushing}, {Kirkpatrick},
  {Mace}, {Gelino}, {Faherty}, {Fajardo-Acosta}, \&
  {Sheppard}}]{2014AJ....147...34S}
{Schneider}, A.~C., {Cushing}, M.~C., {Kirkpatrick}, J.~D., {et~al.} 2014, \aj,
  147, 34

\bibitem[{{Scholz} \& {Meusinger}(2002)}]{2002MNRAS.336L..49S}
{Scholz}, R.~D. \& {Meusinger}, H. 2002, \mnras, 336, L49

\bibitem[{{Sch{\"o}nrich} {et~al.}(2010){Sch{\"o}nrich}, {Binney}, \&
  {Dehnen}}]{2010MNRAS.403.1829S}
{Sch{\"o}nrich}, R., {Binney}, J., \& {Dehnen}, W. 2010, \mnras, 403, 1829

\bibitem[{{Skrutskie} {et~al.}(2006){Skrutskie}, {Cutri}, {Stiening},
  {Weinberg}, {Schneider}, {Carpenter}, {Beichman}, {Capps}, {Chester},
  {Elias}, {Huchra}, {Liebert}, {Lonsdale}, {Monet}, {Price}, {Seitzer},
  {Jarrett}, {Kirkpatrick}, {Gizis}, {Howard}, {Evans}, {Fowler}, {Fullmer},
  {Hurt}, {Light}, {Kopan}, {Marsh}, {McCallon}, {Tam}, {Van Dyk}, \&
  {Wheelock}}]{2006AJ....131.1163S}
{Skrutskie}, M.~F., {Cutri}, R.~M., {Stiening}, R., {et~al.} 2006, \aj, 131,
  1163

\bibitem[{{Smart} {et~al.}(2017){Smart}, {Marocco}, {Caballero}, {Jones},
  {Barrado}, {Beam{\'{\i}}n}, {Pinfield}, \& {Sarro}}]{2017MNRAS.469..401S}
{Smart}, R.~L., {Marocco}, F., {Caballero}, J.~A., {et~al.} 2017, \mnras, 469,
  401

\bibitem[{{Stamer} \& {Inutsuka}(2019)}]{2019MNRAS.488.2644S}
{Stamer}, T. \& {Inutsuka}, S.-i. 2019, \mnras, 488, 2644

\bibitem[{{Stauffer} {et~al.}(2015){Stauffer}, {Cody}, {McGinnis}, {Rebull},
  {Hillenbrand}, {Turner}, {Carpenter}, {Plavchan}, {Carey}, {Terebey},
  {Morales-Calder{\'o}n}, {Alencar}, {Bouvier}, {Venuti}, {Hartmann}, {Calvet},
  {Micela}, {Flaccomio}, {Song}, {Gutermuth}, {Barrado}, {Vrba}, {Covey},
  {Padgett}, {Herbst}, {Gillen}, {Lyra}, {Medeiros Guimaraes}, {Bouy}, \&
  {Favata}}]{2015AJ....149..130S}
{Stauffer}, J., {Cody}, A.~M., {McGinnis}, P., {et~al.} 2015, \aj, 149, 130

\bibitem[{{Stephens} {et~al.}(2009){Stephens}, {Leggett}, {Cushing}, {Marley},
  {Saumon}, {Geballe}, {Golimowski}, {Fan}, \& {Noll}}]{2009ApJ...702..154S}
{Stephens}, D.~C., {Leggett}, S.~K., {Cushing}, M.~C., {et~al.} 2009, \apj,
  702, 154

\bibitem[{{Tinney} \& {Reid}(1998)}]{1998MNRAS.301.1031T}
{Tinney}, C.~G. \& {Reid}, I.~N. 1998, \mnras, 301, 1031

\bibitem[{{Veras} \& {Raymond}(2012)}]{2012MNRAS.421L.117V}
{Veras}, D. \& {Raymond}, S.~N. 2012, \mnras, 421, L117

\bibitem[{{Whitworth} \& {Stamatellos}(2006)}]{2006A&A...458..817W}
{Whitworth}, A.~P. \& {Stamatellos}, D. 2006, \aap, 458, 817

\bibitem[{{Whitworth} \& {Zinnecker}(2004)}]{2004A&A...427..299W}
{Whitworth}, A.~P. \& {Zinnecker}, H. 2004, \aap, 427, 299

\bibitem[{{Wilson} {et~al.}(2014){Wilson}, {Rajan}, \&
  {Patience}}]{2014A&A...566A.111W}
{Wilson}, P.~A., {Rajan}, A., \& {Patience}, J. 2014, \aap, 566, A111

\bibitem[{{Wright} {et~al.}(2010){Wright}, {Eisenhardt}, {Mainzer}, {Ressler},
  {Cutri}, {Jarrett}, {Kirkpatrick}, {Padgett}, {McMillan}, {Skrutskie},
  {Stanford}, {Cohen}, {Walker}, {Mather}, {Leisawitz}, {Gautier}, {McLean},
  {Benford}, {Lonsdale}, {Blain}, {Mendez}, {Irace}, {Duval}, {Liu}, {Royer},
  {Heinrichsen}, {Howard}, {Shannon}, {Kendall}, {Walsh}, {Larsen}, {Cardon},
  {Schick}, {Schwalm}, {Abid}, {Fabinsky}, {Naes}, \& {Tsai}}]{WISE}
{Wright}, E.~L., {Eisenhardt}, P.~R.~M., {Mainzer}, A.~K., {et~al.} 2010, \aj,
  140, 1868

\bibitem[{{Zahnle} \& {Marley}(2014)}]{2014ApJ...797...41Z}
{Zahnle}, K.~J. \& {Marley}, M.~S. 2014, \apj, 797, 41

\end{thebibliography}

\begin{appendix}

  \section{Alternative colour-absolute magnitude diagrams}
\label{appendix:altCAMDs}

\begin{figure*}[ht]
  \centering
  \includegraphics[width=\textwidth]{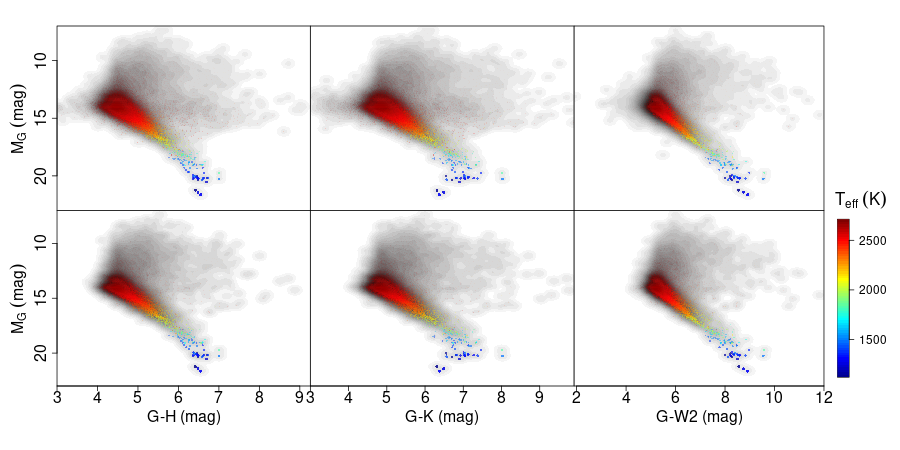}
  \caption{ As Fig. \ref{fig:camds-dense} for additional colour
    indices including the \gaia \gmag\ band magnitude and 2MASS $H$,
    $K$ or AllWISE $W_2$. A kernel density estimate is shown using a
    grey scale.}
  \label{fig:camds-dense2}
\end{figure*}

\begin{figure*}[ht]
  \centering
  \includegraphics[width=\textwidth]{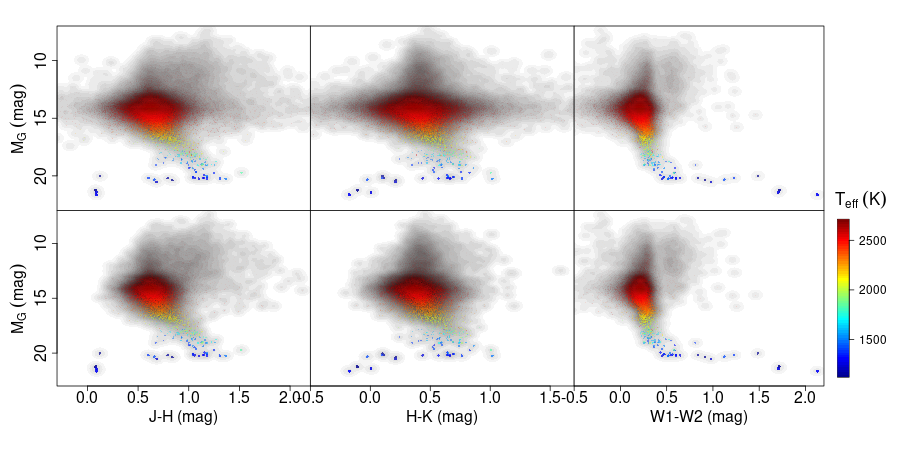}
  \caption{As Fig. \ref{fig:camds-dense} for additional colour indices
    excluding the \gaia\ \gmag\ band magnitude. A kernel density
    estimate is shown using a grey scale.}
  \label{fig:camds-dense3}
\end{figure*}

 This Sect. includes two additional colour-absolute magnitude diagrams
 to provide a complementary view of Fig. \ref{fig:camds-dense}. In
 Fig. \ref{fig:camds-dense2} we show in the $x$ axis alternative
 colour indices constructed from the $G$ band and the 2MASS $H$ and
 $K$ bands (instead of $J$; left and middle plots), and the WISE $W_2$
 band (instead of $W_1$; right plot). In Fig. \ref{fig:camds-dense3}
 we show colour indices combining 2MASS and WISE bands but not the
 \gaia $G$ band. In the left and middle panels, the well known elbow
 towards bluer colour indices is visible. This blue turn for the
 latest spectral types is due to the appearance of methane absorption
 bands in the $H$ and $K$ bands at the L/T spectral type transition
 and the silicate clouds transition from above to below the
 photosphere resulting in a $J$ band brightening.  We interpret the
 larger scatter of the $J-H$ and $H-K$ colour indices as due to their
 comparatively short baseline and the relative faintness of the
 UCDs. Furthermore, these colours are in the peak of energy
 distribution for the hotter UCDs and the behaviour of water vapour
 absorption across the $H$-band can produce additional scatter
 \citep[see for example][]{2022A&A...657A.129A}, in contrast to the
 relative insensitivity of bands like $W1$ and $W2$ which are on the
 Rayleigh-Jeans tail of the energy distribution.

\FloatBarrier

\section{Details of the probabilistic model}\label{appendix:ModelConsiderations}

Inference with the probabilistic model described in Sect. \ref{sec:LF}
was carried out using Tensorflow libraries
\citep{tensorflow2015-whitepaper, TensorflowProbability}. These
libraries adopt the \emph{single-program multiple-data} paradigm
(SPMD). SPMD allows abstractions for scaling the code to
configurations such as TPU (Tensor Processing Unit) pods or 
clusters of GPUs (Graphics Processing Units) that allowed us to
compute the likelihood in a distributed computing framework. This was
required due to the complexity of the hierarchical Bayesian model and
the large number of parameters and observations.

Posterior distributions have been characterised by drawing samples
using the \emph{No U-Turn Sampler} (NUTS) algorithm which is an
adaptive variant of the Hamiltonian Monte Carlo (HMC) method for MCMC
\citep{hoffman2014no}.

\subsection{The modified PERT distribution}\label{appendix:PERT}

Several of the prior probabilities for parameters of the model
described in Sect. \ref{sec:LF} were defined as a modified PERT
distribution. It is defined by four parameters, usually referred to as
$low$, $high$, $peak$ and $temperature$. It has a compact support in
an interval between the $low$ and $high$ values and a most probable
value at the $peak$ which usually indicates the expert's most frequent
prediction. The $temperature$ parameter controls the sharpness of the
peak. Specifically, a modified PERT distribution has the shape:
$${\rm PERT} ({\rm low}, {\rm peak}, {\rm high}, {\rm temperature}) \equiv \, \mu + \sigma \cdot {\rm Beta}(\alpha, \beta), $$
\noindent where
\begin{eqnarray} \label{eq:PERT_alpha_beta}
\mu &=& {\rm low}, \nonumber \\ 
\sigma &=& {\rm high - low}, \nonumber \\                    
\alpha &=& 1 + {\rm temperature} \cdot \cfrac{\rm peak - low}{\rm high - low}, \\
\beta &=& 1 + {\rm temperature} \cdot \cfrac{\rm high - peak}{\rm high - low}, \nonumber
\end{eqnarray}
\noindent with ${\rm temperature} > 0$. 

\begin{figure}[ht]
  \centering
  \includegraphics[width=\linewidth]{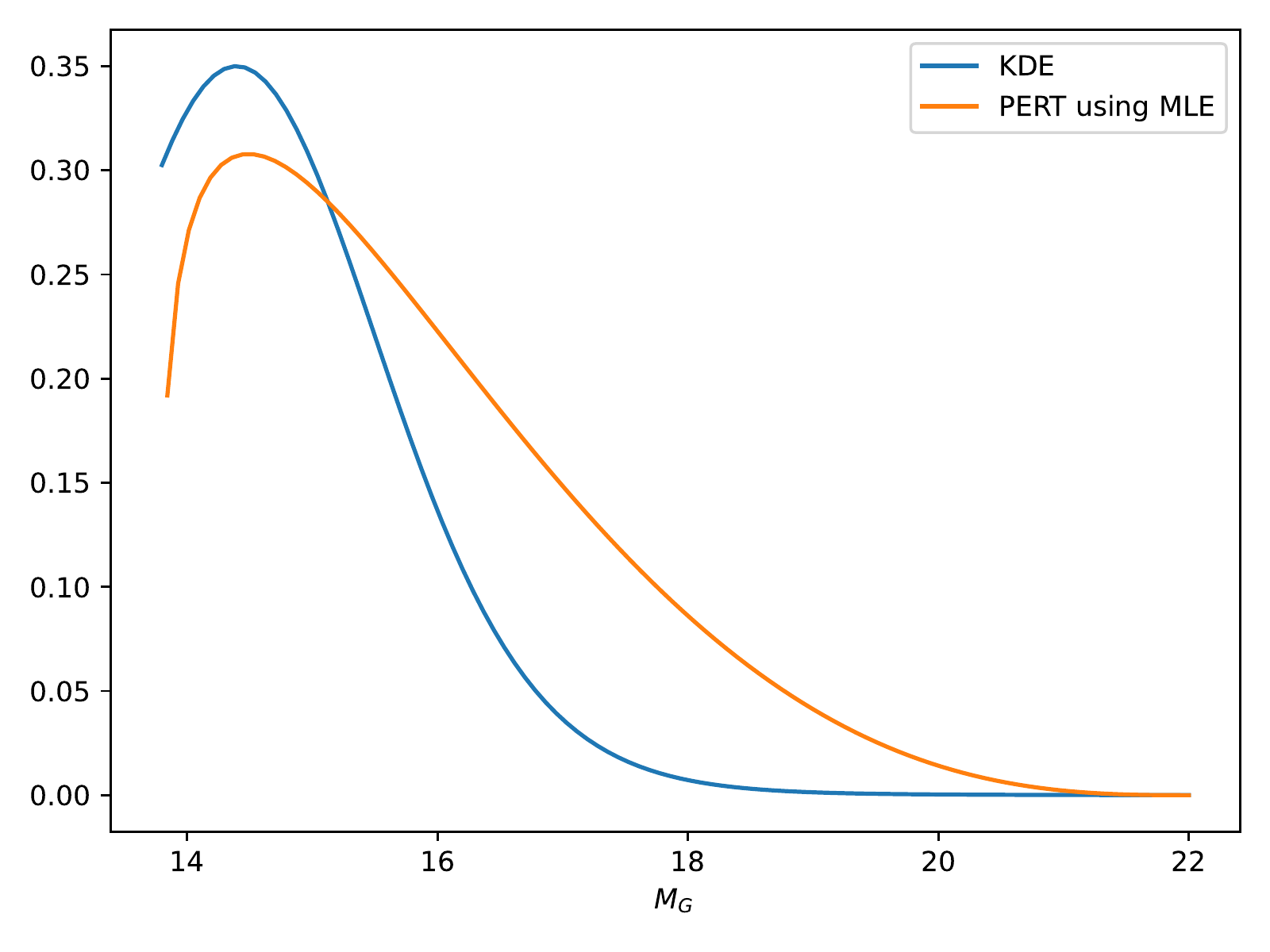}
  \caption{Modified PERT distribution for the $M_G$ prior and the
    kernel density estimate of the observed $M_G$ distributions
    derived using the naive inversion of the parallax.}
  \label{fig:PERT_MG_02}
\end{figure}

\subsection{Approximate Gaussian Process}\label{appendix:GP}

In this paper we will use the approximation proposed
by~\cite{Higdon2020}, for constructing a Gaussian process $f(x)$ over
a general region $x \in \mathcal{X}$ by convolving a continuous white
noise process $\beta (x)$ with a smoothing kernel $k(x)$ so that
\begin{equation}\label{eq:GP_convolved}
f(x) = \int_{\mathcal{X}} k(u-x) \beta(u) \, \text{d} u,
\text{ for } x \in \mathcal{X}.    
\end{equation}
In~\cite{Higdon2020} this integral is approximated for a basic spatial
model. In brief, let $y_1, y_2, \dots y_n$ be data recorded over
spatial locations $x_1, x_2, \dots x_n$ in $\mathcal{X}$, and consider
a simple spatial model such as
\begin{equation}\label{eq:simple_spatial_model}
y = \mu + z + \epsilon,    
\end{equation}
\noindent where the elements of $z = (z_1, z_2, \dots z_n)^T$ are the
restriction of the Gaussian process $f(x)$ to the data locations $x_1,
x_2, \dots, x_n$. In this case, the Gaussian process $f(x)$ has zero
mean and, instead of being defined by its covariance function, it is
determined by the latent process $\beta(x)$ and the smoothing kernel
$k(x)$.

The latent process $\beta(x)$ is restricted to be nonzero at spatial
locations $\omega_1, \omega_2, \dots \omega_m$, and in $\mathcal{X}$,
and we consider $\beta_j = \beta(\omega_j), \, j=1, \dots, m$.  Each
$\beta_j$ is modelled as an independent draw from a $\mathcal{N}(0,
\sigma_x)$ distribution. Given Eq.  ~\ref{eq:GP_convolved} the
continuous Gaussian process is approximated as
\begin{equation}\label{eq:GP_approx}
    f(x) = \sum_{j=1}^m \beta_j k(x-\omega_j),
\end{equation}
\noindent where $k(\cdot - \omega_j)$ is a kernel centred at
$\omega_j$. In this paper we choose $k(\cdot)$ to be a Gaussian
density which implies a smooth radially symmetric kernel. Therefore,
equation \ref{eq:simple_spatial_model} results in the linear model
\begin{equation}
    y = \mu\bm{1}_n + K \bm{\beta} + \epsilon
\end{equation}
\noindent where $\bm{1}_n$ is the $n$-vector of 1's, and the elements
of $K$ are given by
\begin{eqnarray*}
K_{ij} &=&  k(x_i-\omega_j) \\
\bm{\beta} & \sim & \mathcal{N} (0, \sigma_x I_m)  \\
\bm{\epsilon} & \sim & \mathcal{N} (0, \sigma_{\epsilon} I_n).
\end{eqnarray*}
From a statistical point of view this model is a basic mixed effects
model that can be included as part of our Bayesian model.

\FloatBarrier

\newpage
\section{Stellar associations complementary material}\label{appendix:clusters}

In this Sect. we include additional material with details about the
cross match between HMAC and BANYAN groups
(Sect. \ref{appendix:clusters-cm}) and several diagrams (sky
positions, CAMD, tangential velocities and histograms of the inverse
of the parallax) showing the properties of the HMAC clusters without
members in common with the BANYAN groups in
Sect. \ref{appendix:clusters-hmaconly}.

\subsection{Intersection between HMAC and BANYAN clusters}
\label{appendix:clusters-cm}

Table \ref{appendix:clusters-cm} shows the number of sources in common
between BANYAN and HMAC groups. The leftmost column includes numerical
identifiers of the HMAC clusters.

\begin{table*}

\setlength{\tabcolsep}{3pt}
    \caption{\label{tab:clusters-table} Number of sources in common
      between the BANYAN groups (top row) and the clusters identifiers
      obtained using HMAC (leftmost column).}
    \begin{tabular}{
    ccccccccccccccccccccccccccc}
        \hline\hline
&ARG&CRA&EPSC&IC2391&IC2602&LCC&OCT&PL8&PLE&ROPH&TAU&UCL&UCRA&USCO\\
\hline
2&0&0&0&0&0&0&0&0&0&59&0&117&0&469\\
7&0&0&0&0&0&2&0&0&0&0&0&139&0&2\\
8&0&0&0&0&0&0&0&0&0&0&0&103&0&0\\
9&0&0&0&0&0&0&17&0&0&0&0&0&0&0\\
10&0&0&5&0&0&106&0&0&0&0&0&1&0&0\\
11&0&0&0&0&0&33&0&0&0&0&0&68&0&0\\
15&0&0&0&0&0&0&0&0&70&0&0&0&0&0\\
18&0&0&0&0&0&64&0&0&0&0&0&1&0&0\\
19&0&7&0&0&0&0&0&0&0&0&0&0&44&0\\
20&17&0&0&19&0&0&0&0&0&0&0&0&0&0\\
21&0&0&0&0&0&0&0&0&0&0&52&0&0&0\\
22&0&0&0&0&0&1&0&16&0&0&0&0&0&0\\
24&0&0&0&0&10&0&0&2&0&0&0&0&0&0\\
25&0&0&0&0&0&0&0&0&0&0&17&0&0&0\\
35&0&0&0&0&0&0&0&0&0&0&0&12&0&0\\
46&0&0&0&0&2&0&0&0&0&0&0&0&0&0\\
47&4&0&0&0&0&0&0&0&0&0&0&0&0&0\\
58&0&0&0&0&0&0&0&0&0&0&5&0&0&0\\
80&0&0&0&0&0&0&0&0&0&0&0&4&0&0\\

        \hline   
    \end{tabular}
\end{table*}

\subsection{HMAC clusters without members in any of the BANYAN groups}
\label{appendix:clusters-hmaconly}

\begin{figure*}[th]
  \centering
  \includegraphics[scale=0.30]{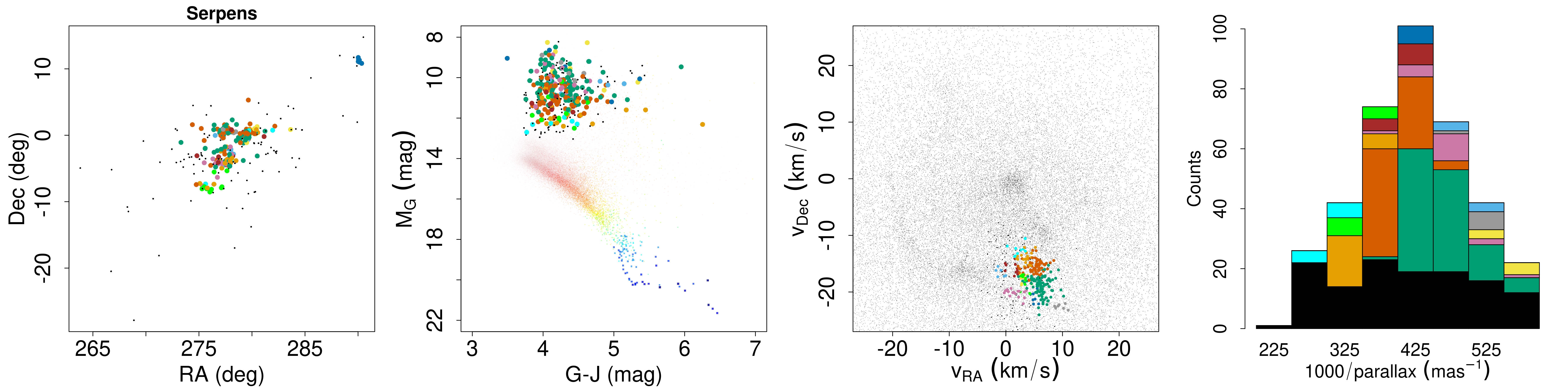}
  \caption{Properties of one (level 4) HMAC cluster tentatively
    identified with Serpens (but see text) with no sources in common
    with BANYAN associations. From left to right: equatorial
    coordinates with various colours separating subclusters at a
    hierarchy level lower (and hence finer) than 4; CAMD (points
    represent the full set of UCD candidates in the catalogue and
    circles identify the HMAC group at level 4; black dots represent
    sources not assigned to any subcluster in the lower level);
    tangential velocities; and the stacked histogram of the inverse of
    the parallax in all subclusters.}
  \label{fig:hmac-only-610}
\end{figure*}

\begin{figure*}[th]
  \centering
  \includegraphics[scale=0.3]{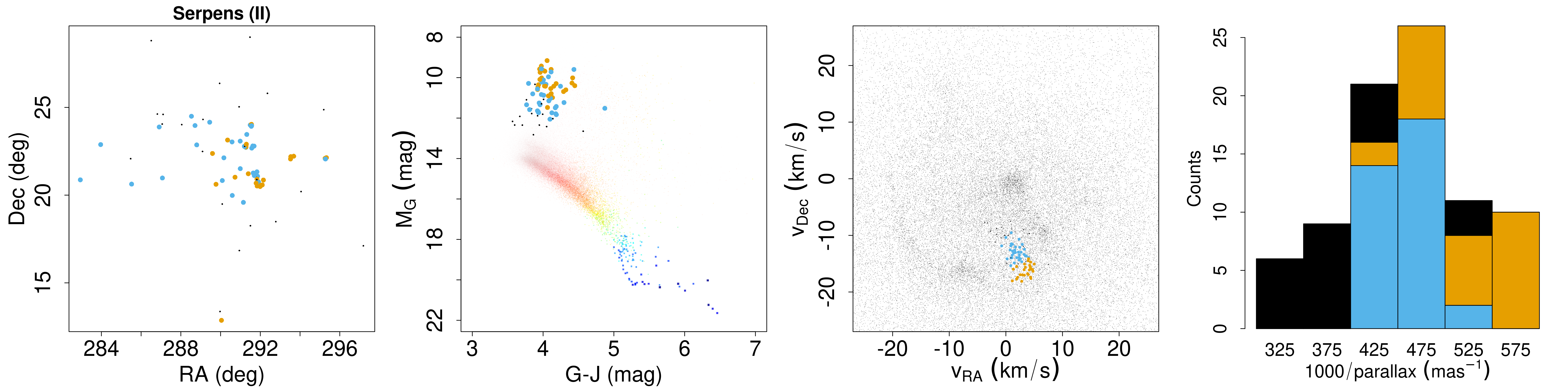}
  \caption{As Figure \ref{fig:hmac-only-610} but for Serpens (II)}
  \label{fig:hmac-only-9430}
\end{figure*}

\begin{figure*}[th]
  \centering
  \includegraphics[scale=0.3]{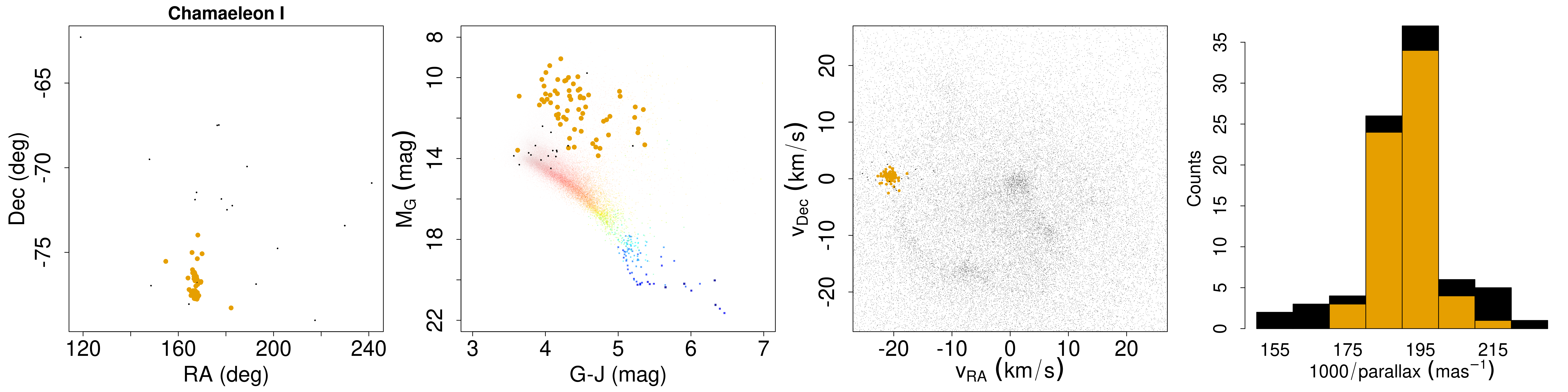}
  \caption{As Figure \ref{fig:hmac-only-610} but for Chamaleon {\sc I}.}
  \label{fig:hmac-only-774}
\end{figure*}

\begin{figure*}[th]
  \centering
  \includegraphics[scale=0.3]{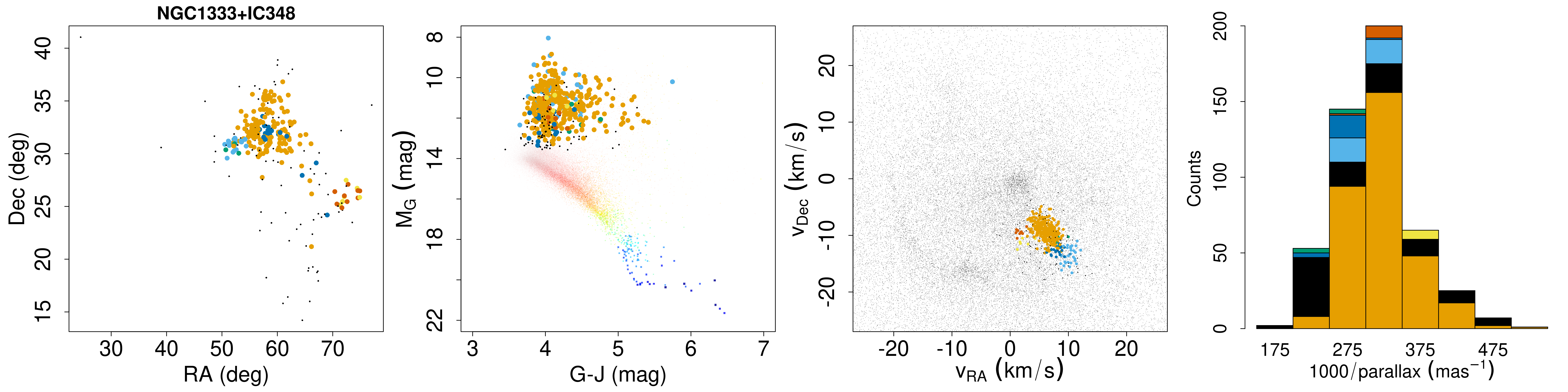}
  \caption{As Figure \ref{fig:hmac-only-610} but for the Perseus region.} 
  \label{fig:hmac-only-795}
\end{figure*}

\begin{figure*}[th]
  \centering
  \includegraphics[scale=0.3]{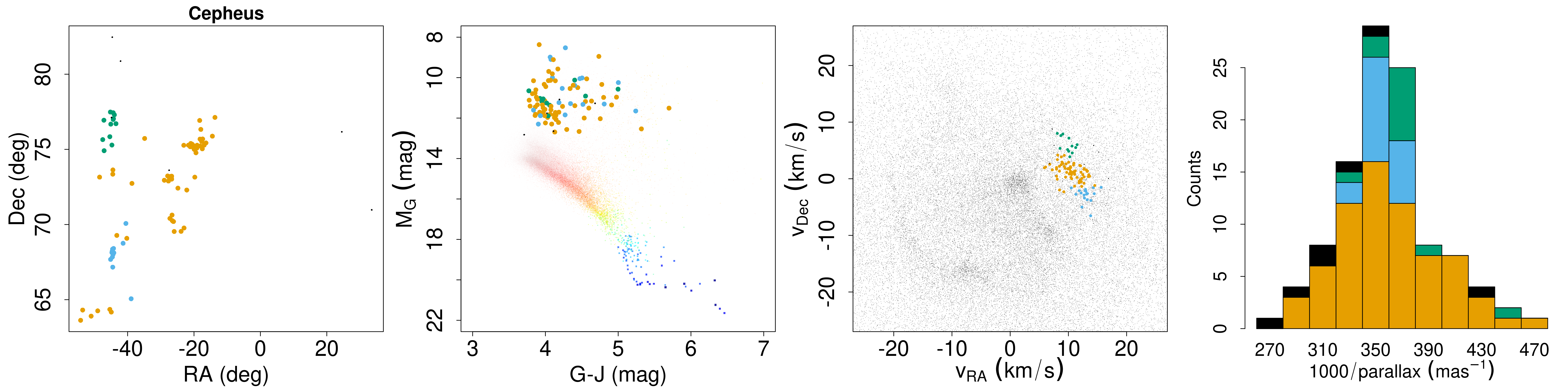}
  \caption{As Figure \ref{fig:hmac-only-610} but for the Cepheus region}
  \label{fig:hmac-only-897}
\end{figure*}

\begin{figure*}[th]
  \centering
  \includegraphics[scale=0.3]{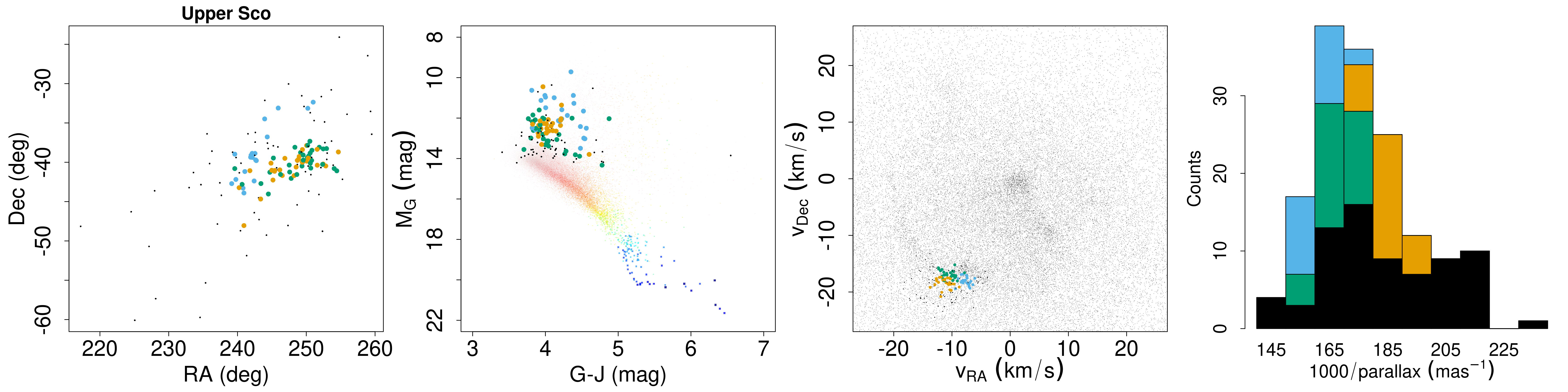}
  \caption{As Figure \ref{fig:hmac-only-610} but for Upper Scorpio}
  \label{fig:hmac-only-932}
\end{figure*}

\begin{figure*}[th]
  \centering
  \includegraphics[scale=0.3]{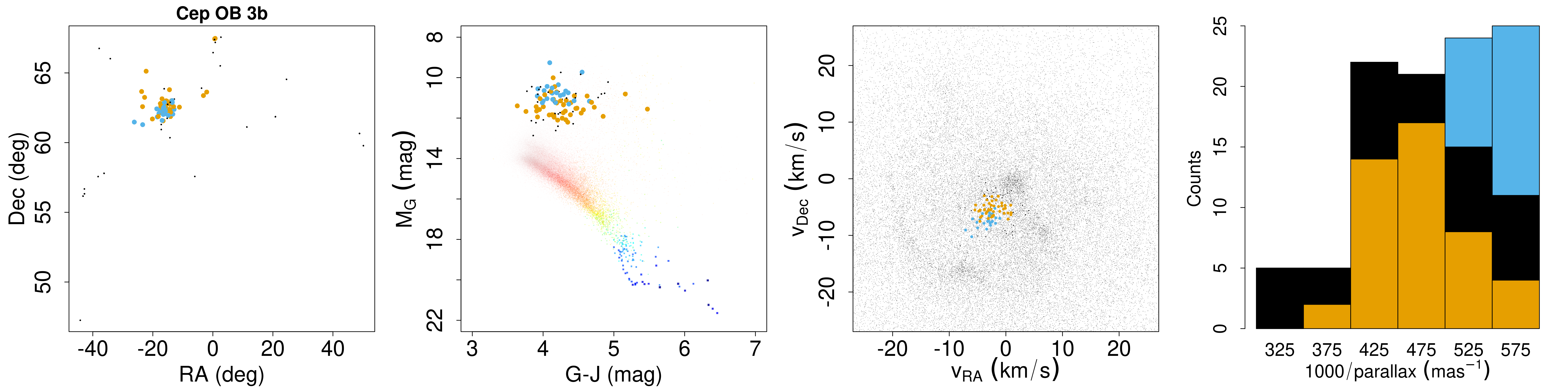}
  \caption{As Figure \ref{fig:hmac-only-610} but for Cep OB 3b}
  \label{fig:hmac-only-1085}
\end{figure*}

\begin{figure*}[th]
  \centering
  \includegraphics[scale=0.3]{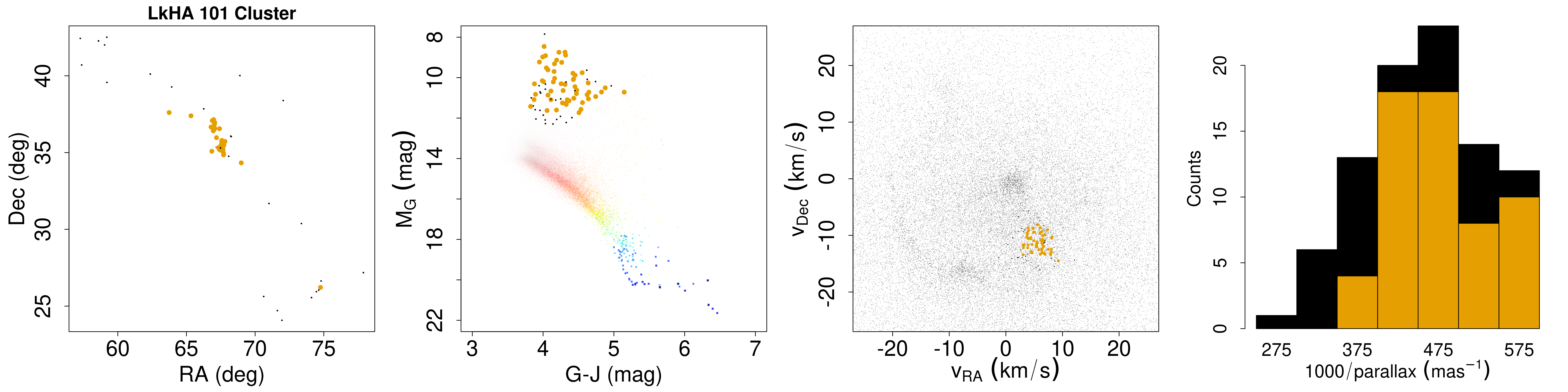}
  \caption{As Figure \ref{fig:hmac-only-610} but for the LkHA 101 Cluster}
  \label{fig:hmac-only-1166}
\end{figure*}

\begin{figure*}[th]
  \centering
  \includegraphics[scale=0.3]{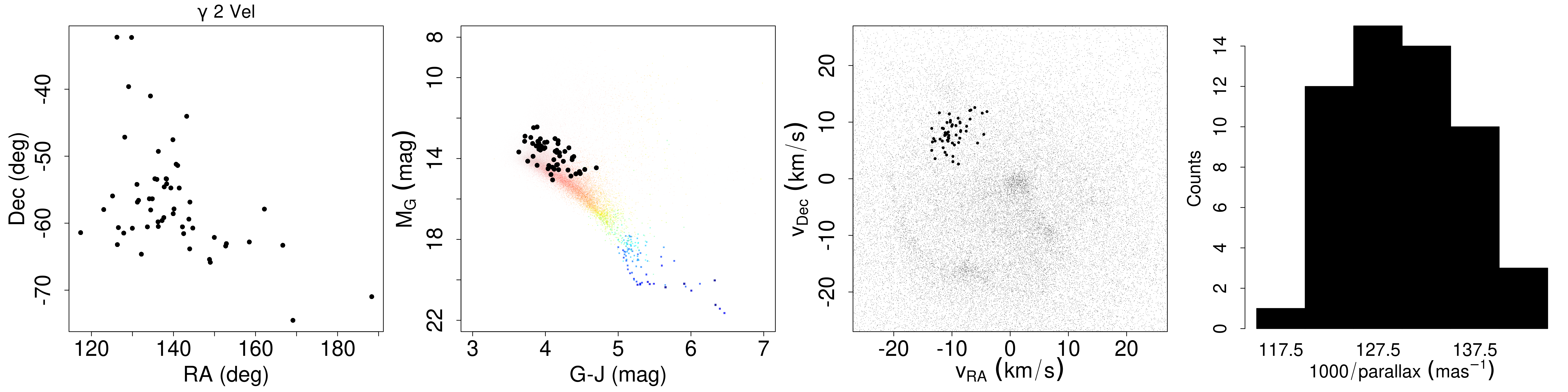}
  \caption{As Figure \ref{fig:hmac-only-610} but for $\gamma$2 Vel (I)}
  \label{fig:hmac-only-1153}
\end{figure*}

\begin{figure*}[th]
  \centering
  \includegraphics[scale=0.3]{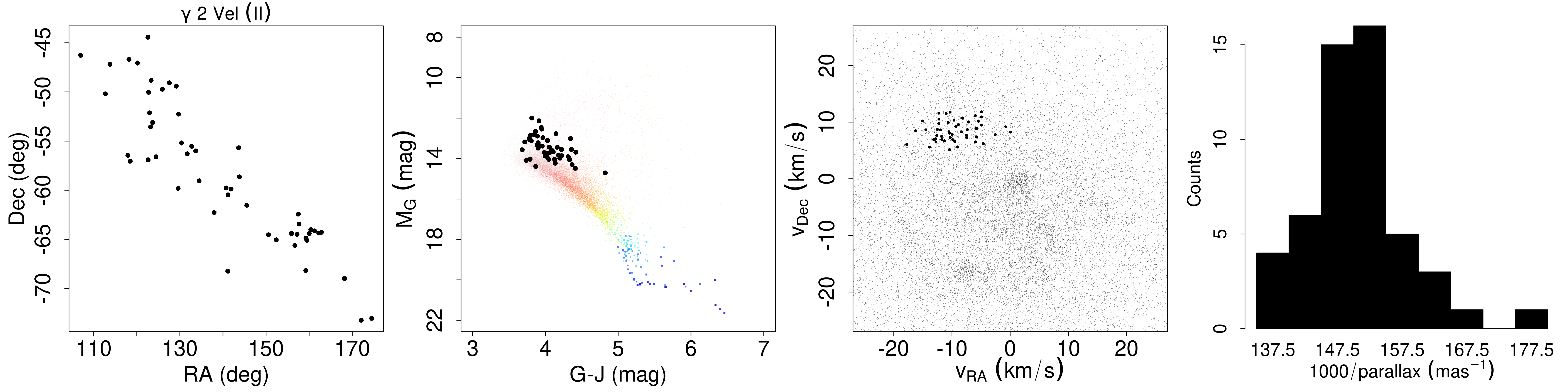}
  \caption{As Figure \ref{fig:hmac-only-610} but for $\gamma$2 Vel (II)}
  \label{fig:hmac-only-2566}
\end{figure*}

\begin{figure*}[th]
  \centering
  \includegraphics[scale=0.3]{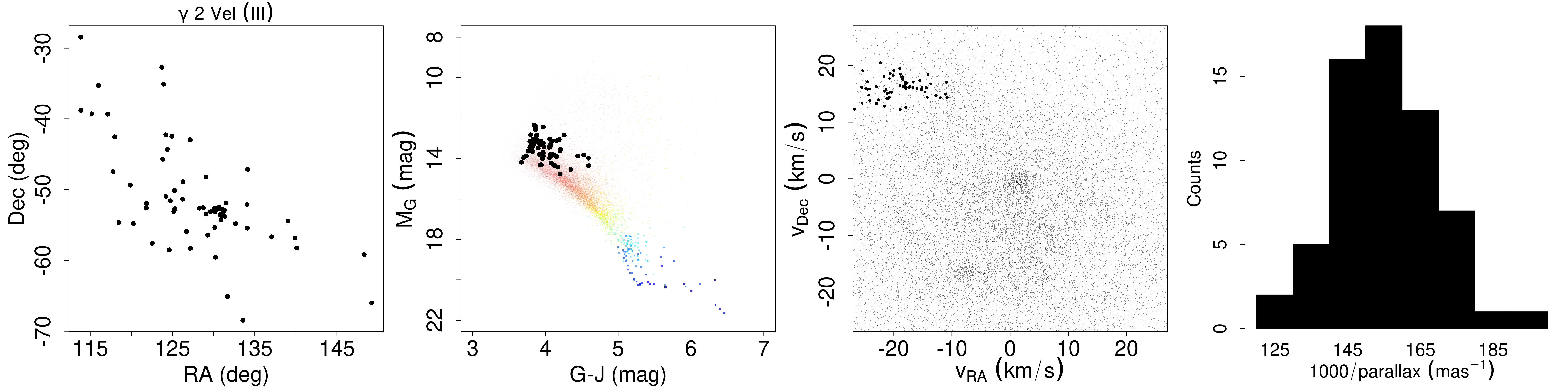}
  \caption{As Figure \ref{fig:hmac-only-610} but for $\gamma$2 Vel (III)}
  \label{fig:hmac-only-3564}
\end{figure*}

\begin{figure*}[th]
  \centering
  \includegraphics[scale=0.3]{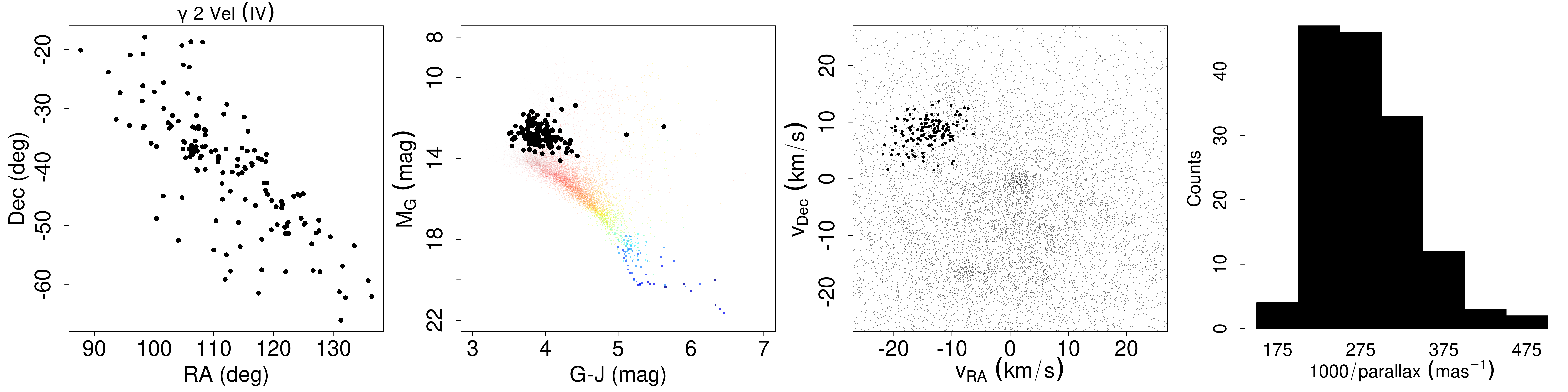}
  \caption{As Figure \ref{fig:hmac-only-610} but for $\gamma$2 Vel (IV)}
  \label{fig:hmac-only-4565}
\end{figure*}

\begin{figure*}[th]
  \centering
  \includegraphics[scale=0.3]{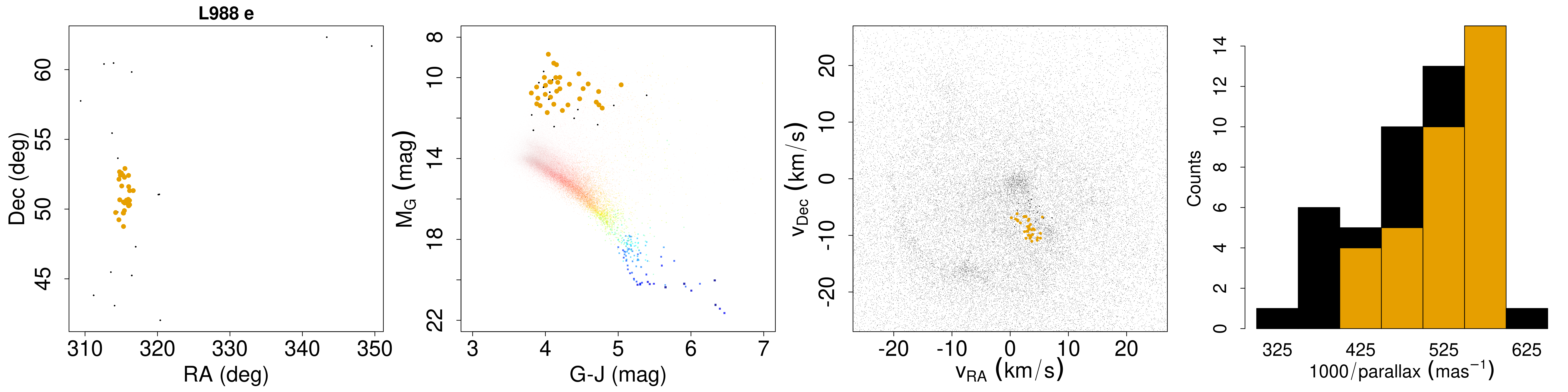}
  \caption{As Figure \ref{fig:hmac-only-610} but for L988 e}
  \label{fig:hmac-only-1561}
\end{figure*}

\begin{figure*}[th]
  \centering
  \includegraphics[scale=0.3]{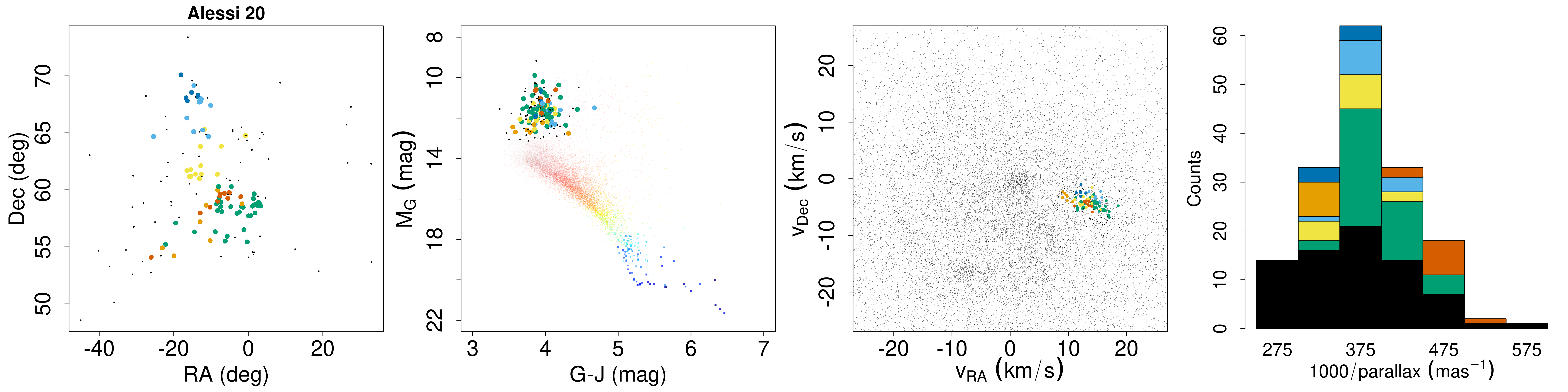}
  \caption{As Figure \ref{fig:hmac-only-610} but for the region near Alessi 20}
  \label{fig:hmac-only-8984}
\end{figure*}

\begin{figure*}[th]
  \centering
  \includegraphics[scale=0.3]{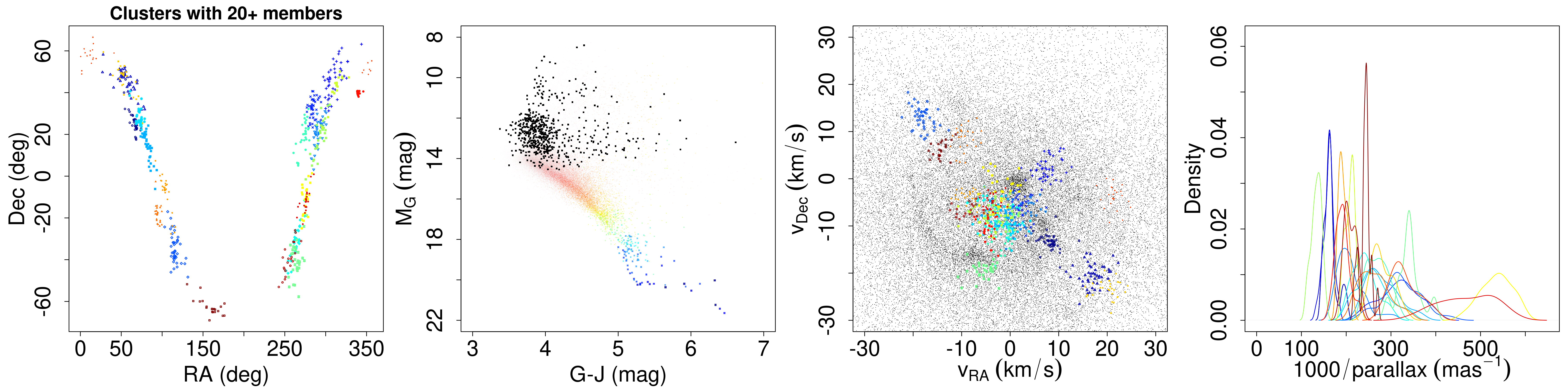}
  \caption{As Figure \ref{fig:hmac-only-610} for the rest of clusters
    with more than 20 members. We use different colours for the
    various clusters except for the CAMD to avoid confusion with the
    colour code reflecting \teff along the Main Sequence.}
  \label{fig:hmac-only-restGT20}
\end{figure*}

Figures \ref{fig:hmac-only-610}--\ref{fig:hmac-only-restGT20} show the
distribution in equatorial coordinates (left), the CAMD (middle left),
the tangential velocities (middle right), and the histogram of the
inverse of the parallax (right) for several HMAC clusters without
members in any BANYAN groups. Tentative identifications are included
in the top of the left plot and in the captions.

\subsection{Residual overdensities.}

\begin{figure}[th]
  \centering
  \includegraphics[scale=0.5]{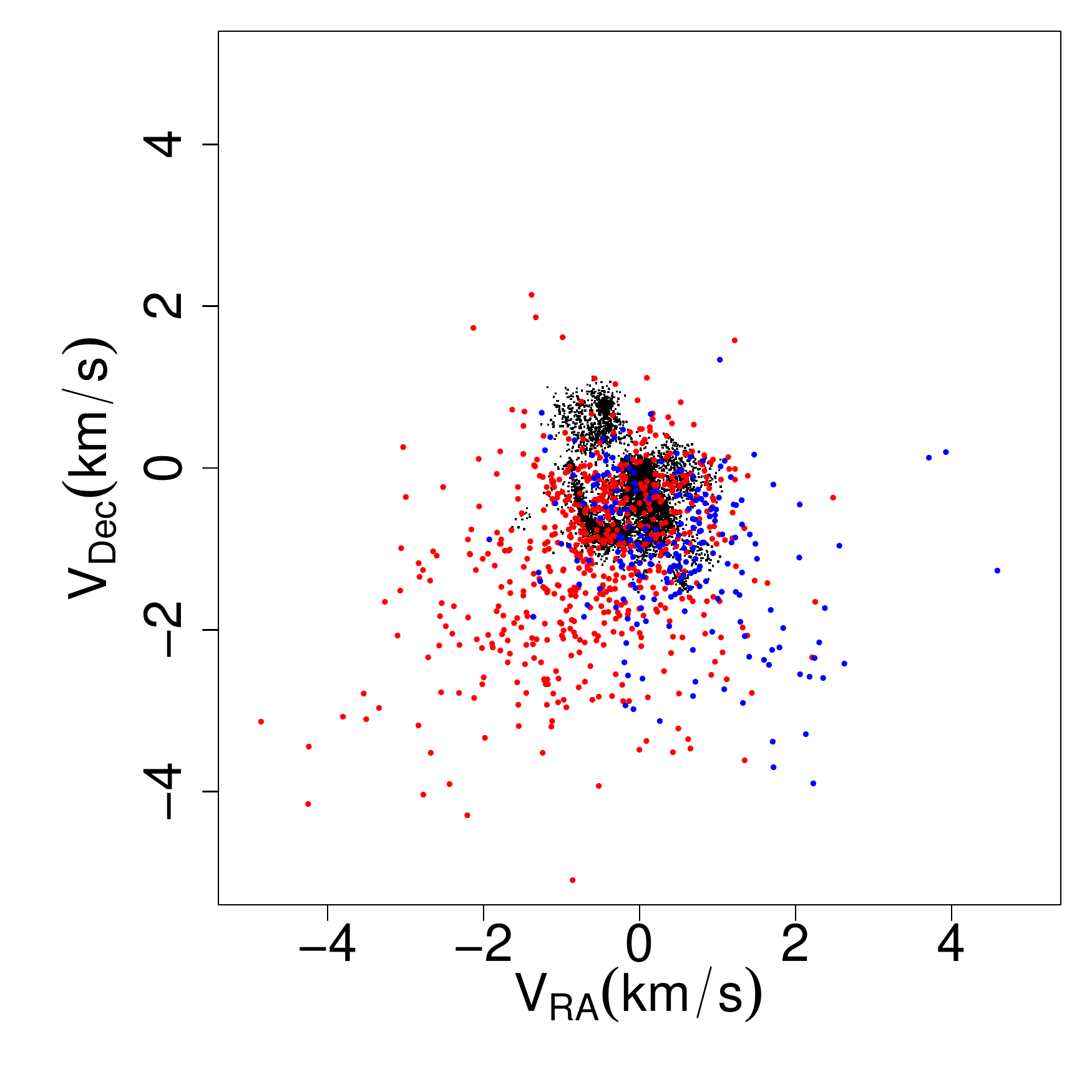}
  \caption{Distribution in the space of tangential velocities of the
    sources in residual overdensities shown in the lower right panel
    of Figure \ref{fig:lb} using the same colour code.}
  \label{fig:resClusterstanvel}
\end{figure}

\begin{figure}[th]
  \centering
  \includegraphics[scale=0.5]{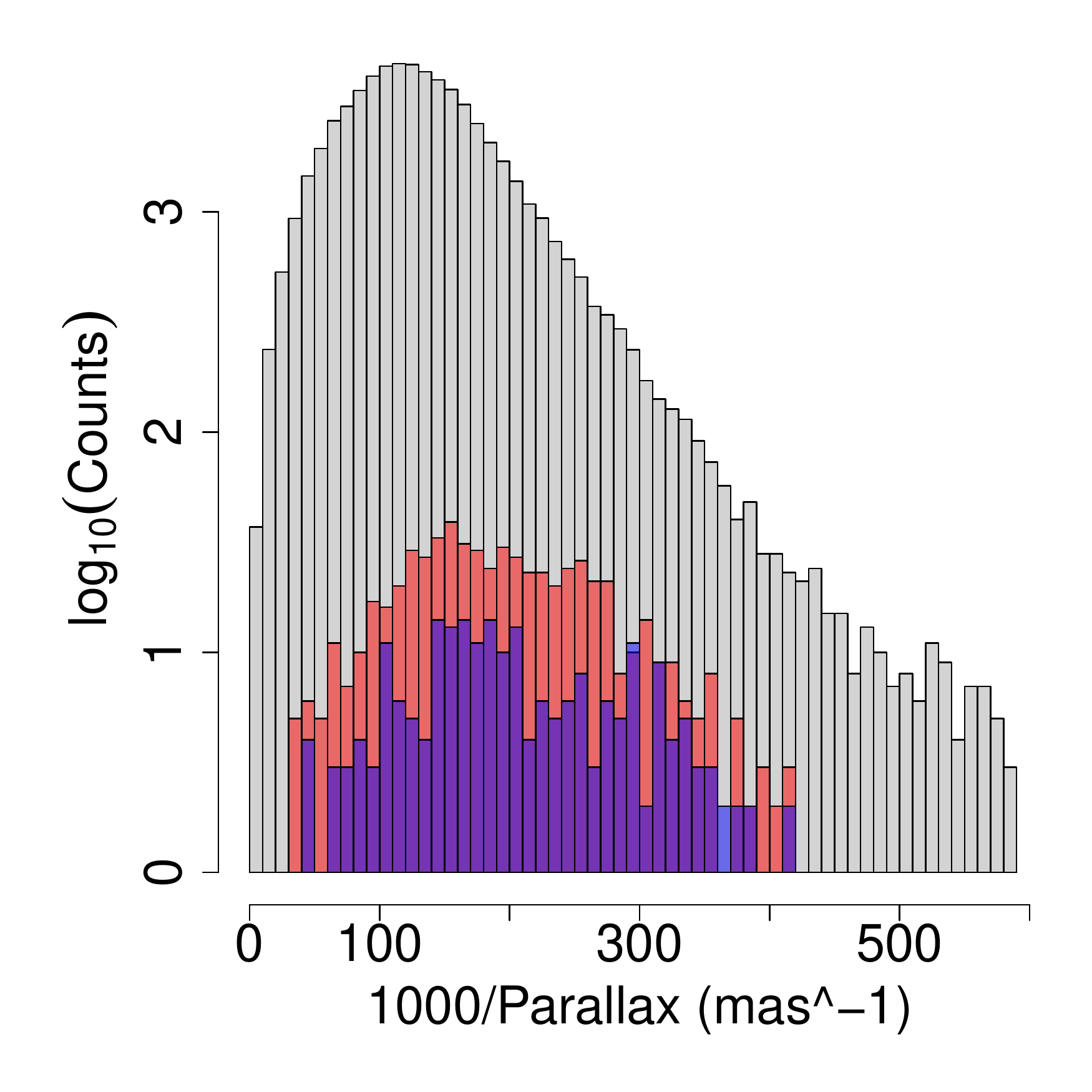}
  \caption{Histogram of the inverse of the parallax of candidate UCDs
    (white) and of sources in the residual overdensities marked in red
    and blue in the lower right panel of Figure \ref{fig:lb} using the
    same colour code.}
  \label{fig:resClustersParallaxes}
\end{figure}

\begin{figure}[th]
  \centering
  \includegraphics[scale=0.5]{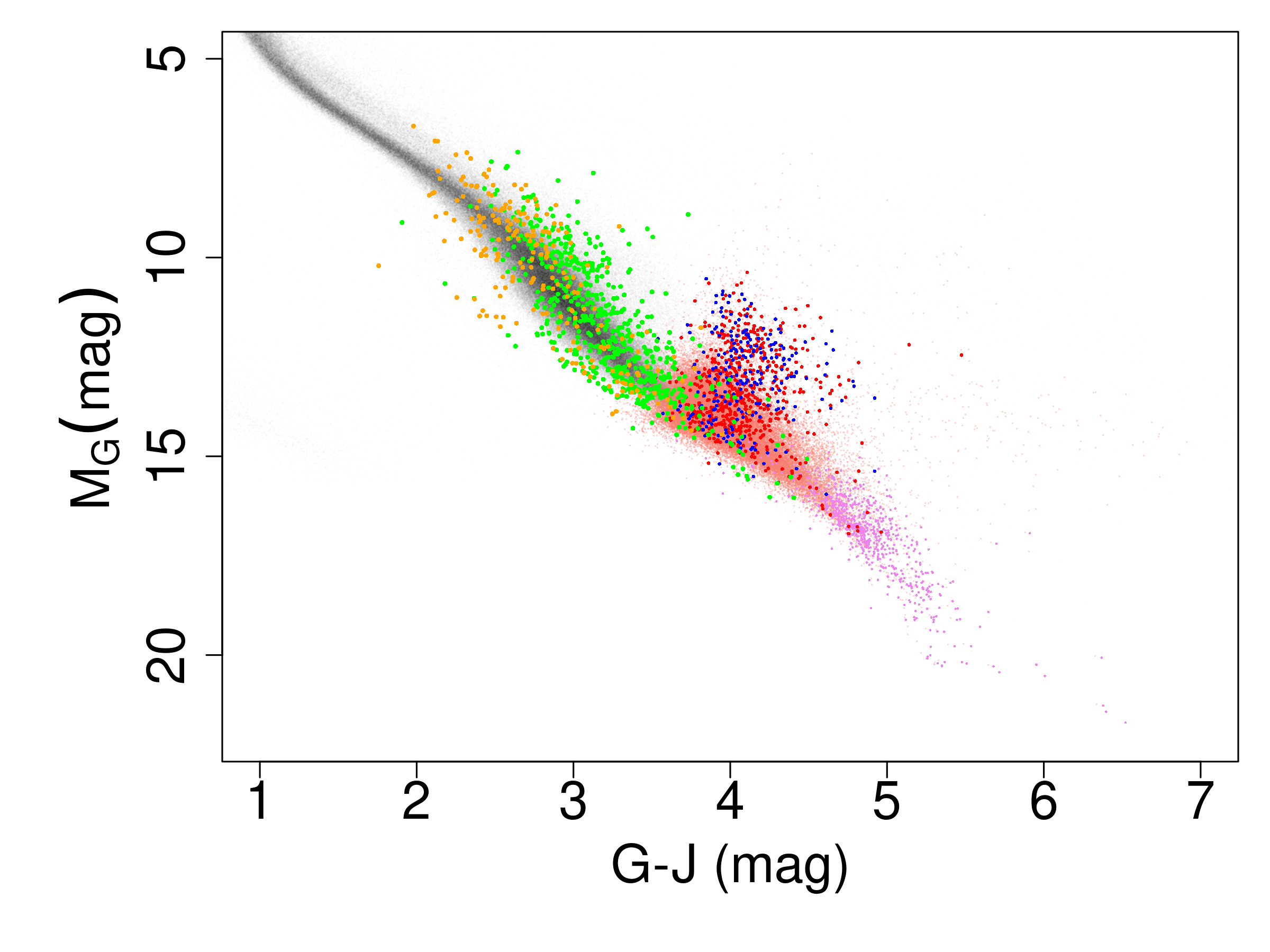}
  \caption{Distribution of sources in the residual overdensities
    marked in red and blue in the lower right panel of Figure
    \ref{fig:lb} in the CAMD diagram of $(G-J)$ and $M_G$. It shows
    the Main Sequence derived from the Gaia Catalogue of Nearby Stars
    as black dots (using transparency); the list of sources in our
    catalogue of UCDs in HMAC groups of less than 10 members (salmon
    dots with transparency); the GUCDS (violet dots); and the position
    of the sources in the blue and red overdensities after dereddening
    using the Planck GNILC \citep{2016A&A...596A.109P} dust map as
    orange and green dots respectively.}
  \label{fig:resClustersCAMD}
\end{figure}

Figures \ref{fig:resClusterstanvel}--\ref{fig:resClustersCAMD} include
additional diagnostic plots for the interpretation of the residual
overdensities shown in Fig. \ref{fig:lb}.

\end{appendix}

\end{document}